\RequirePackage{fix-cm}
\documentclass[smallextended]{svjour3}       
\smartqed  
\usepackage{graphicx}
\usepackage{natbib}
\bibpunct{(}{)}{;}{a}{}{,}
\newcommand{\xray}{\hbox{X-ray}}
\newcommand{\xrays}{\hbox{X-rays}}
\newcommand{\soltan}{So\l tan}

\newcommand{\asca}{{\it ASCA\/}}
\newcommand{\astroh}{{\it ASTRO-H\/}}
\newcommand{\athena}{{\it Athena\/}}
\newcommand{\chandra}{{\it Chandra\/}}
\newcommand{\erosita}{{\it eROSITA\/}}
\newcommand{\herschel}{{\it Herschel\/}}
\newcommand{\hexp}{{\it HEX-P\/}}
\newcommand{\hst}{{\it HST\/}}
\newcommand{\integral}{{\it INTEGRAL\/}}
\newcommand{\nustar}{{\it NuSTAR\/}}
\newcommand{\rosat}{{\it ROSAT\/}}

\newcommand{\sax}{{\it BeppoSAX\/}}
\newcommand{\smartx}{{\it SMART-X\/}}
\newcommand{\spitzer}{{\it Spitzer\/}}
\newcommand{\swift}{{\it Swift\/}}
\newcommand{\wfxt}{{\it WFXT\/}}
\newcommand{\xmm}{{\it XMM-Newton\/}}

\newcommand{\simgt}{\lower 2pt \hbox{$\, \buildrel {\scriptstyle >}\over {\scriptstyle \sim}\,$}}
\newcommand{\simlt}{\lower 2pt \hbox{$\, \buildrel {\scriptstyle <}\over {\scriptstyle \sim}\,$}}
 
\newcommand{\aox}{$\alpha_{\rm ox}$}

\begin{document}

\title{Cosmic X-ray surveys of distant active galaxies
}
\subtitle{The demographics, physics, and ecology of growing supermassive black holes}


\author{W.N. Brandt         \and
        D.M. Alexander 
}


\institute{W.N. Brandt \at
           Department of Astronomy and Astrophysics and the Institute for Gravitation and the Cosmos; 
           The Pennsylvania State University; 
           525 Davey Lab; 
           University Park, PA 16802; USA \\
           Tel.: +1 814 865 3509 \\
           \email{niel@astro.psu.edu}          
           \and
           D.M. Alexander \at
           Department of Physics;
           Durham University;
           Durham DH1 3LE; UK \\
           Tel.: +44 191 3343594 \\
           \email{d.m.alexander@durham.ac.uk}}         

\date{Received: date / Accepted: date}

\maketitle

\begin{abstract}
We review results from cosmic \xray\ surveys of active galactic nuclei (AGNs) over 
the past $\approx 15$~yr that have dramatically improved our understanding of 
growing supermassive black holes in the distant universe. First, 
we discuss the utility of such surveys for AGN investigations and the capabilities 
of the missions making these surveys, emphasizing \chandra, \xmm, and \nustar. 
Second, we briefly describe the main cosmic \xray\ surveys, the essential roles 
of complementary multiwavelength data, and how AGNs are selected from these 
surveys. We then review key results from these surveys on the AGN population 
and its evolution (``demographics''), the physical processes operating in AGNs 
(``physics''), and the interactions between AGNs and their environments 
(``ecology''). We conclude by describing some significant unresolved questions 
and prospects for advancing the field. 
\end{abstract}



\section{Introduction}
\label{intro}

\subsection{General utility of X-ray surveys for studies of active galactic nuclei}
\label{generalutility}

Cosmic \xray\ surveys have now achieved sufficient sensitivity and sky coverage
to allow the study of many distant source populations including 
active galactic nuclei (AGNs), 
starburst galaxies, 
normal galaxies, 
galaxy clusters, and 
galaxy groups. 
Among these, AGNs, representing actively growing supermassive black holes 
(SMBHs), dominate the source number counts as well as the received integrated 
\xray\ power. This has led to an impressive literature on the demographics, 
physics, and ecology of distant growing SMBHs found in \xray\ surveys.

The intrinsic \xray\ emission from AGNs largely originates in the 
immediate vicinity of the SMBH. The \xray\ continuum arises via Compton 
up-scattering in an accretion-disk ``corona'' over a broad \xray\ band, 
and also perhaps via accretion-disk emission at low \xray\ energies
\citep[e.g.,][]{mushotzky1993,reynolds2003,fabian2006,turner2009,done2010,gilfanov2014}.  
AGNs hosting powerful jets furthermore often show strong 
jet-linked \xray\ continuum emission
\citep[e.g.,][]{worrall2009,miller2011}.  
This intrinsic \xray\ emission may then interact with matter throughout 
the nuclear region to produce, via Compton ``reflection'' and scattering, 
more distributed \xray\ emission. 
In some cases, when the intrinsic \xrays\ are obscured, such 
reflected/scattered emission may dominate the observed luminosity. 

\begin{figure}
\includegraphics[scale=0.6,angle=270]{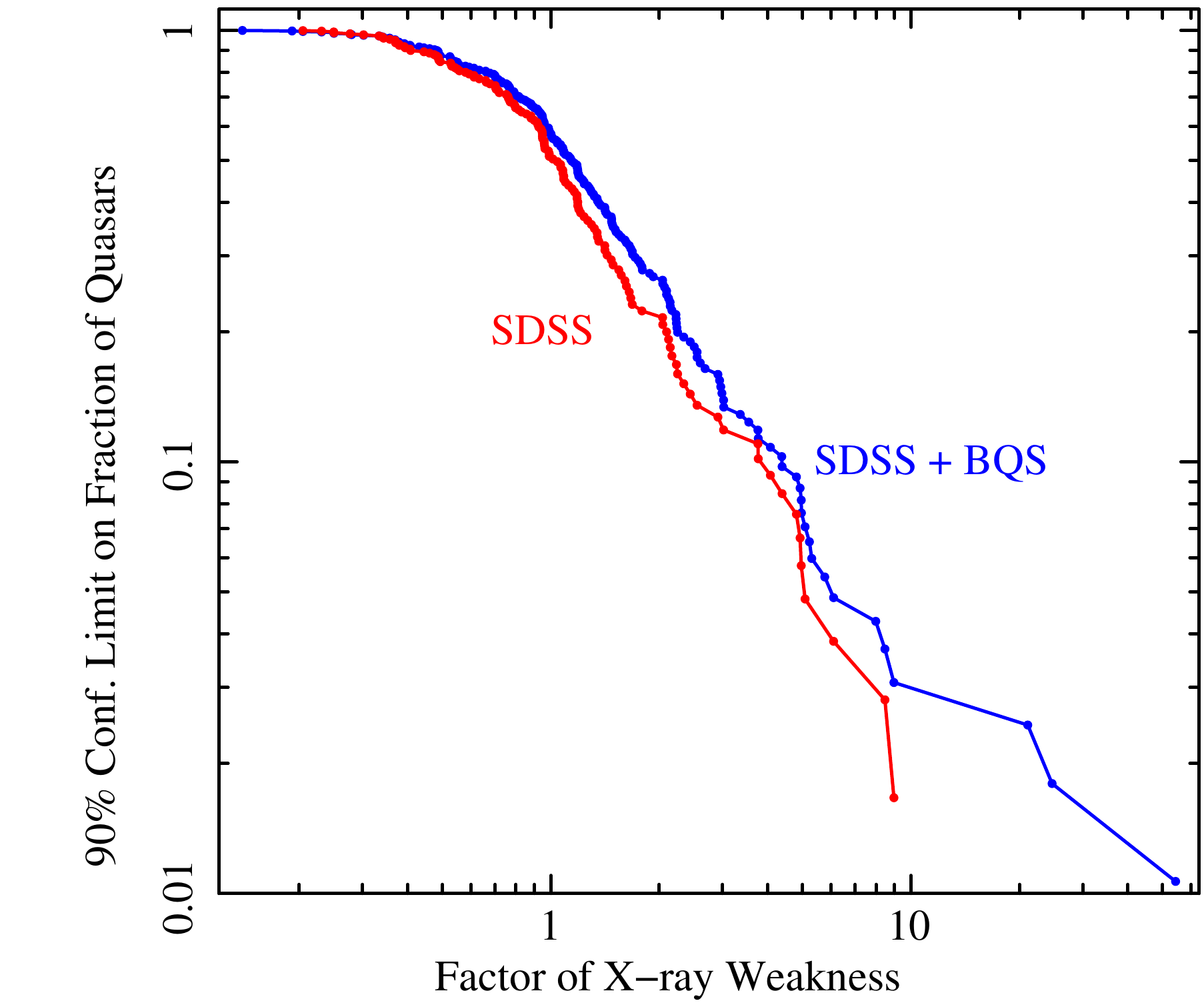}
\caption{Upper limit at 90\% confidence on the fraction of Sloan
Digital Sky Survey (SDSS; red curve) and SDSS+Bright Quasar Survey 
(BQS; blue curve) radio-quiet quasars that are \xray\ weak by a given 
factor. The factor of \xray\ weakness is computed relative to expectations 
based on optical/UV luminosity (see Section~\ref{aox-results}), where a 
value of unity represents the average quasar. Broad Absorption 
Line quasars, which are known often to have heavy \xray\ absorption, 
have been excluded when making this plot. Note that quasars that are
\xray\ weak by a factor of 10 represent $\simlt 3$\% of the population. 
Adapted from \citet{gibson2008}.}
\label{fig-gibson2008-xray-weakness}       
\end{figure}

\begin{figure}
\includegraphics[scale=0.45,angle=0]{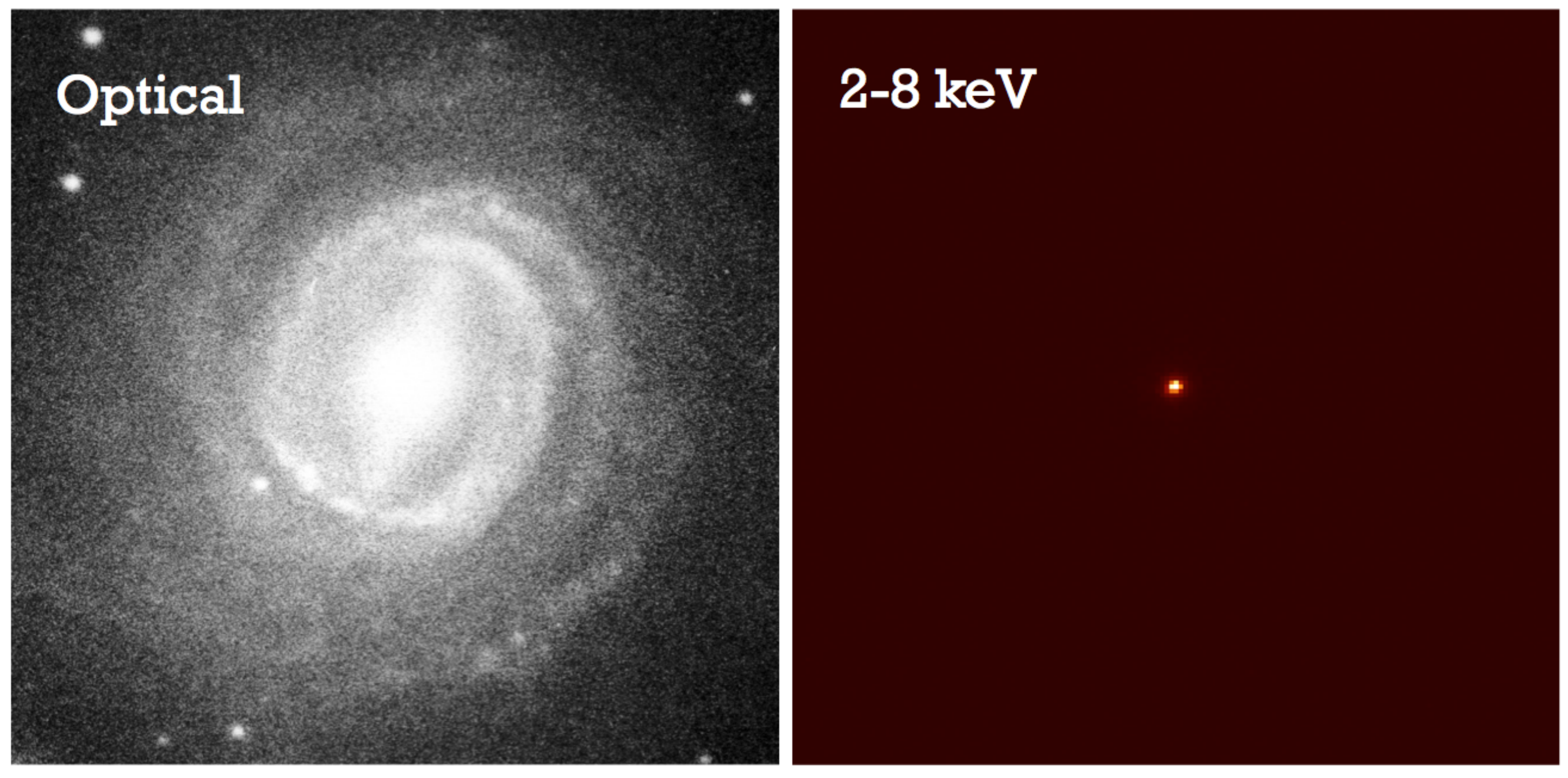}
\caption{Optical and \chandra\ \hbox{2--8~keV} images of a local
active galaxy (NGC~3783); each image is 1.5~arcmin on a side. 
Note the substantial host-galaxy starlight competing 
with the AGN light in the optical band while, in 
the \xray\ band, only the AGN light is apparent. The large 
contrast in the \xray\ band between AGN light and starlight helps 
greatly with the identification of distant AGNs.}
\label{fig-ngc3783-opt-xray}       
\end{figure}

Cosmic \xray\ surveys of AGNs offer considerable utility for several 
reasons: 

\begin{enumerate}

\item
X-ray emission appears to be nearly universal from the 
luminous AGNs that dominate SMBH growth in the Universe. 
When AGNs have been reliably 
identified using optical, infrared, and/or radio 
techniques, they almost always also show \xray\ AGN signatures
(e.g., see Fig.~\ref{fig-gibson2008-xray-weakness} and  
\citealp[][]{avni1986,brandt2000,mushotzky2004,gibson2008}).
Thus, the intrinsic \xray\ emission from the accretion disk and its 
corona empirically appears robust, even if its detailed nature is
only now becoming clear \citep[e.g.,][]{done2010,schnittman2013}.  
This point is discussed further in Section~\ref{agns-missed} and 
Section~\ref{aox-results}. 

\item 
X-ray emission is penetrating with reduced absorption bias. 
The high-energy \xray\ emission observed from AGNs is capable of
directly penetrating through substantial columns with 
hydrogen column densities of 
\hbox{$N_{\rm H}=10^{21}$--10$^{24.5}$}~cm$^{-2}$
\citep[e.g.,][and references therein]{wilms2000}.\footnote{For
purposes of basic comparison, the column density through your
hand is \hbox{$N_{\rm H}\sim 10^{23}$~cm$^{-2}$}, while that through
your chest is \hbox{$N_{\rm H}\sim 10^{24}$~cm$^{-2}$}
(with significant variation depending upon the amount of 
bone intercepted).} 
This is critically important, since the majority of AGNs in the
Universe are now known to be absorbed by such column densities
(see Section~\ref{obscuration}). \xray\ surveys thus aid greatly
in identifying the majority AGN populations and, moreover, in 
allowing their underlying luminosities to be assessed reliably
(in a regime where optical/UV luminosity indicators are 
generally unreliable). Only in the highly Compton-thick regime 
(\hbox{$N_{\rm H}\gg 1/\sigma_{\rm T}$}, corresponding to 
\hbox{$N_{\rm H}\gg 1.5\times 10^{24}$~cm$^{-2}$}) 
does direct transmission become impossible, but here one
can still investigate the (much fainter) \xrays\ that are 
reflected or scattered around the absorber
\citep[e.g.,][]{comastri2004,georgantopoulos2013}. An additional 
relevant advantage of \xray\ studies is that, as one studies
objects at increasing redshift in a fixed observed-frame 
band, one gains access to increasingly penetrating rest-frame
emission (i.e., higher rest-frame energies are probed); note 
the opposite generally applies in the
optical and UV bandpasses where dust-reddening 
effects increase toward shorter wavelengths
\citep[e.g.,][]{cardelli1989}. 

\item 
X-rays have low dilution by host-galaxy starlight (i.e., emission 
at any wavelength associated with stellar processes). AGNs
generally have much higher ratios of $L_{\rm X}/L_{\rm Opt}$ 
and thus $f_{\rm X}/f_{\rm Opt}$ than 
stars \citep[e.g.,][]{maccacaro1988}. 
Thus, \xrays\ provide excellent contrast between
SMBH accretion light and starlight (see Fig.~\ref{fig-ngc3783-opt-xray}), 
allowing one to construct pure samples of AGNs even down to relatively 
modest luminosities. This aspect of \xray\ surveys is critical, 
for example, at high redshift where it is often unfeasible, 
at any wavelength, to resolve spatially the AGN light from 
host starlight. For weak or highly obscured AGNs, such dilution 
by host starlight can make AGNs difficult to separate from galaxies 
in the optical/UV regime 
\citep[e.g.,][]{moran2002,hopkins2009}. 

\item 
The \xray\ spectra of AGNs are rich with diagnostic potential
that can be exploited when sufficient source counts are 
collected. At a basic level, the distinctive \xray\ spectral 
characteristics of AGNs can often aid with their 
identification, improving still further the purity of AGN 
samples (see the previous point). Furthermore, measurements
of low-energy photoelectric absorption cut-offs, underlying continuum
shapes, Compton reflection continua, fluorescent line
emission (e.g., from the iron~K$\alpha$ transition), and 
absorption edges (e.g., the iron~K edge) can 
diagnose system luminosity, obscuration level, nuclear 
geometry, disk/corona conditions, and Eddington ratio
($L_{\rm Bol}/L_{\rm Edd}$). 

\end{enumerate}

While these basic points of utility have led to great success
for the enterprise of \xray\ surveys, such surveys do have 
their shortcomings; e.g., in the regime of highly Compton-thick
absorption or in cases of intrinsically \xray\ weak AGNs
(see Section~\ref{agns-missed}). Thus,
when possible, it is critical to complement \xray\ surveys with 
suitably matched multiwavelength surveys in the area of sky
under study. These can help considerably in filling the small 
chinks in the armor of \xray\ surveys, thereby allowing nearly 
complete identification of all significant SMBH growth. 

\subsection{The survey capabilities of relevant distant-universe missions:
\chandra, \xmm, and \nustar}
\label{surveycapabilities}

In this review, we will describe some of the main discoveries 
on AGNs coming from the intensive activity in \xray\ \hbox{(0.5--100~keV)} 
surveys research over the past 15~yr, mainly focusing on missions that can 
make sensitive ``blank-field'' surveys of typical AGNs in the distant 
\hbox{($z=0.1$--5)} universe. Our emphasis will thus be on results 
from 
the {\it Chandra X-ray Observatory\/} (hereafter \chandra; \citealp[e.g.,][]{weisskopf2000}),
the {\it X-ray Multi-Mirror Mission\/} (hereafter \xmm; \citealp[e.g.,][]{jansen2001}), and
the {\it Nuclear Spectroscopic Telescope Array\/} (hereafter \nustar; \citealp[e.g.,][]{harrison2013}). 
The new results from these missions rest squarely upon a rich heritage of \xray\ survey studies 
with several superb earlier \xray\ missions, as briefly described in, e.g., Section~\ref{pre-2000}. 
We will also introduce more local results from, e.g., 
the {\it Swift Gamma-Ray Burst Explorer\/} 
(hereafter \swift; \citealp[e.g.,][]{gehrels2004}) and   
the {\it International Gamma-Ray Astrophysics Laboratory\/} 
(hereafter \integral; \citealp[e.g.,][]{winkler2003}) 
where they connect strongly with results from the distant 
universe, although these critical local investigations are not our 
primary focus (and deserve to be the subject of an entirely separate
dedicated review). 


In terms of basic survey capability, both \chandra\ (launched in 
1999 July) and \xmm\ (launched in 1999 December) provide 
\xray\ {\it spectroscopic imaging\/} over broad bandpasses (\hbox{0.3--8~keV} 
and \hbox{0.2--10~keV}, respectively) and over respectable fields
of view (290~arcmin$^2$ for \chandra\ ACIS-I, and 720~arcmin$^2$ 
for \xmm\ EPIC-pn).\footnote{Note that many \xray\ detectors, 
including those used on \chandra\ and \xmm\ to perform cosmic 
surveys, {\it simultaneously\/} obtain imaging, spectral, and 
timing data for the collected photons
(\citealp[e.g.,][]{struder2001,turner2001,garmire2003}). 
Such \xray\ observations are 
qualitatively different from those generally 
taken in the optical/infrared where, e.g., imaging and 
spectroscopy are largely distinct.} 
Their imaging point spread functions are 
excellent (an on-axis half-power diameter of 0.84~arcsec for \chandra) 
or good (15~arcsec for \xmm), though these degrade significantly 
with increasing off-axis angle. Their most sensitive surveys reach about 
\hbox{80--400} times deeper than those of previous \xray\ missions, and 
excellent source positions (accurate to \hbox{0.5--4~arcsec}) allow effective 
multiwavelength follow-up studies even at the faintest \xray\ fluxes. 
Typical survey projects with \chandra\ and \xmm\ generate
hundreds-to-thousands of detected AGNs, allowing powerful statistical 
studies of source populations. Furthermore, systematic public data 
archiving practices allow effective survey combination, so that 
source populations spanning wide ranges of luminosity and redshift 
can be studied together. 

\nustar\ is a more recently launched mission (2012 June) that is 
now transforming surveys of the \xray\ universe above 10~keV. It is
the first focusing high-energy \hbox{(3--79~keV)} \xray\ observatory 
in orbit; for \xray\ surveys of the generally faint sources in the 
distant universe, \nustar\ is most effective up to $\approx 24$~keV
(at higher energies, rising background levels and dropping photon
collecting area limit its sensitivity to faint sources). 
This coverage of high energies 
equates to reduced absorption bias even relative to \chandra\ and \xmm.
The reduced bias is particularly key at $z\simlt 1$ where the 
rest-frame energies covered by \chandra\ and \xmm\ are still
modest. The \nustar\ field of view for spectroscopic imaging
is 140~arcmin$^2$, and its imaging point 
spread function has an on-axis half-power diameter of 58~arcsec
but with a sharp core having a full width at half maximum of 18~arcsec
(this is about an order of magnitude improvement compared to the 
imaging capabilities of previous coded-mask instruments in orbit; 
\citealp[e.g.,][]{tuel10bat}). 
Its sensitivity in probing the hard \xray\ sky is about two orders 
of magnitude better than previous collimated or coded-mask 
instruments, making it the first genuine surveyor of the distant 
universe from \hbox{10--24~keV}. The first publications from the 
extensive \nustar\ survey programs are presently appearing with 
more in preparation; these 
have been delivered by the members of the \nustar\ survey teams, 
although the underlying data will be made public for investigations 
by the whole astronomical community. Many additional \nustar\ results 
are expected in coming years.

\subsection{Structure of this review, other relevant reviews, and definitions}
\label{structureofthisreview}

As noted above, there is a vast literature on the demographics, 
physics, and ecology of distant growing SMBHs found in \xray\ surveys.
Indeed, more than 500 papers on this subject have been published over 
the past 15~yr based on surveys with \chandra, \xmm, and 
\nustar\ alone. Our primary aim here is to describe briefly some of the
main discoveries coming from these efforts to make them 
more accessible to interested researchers and students. We note in 
advance that, owing to the vastness of the literature, it will not
be possible to cover all relevant work; our apologies in advance if 
we could not cover your favorite result or paper.

The structure for the rest of this review will be the following: 

\begin{itemize}

\item
In Section~2, we will review the main \xray\ surveys of the distant
universe and their supporting observations. We will also describe 
how AGNs are selected effectively using the \xray\ and multiwavelength
data. 

\item 
Section~3 (``AGN demographics'') 
will cover demographic results for distant \xray\ selected AGNs, 
focusing on AGN evolution over cosmic time. This will also 
include brief discussions of 
the population of AGNs missed in cosmic \xray\ surveys, 
the \soltan\ and related arguments, and 
the environmental dependence of AGN evolution. 

\item 
Section~4 (``AGN physics'')
will describe insights on the physical processes operating
in AGNs that have come from \xray\ surveys. 

\item 
Section~5 (``AGN ecology'')
will describe what \xray\ surveys have revealed about interactions 
between growing SMBHs and their environments (mainly their host 
galaxies). This section will also discuss the relative radiative
output from SMBHs and stars over cosmic time. 

\item 
In Section~6, we will outline key outstanding questions.  
We will also describe prospects for advancing the field both in the 
near and longer terms using both \xray\ and multiwavelength follow-up 
facilities. 

\end{itemize}

\noindent
We note that Sections 3, 4, and 5 cover strongly inter-related themes, 
and that there is inevitably a degree of subjectivity in assigning some 
results to a single one of these sections. Nevertheless, this
structure is useful for basic organizational purposes. 

Over the past 15~yr, a number of other relevant in-depth reviews have 
been prepared that have some overlap with the topics discussed here. 
These include
\citet{hasinger2000},
\citet{gilli2004}, 
\citet{brandt2005}, 
\citet{urry2007},
\citet{hickox-news2009},
\citet{brandt2010},
\citet{alexander2012},
\citet{treister-review2012},
\citet{kormendy2013},
\citet{merloni2013}, 
\citet{shankar2013}, and  
\citet{gilfanov2014}.
We encourage interested readers to consult these reviews as well,
noting that they generally emphasize somewhat different topics
than those emphasized here. 
We also refer interested readers to the chapters in the 
Astrophysics and Space Science Library volume 
titled ``Supermassive Black Holes in the Distant Universe''
\citep{barger2004}. 

Throughout this review we shall adopt J2000 coordinates and
a standard cosmology with 
\hbox{$H_0=70$~km~s$^{-1}$~Mpc$^{-1}$}, 
\hbox{$\Omega_{\rm M}=0.3$}, and 
\hbox{$\Omega_{\Lambda}=0.7$}. 
When quoting effective hydrogen column densities estimated 
from \xray\ spectral analyses, we will adopt the cosmic 
abundances of \citet{Anders1989}.
When referring to \xray\ obscured AGNs, we will be considering
systems with $N_{\rm H}\simgt 10^{22}$~cm$^{-2}$ unless noted
otherwise. This threshold value is commonly adopted though 
admittedly somewhat arbitrary, being close to the typical
maximum absorption expected from a galactic disk. It is
also broadly consistent with the division in \xray\ 
absorption level between optically obscured and optically
unobscured AGNs. 
When referring to highly \xray\ obscured AGNs, we will 
generally mean systems with column densities at least a 
factor of $\approx 50$ greater; i.e., 
$N_{\rm H}\simgt 5\times 10^{23}$~cm$^{-2}$. 



\section{The main cosmological X-ray surveys, their supporting observations, and AGN selection}
\label{surveys-supporting-selection}


\subsection{Description of the current main cosmic X-ray surveys of distant AGNs}
\label{main-surveys}

The capabilities of \chandra, \xmm, and \nustar\ (see Section~\ref{surveycapabilities})
have led to a substantial number of \xray\ surveys being conducted of the distant
universe. These include targeted surveys, both deep and wide, where a sky area of 
particular interest is observed; e.g., a field already having excellent multiwavelength 
data or a field containing a notable object such as a high-redshift protocluster. 
Furthermore, these include serendipitous surveys that investigate the serendipitous 
sources detected in a number of fields observed for other reasons.


\begin{table}
\caption{Selected Extragalactic X-ray Surveys with \chandra, \xmm, and \nustar}
\label{surveystable}      
\begin{tabular}{lrrl}
\hline\noalign{\smallskip}
Survey                        & Rep. Eff.   & Solid Angle  & Representative \\
Name                          & Exp. (ks)   & (arcmin$^2$) & Reference      \\
\noalign{\smallskip}\hline\noalign{\smallskip}
\noalign{\centering \chandra\ (0.3--8~keV)}
\noalign{\smallskip}\hline\noalign{\smallskip}
\chandra\ Deep Field-South (CDF-S)       & 3870        &  465      & \citet{xue2011}               \\
\chandra\ Deep Field-North (CDF-N)       & 1950        &  448      & \citet{alexander2003}         \\          
AEGIS-X Deep                             &  800        &  860      & \citet{goulding2012}          \\
SSA22 protocluster                       &  392        &  330      & \citet{lehmer2009}            \\
HRC Lockman Hole                         &  300        &  900      & PI: S.S. Murray               \\
Extended CDF-S (E-CDF-S)                 &  250        & 1,128     & \citet{lehmer2005}            \\
AEGIS-X                                  &  200        & 2,412     & \citet{laird2009}             \\
Lynx                                     &  185        &  286      & \citet{stern2002}             \\           
LALA Cetus                               &  174        &  297      & \citet{wang2007}              \\
LALA Bo\"otes                            &  172        &  346      & \citet{wang2004}              \\
C-COSMOS and COSMOS-Legacy               &  160        & 6,120     & \citet{elvis2009}             \\
SSA13                                    &  101        &  345      & \citet{barger2001ssa13}       \\
Abell~370                                &   94        &  345      & \citet{barger2001a370}        \\
3C~295                                   &   92        &  274      & \citet{delia2004}             \\
ELAIS N1+N2                              &   75        &  590      & \citet{manners2003}           \\
WHDF                                     &   72        &  286      & \citet{bielby2012}            \\
CLANS (Lockman Hole)                     &   70        & 2,160     & \citet{trouille2008}          \\
SEXSI$^{\rm a}$                           &   45        & 7,920     & \citet{harrison2003}          \\
CLASXS (Lockman Hole)                    &   40        & 1,620     & \citet{trouille2008}          \\
13~hr Field                              &   40        &  710      & \citet{mchardy2003}           \\
ChaMP$^{\rm a}$                            &   25        & 34,560    & \citet{kim2007}               \\
XDEEP2 Shallow                           &   15        & 9,432     & \citet{goulding2012}          \\
\chandra\ Source Catalog (CSC)$^{\rm a}$  &   13        & 1,150,000 & \citet{evans2010}             \\
Stripe 82X---\chandra$^{\rm a}$           &    9        & 22,320     & \citet{lamassa2013chandra}    \\
NDWFS XBo\"otes                         &    5        & 33,480     & \citet{murray2005}            \\
\noalign{\smallskip}\hline\noalign{\smallskip}
\noalign{\centering \xmm\ (0.2--12~keV)}
\noalign{\smallskip}\hline\noalign{\smallskip}
\chandra\ Deep Field-South (CDF-S)  & 2820        &  830             & \citet{ranalli2013}         \\
Lockman Hole                        &  640        &  710             & \citet{brunner2008}         \\
\chandra\ Deep Field-North (CDF-N)  &  180        &  752             & \citet{miyaji2003}          \\
13~hr Field                         &  120        &  650             & \citet{loaring2005}         \\
ELAIS-S1                            &   90        & 2,160            & \citet{puccetti2006}        \\
Groth-Westphal                      &   81        &  730             & \citet{miyaji2004}          \\
COSMOS                              &   68        & 7,670            & \citet{cappelluti2009}      \\
Subaru \xmm\ Deep Survey (SXDS)     &   40        & 4,100            & \citet{ueda2008}            \\
Marano Field                        &   30        & 2,120            & \citet{lamer2003}           \\
HELLAS2XMM$^{\rm a}$                 &   25        & 10,440           & \citet{baldi2002}           \\
XMM-LSS XMDS                        &   23        & 3,600            & \citet{chia2005}            \\
3XMM$^{\rm a}$                       &   15        & 2,300,000        & \citet{watson2012}          \\
Stripe 82X---\xmm$^{\rm a}$           &   15        & 37,800           & \citet{lamassa2013xmm}      \\
XMM-LSS                             &   10        & 39,960           & \citet{chia2013}            \\
XMM-XXL                             &   10        & 180,000          & \citet{pierre2012}          \\
Stripe 82X---\xmm\ Targeted         &    8        & 129,600          & PI: C.M. Urry               \\
\xmm\ Slew Survey (XMMSL1)$^{\rm a}$ & 0.006       & $8\times 10^7$   & \citet{warwick2012}         \\
\noalign{\smallskip}\hline\noalign{\smallskip}
\noalign{\centering \nustar\ (3--24~keV)}
\noalign{\smallskip}\hline\noalign{\smallskip}
Extended CDF-S (E-CDF-S)                       & 200         &  1,100  &  Mullaney et~al, in prep   \\
AEGIS-X                                        & 270         &    860  &  Aird et~al, in prep       \\
COSMOS                                         & 65          &  6,120  &  Civano et~al, in prep     \\
Serendipitous Survey$^{\rm a}$                   & 22          & 19,000  &  \citet{alexander2013}      \\
\noalign{\smallskip}\hline
\end{tabular}
$^{\rm a}$Serendipitious survey; see Section~\ref{main-surveys} for brief discussion 
regarding such surveys. 
\end{table}


Some selected extragalactic surveys conducted with \chandra, \xmm, and \nustar\ 
are listed in Table~\ref{surveystable}. A number of aspects of this table 
deserve note: 

\begin{enumerate}

\item
These surveys often have wide ranges of sensitivity across their associated 
solid angles due to, e.g., differing satellite pointing strategies and 
instrumental effects. The serendipitous surveys (e.g., ChaMP, SEXSI, CSC, HELLAS2XMM, 
3XMM) particularly stand out in this regard, often being made up of observations 
differing by an order of magnitude or more in exposure time. 

\item 
The solid angles quoted generally represent the total sky coverage at 
bright \xray\ flux limits.

\item 
When listing \xmm\ exposure times, we have attempted to remove time 
intervals affected by strong background flaring; such intervals are
generally not useful for surveys of faint cosmic sources. 

\item 
The listed exposure times are for a single \xray\ telescope and focal
plane module (FPM). Note that \xmm\ and \nustar\ have three
and two simultaneously operating telescopes/FPMs, respectively. 

\item
Some of the surveys have overlap of their solid angles of coverage
(e.g., \hbox{CDF-S} vs.\ \hbox{E-CDF-S}; 
\hbox{AEGIS-X} Deep vs.\ \hbox{AEGIS-X};
XDEEP2 Shallow vs.\ Stripe~82X---\chandra; 
SXDS vs.\ \hbox{XMM-LSS} XMDS vs.\ \hbox{XMM-LSS} vs.\ \hbox{XMM-XXL}). 

\item
Some of these surveys are still increasing in solid angle and/or depth. 
For example, many of the serendipitous surveys continue to grow as
more \xray\ observations are performed, and the \chandra\ CDF-S 
survey is presently being raised to a 7~Ms exposure. 

\item
Some of the listed surveys have been conducted in multiple epochs spanning 
up to $\approx 15$~yr, allowing assessments of long-timescale \xray\ 
variability for the detected sources
(\citealp[e.g.,][]{paolillo2004,papadakis2008,young2012,lanzuisi2014}).  

\end{enumerate}

\begin{figure}
\includegraphics[scale=0.6,angle=270]{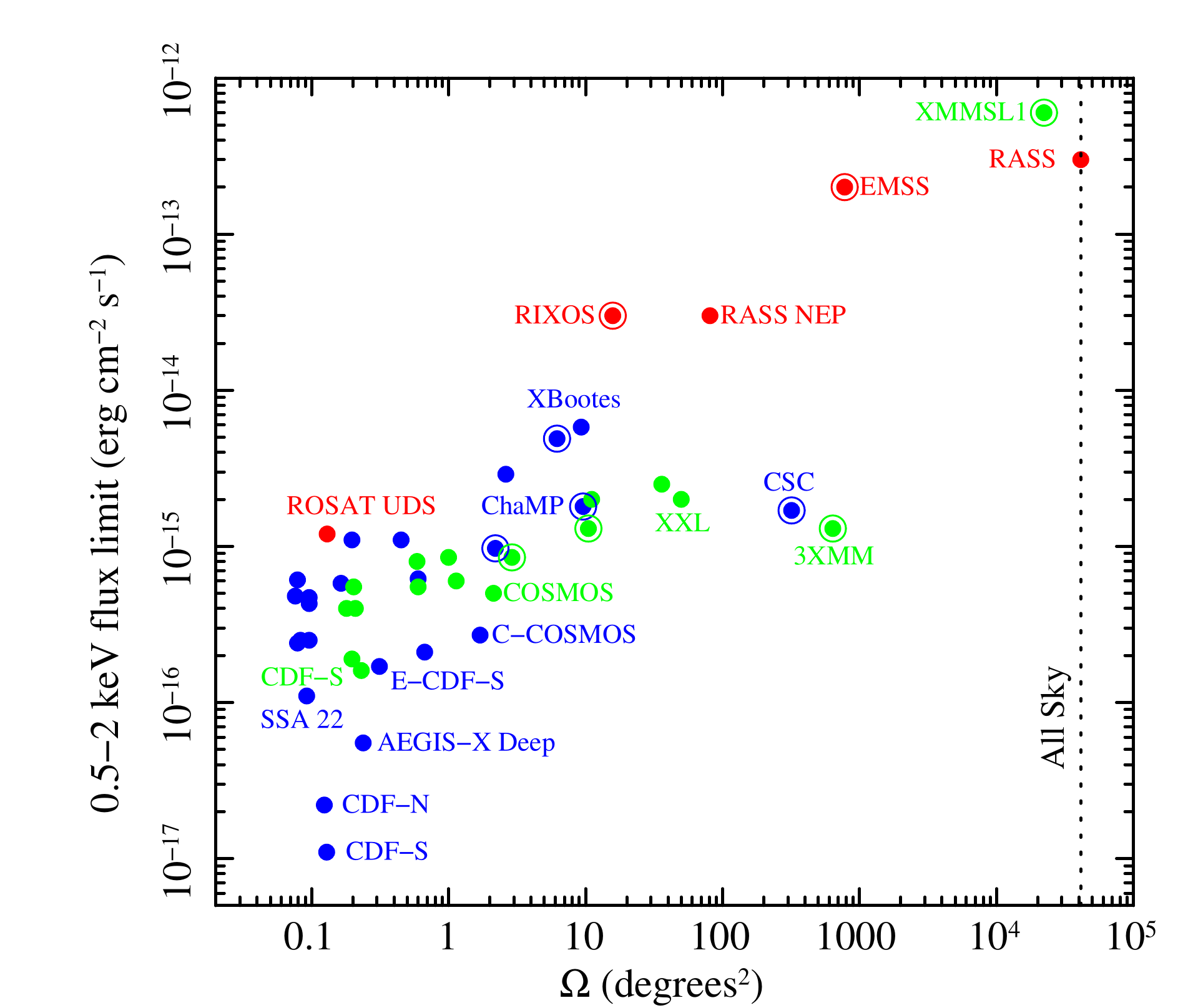}
\includegraphics[scale=0.6,angle=270]{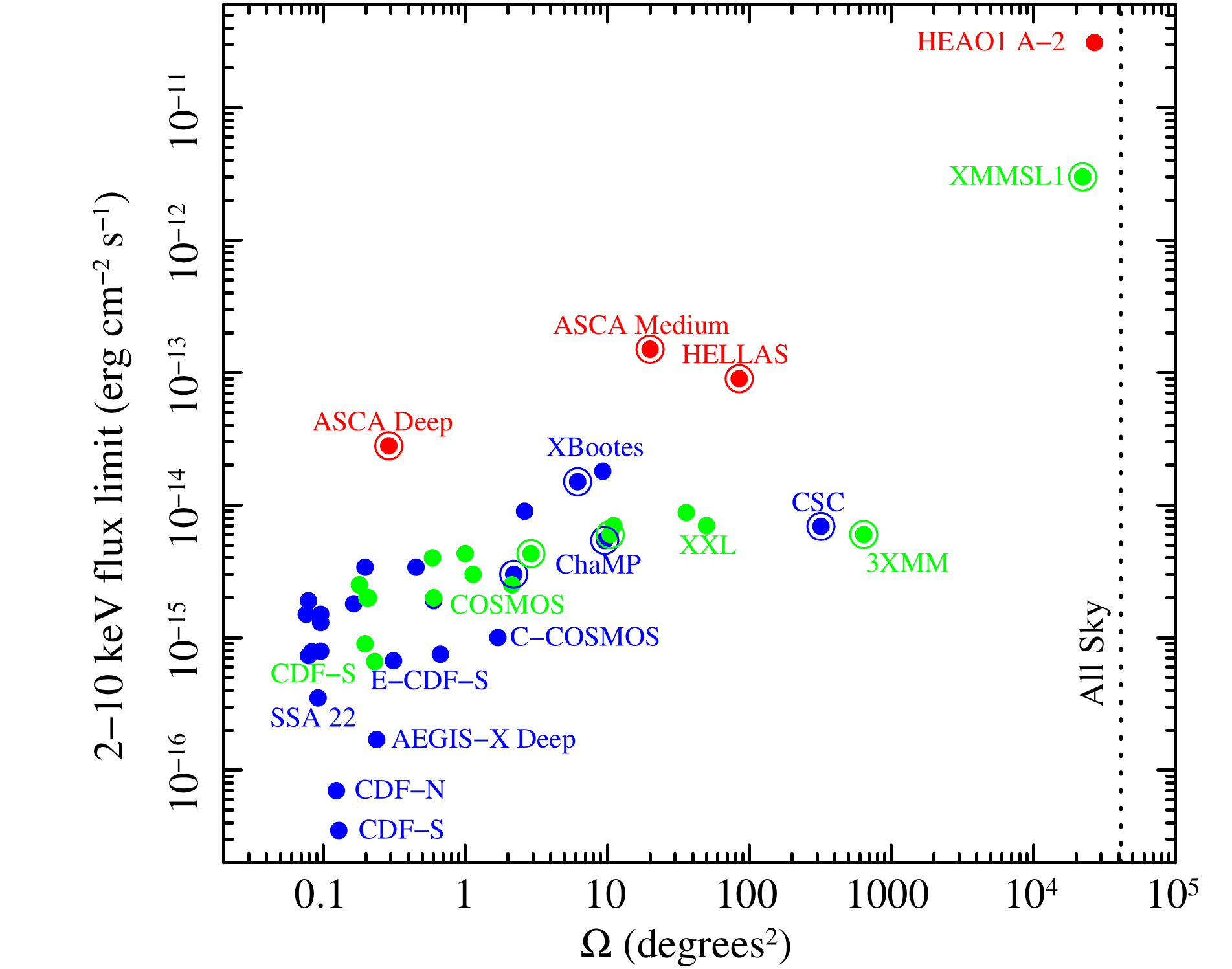}
\caption{Solid angle of sky coverage vs.\ sensitivity in both the \hbox{0.5--2~keV} 
and \hbox{2--10~keV} bands for the surveys in Table~\ref{surveystable} from 
\chandra\ (blue) and \xmm\ (green). For comparison purposes, a few surveys 
from previous \xray\ missions are shown in red. The circles around some of 
the points indicate serendipitous surveys as also denoted in 
Table~\ref{surveystable}. Some of the surveys are labeled by name 
(sometimes abbreviated) in regions where symbol crowding is not too strong. 
The vertical dotted line shows the solid angle for the whole sky. Each of 
the surveys has a range of sensitivity across its solid angle, and different 
authors use somewhat different methodologies for computing and quoting 
sensitivity; this leads to small uncertainties in the precise relative 
locations of the data points.}
\label{fig-omega-depth}       
\end{figure}

\noindent
Interested readers should consult the cited papers in 
Table~\ref{surveystable} for survey-specific details. 
Plots of solid angle of sky coverage vs.\ sensitivity 
in both the \hbox{0.5--2~keV} and \hbox{2--10~keV} bands 
for \chandra\ and \xmm\ surveys are given in 
Figure~\ref{fig-omega-depth}. 

Together, all these surveys cover a broad part of the practically accessible 
sensitivity vs.\ solid-angle ``discovery space'' via the standard ``wedding-cake'' 
design, providing a quite complete understanding of AGN populations in the 
distant universe (though, as discussed in Section~\ref{new-surveys}, there is
still room for important improvements). One persistent limitation of 
\xray\ surveys in general, however, has 
been the lack of sensitive and thoroughly followed-up 
surveys over hundreds-to-thousands of deg$^2$; the widest sensitive
surveys presently are 3XMM and XMM-XXL. This limitation has hindered \xray\ 
constraints upon rare objects, such as the most luminous AGNs in the 
Universe, although targeted \xray\ follow-up studies of such objects
selected at other wavelengths have mitigated this issue to some degree 
(\citealp[e.g.,][]{vignali2005,just2007,stern2014}). \erosita\ 
(\citealp[e.g.,][]{merloni2012}) is expected to improve this situation
significantly in the near future, and dedicated wide-field \xray\ 
telescopes (\citealp[e.g.,][]{murray2013,rau2013}) could make major additional
strides (see Section~\ref{new-xrays} for more detailed discussion). 

\begin{figure}
\includegraphics[scale=0.7,angle=0]{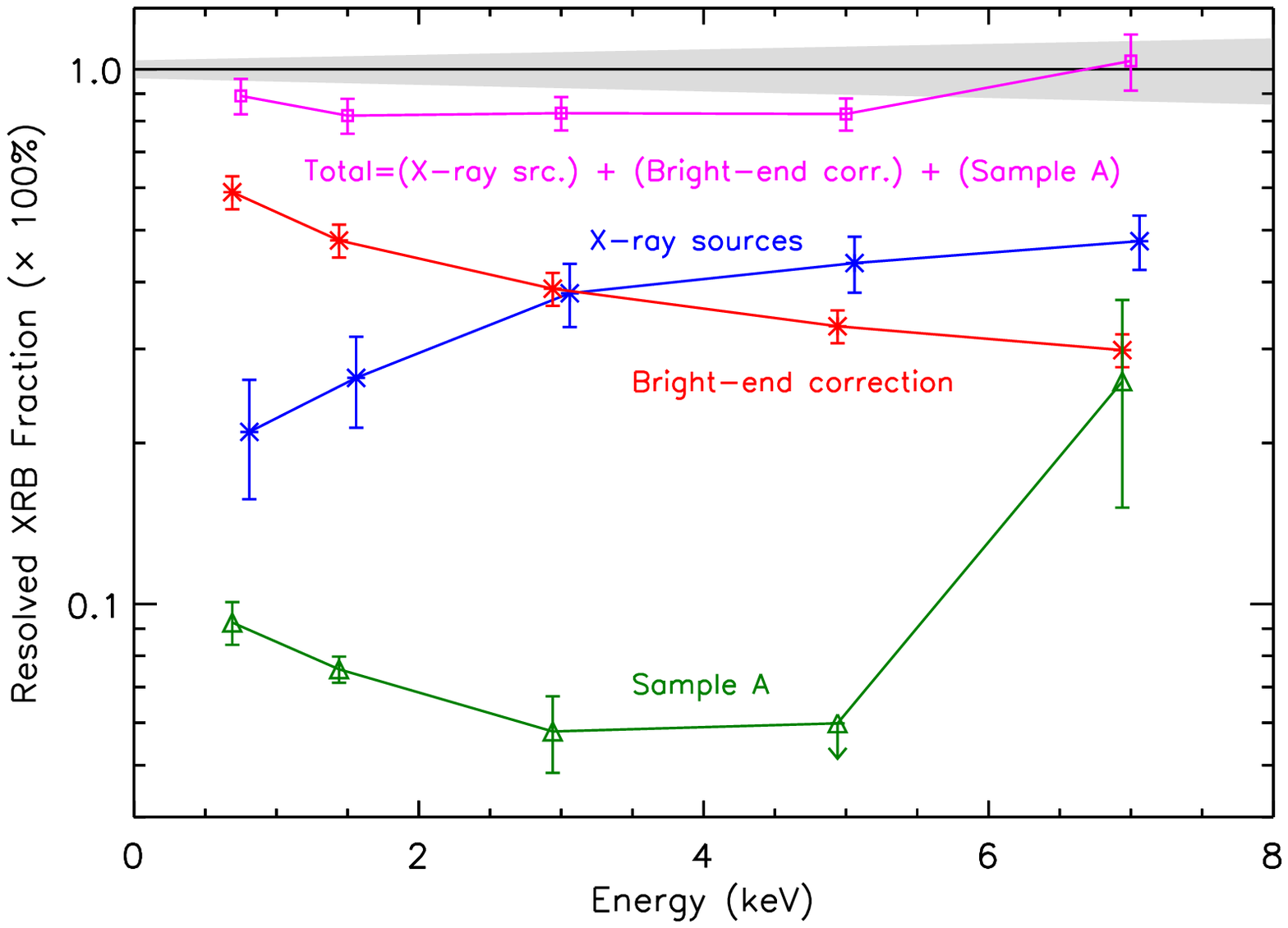}
\caption{The resolved fraction of the CXRB in the 4~Ms \hbox{CDF-S}
observation as a function of energy from \hbox{0.5--8~keV}. 
Shown are
the resolved fraction from the 4~Ms \hbox{CDF-S} sources (blue), 
the ``bright-end correction'' that accounts for bright sources 
too rare to be found within the studied \hbox{CDF-S} field of view (red), and
results from stacking \xray\ photons coincident with 
$z$-band identified galaxies that are not individually \xray\
detected (green, titled ``Sample~A''). The summation of these
three components is also shown (magenta). The total CXRB intensity 
is taken from \citet{hickox2006} with (non-negligible)
uncertainty indicated by the gray area. 
The unresolved CXRB below $\approx 5$~keV, even after
inclusion of the stacked emission coincident with galaxies, is
likely associated with groups/clusters of galaxies. 
Adapted from \citet{xue2012}.}
\label{fig-xue2012-resolved-fraction}       
\end{figure}

Surveys with \chandra\ and \xmm\ have resolved \hbox{$\approx 75$--80\%} 
of the cosmic \xray\ background (CXRB) from \hbox{0.5--10~keV} into 
point sources (e.g., see Section~1.3 of \citealp[][]{brandt2005} for 
further discussion). Up-to-date measurements of the resolved fraction
of the CXRB below 10~keV in discrete energy bands may be found in, 
e.g., \citet{hickox2006}, \citet{lehmer2012}, \citet{xue2012}, and
Ranalli et~al, in prep; Fig.~\ref{fig-xue2012-resolved-fraction} shows 
results from one such analysis. These 
measurements provide useful integral constraints upon 
remaining undetected \xray\ source populations, although they still
lie significantly below the peak of the CXRB at \hbox{20--40~keV}. 
The deepest surveys with \nustar\ are expected to resolve \hbox{30--40\%} 
of the \hbox{8--24~keV} CXRB (\citealp[e.g.,][]{ballantyne2011}; 
J.A. Aird 2014, private communication), reaching closer to its peak.  
This is a large improvement over pre-\nustar\ results, where only
a few percent of the \hbox{10--100~keV} CXRB was resolved
(\citealp[e.g.,][]{krivonos2005,treister-integral2009}).

\subsection{Supporting multiwavelength observations and spectroscopic follow-up}
\label{followup-multiwavelength}

Characterization of the detected \xray\ sources using both multiwavelength photometric
data and spectroscopic observations is crucial for investigating their nature. At the
most basic level, a detected \xray\ source must be matched reliably to a multiwavelength
photometric counterpart so that, e.g., the feasibility of spectroscopic observations can be
determined; such matching is done most effectively using a likelihood-ratio technique 
(\citealp[e.g.,][]{sutherland1992,rutledge2000,ciliegi2003,naylor2013}). 
Counterpart matching is straightforward for the majority of sources in \chandra\ 
surveys owing to the excellent angular resolution of \chandra\ 
(generally providing \hbox{0.5--1.5$^{\prime\prime}$} positions;
\citealp[e.g.,][]{evans2010}), 
although there are genuine matching challenges for the faintest optical, 
near-infrared (NIR; about \hbox{1--5~$\mu$m}), and/or mid-infrared 
(MIR; about \hbox{5--30~$\mu$m}) counterparts. For \xmm\ surveys the 
counterpart matching is more challenging 
(generally \hbox{2--4$^{\prime\prime}$} positions;
\citealp[e.g.,][]{watson2009}), 
although the majority of sources in \xmm\ surveys 
can also be matched to optical/NIR/MIR counterparts. In current \nustar\ surveys
(generally \hbox{10--20$^{\prime\prime}$} positions; \citealp[e.g.,][]{harrison2013}), 
the detected sources are typically first matched to \chandra/\xmm\ (or other \xray) 
sources which then are matched to optical/NIR/MIR counterparts. The \xray\ sources found 
in the surveys listed in Table~\ref{surveystable} span an extremely broad range of 
optical/NIR/MIR flux; e.g., the $I$-band magnitudes for AGNs range from brighter 
than 15th to fainter than 28th magnitude
(\citealp[e.g.,][]{barger2003,szokoly2004,laird2009,brusa2010,luo2010,pineau2011,
rovilos2011,xue2011,civano2012,trichas2012}).  

\begin{figure}
\includegraphics[scale=0.35,angle=0]{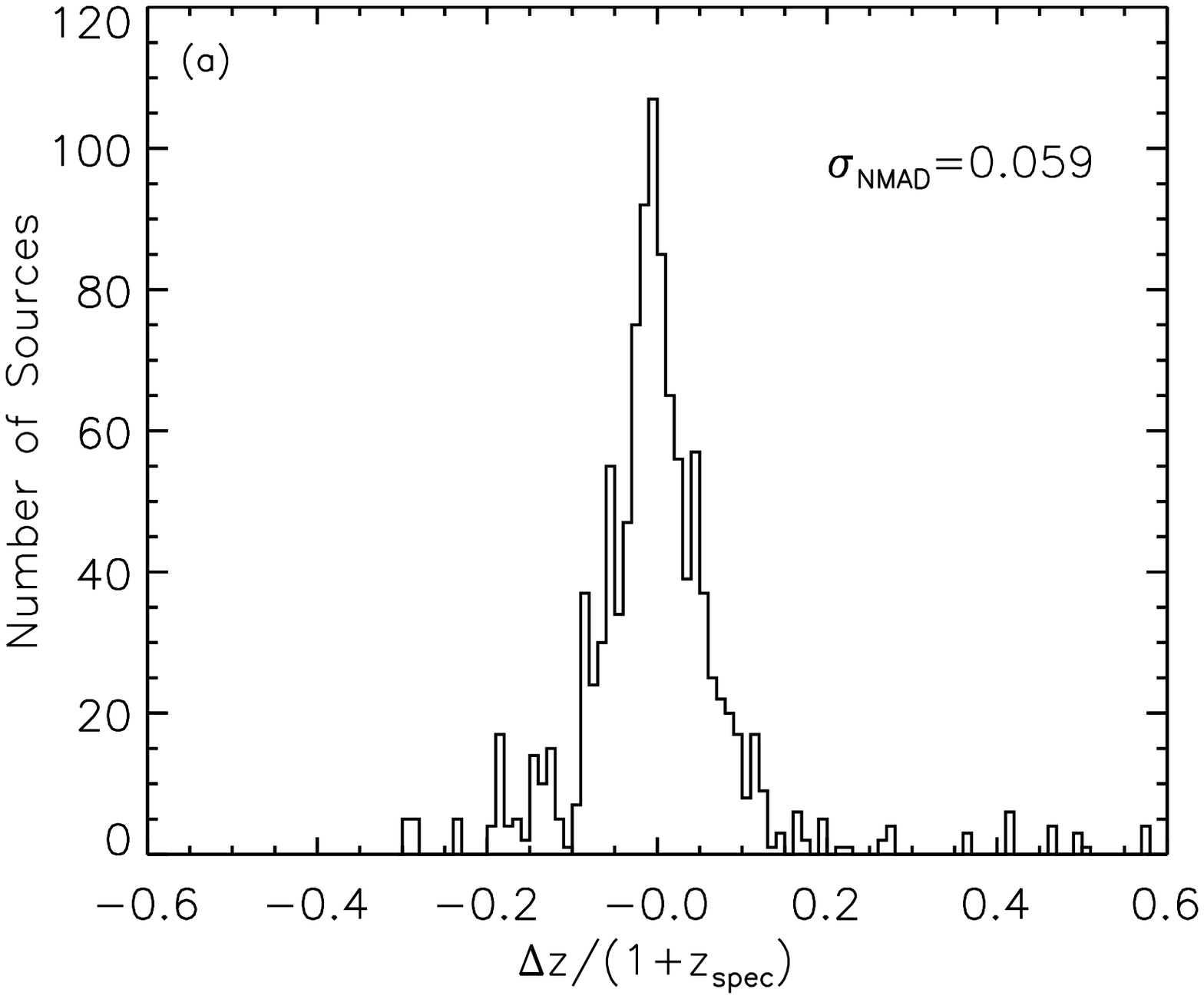}
\hspace*{-0.7cm}
\includegraphics[scale=0.35,angle=0]{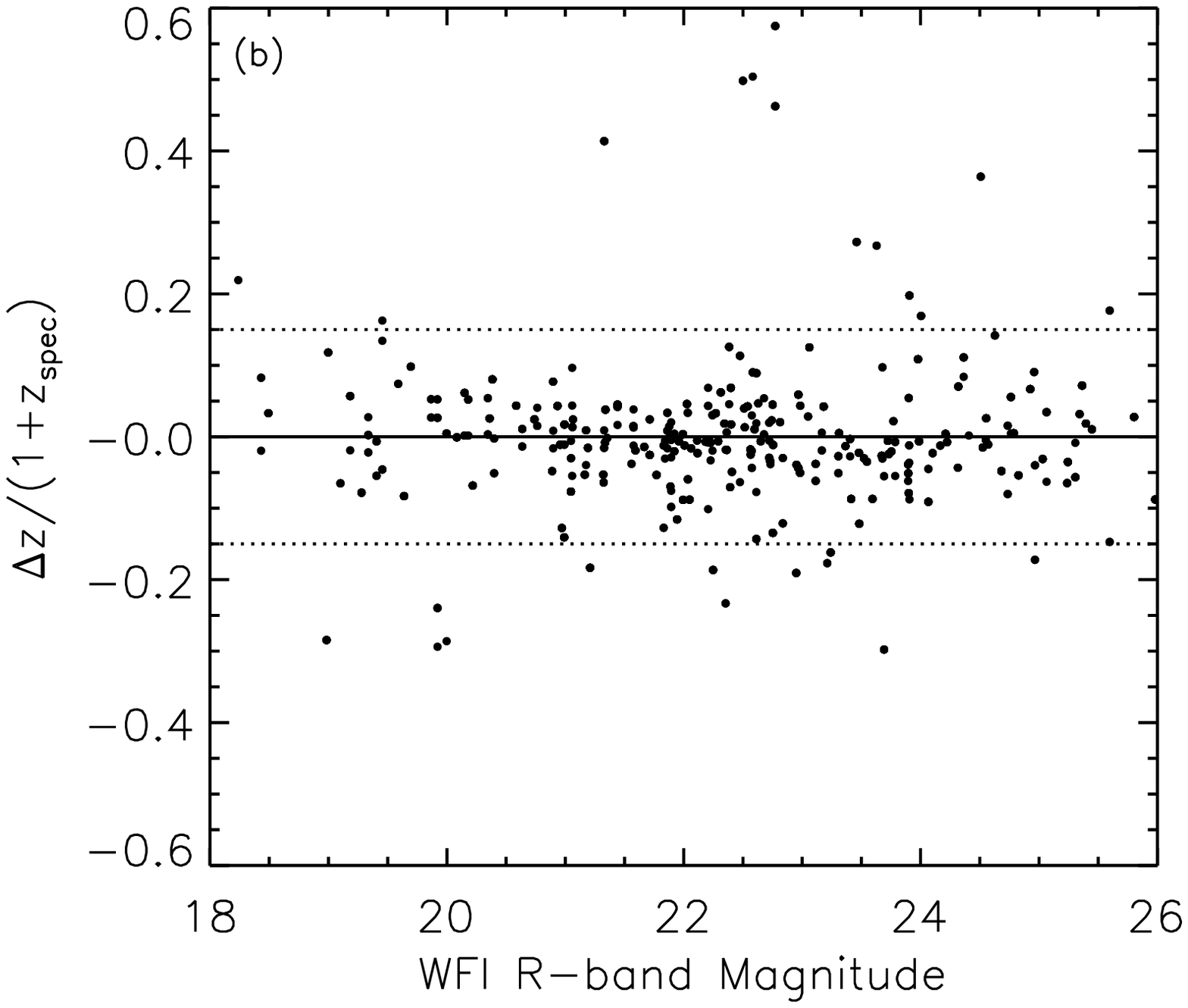}
\caption{(a) Distribution of the photometric redshift (photo-$z$, or $z_{\rm photo}$) 
accuracy, $\Delta z/(1+z_{\rm spec})$, derived from an unbiased blind-test 
sample of \xray\ AGNs in the \hbox{CDF-S} ($z_{\rm spec}$ values are 
spectroscopic redshifts). The typical photo-$z$ accuracy for the sample 
is evaluated with a robust estimator, the normalized median absolute deviation 
($\sigma_{\rm NMAD}$): 
$\sigma_{\rm NMAD}=1.48\times {\rm median}[(|\Delta z-{\rm median}(\Delta z)|)/(1+z_{\rm spec})]$. 
(b) Photo-$z$ accuracy, $\Delta z/(1+z_{\rm spec})$, vs.\ $R$-band magnitude for 
the blind-test sample. The solid line indicates
$z_{\rm photo}=z_{\rm spec}$, and the dotted lines represent relations
of $z_{\rm photo}=z_{\rm spec}\pm0.15(1+z_{\rm spec})$. The photo-$z$
accuracy declines toward faint $R$-band magnitudes, particularly 
when considering the frequency of 
catastrophically incorrect photo-$z$ values. 
Adapted from \citet{luo2010}.}
\label{fig-luo2010-photoz}       
\end{figure}

The \xray\ data from current surveys generally do not allow direct redshift
determination, although there are occasional cases where redshifts can be 
measured based on the strong iron~K$\alpha$ line and/or iron~K edge in \xray\ spectra 
(\citealp[e.g.,][]{iwasawa2012,delmoro2014}). Thus, spectroscopic and/or
photometric determination of redshifts, usually in the optical/infrared, is 
essential for, e.g., calculation of source luminosity, the most meaningful \xray\ 
spectral modeling, and studies of source cosmic evolution. In line with the
broad range of optical/NIR/MIR flux for the counterparts, a wide variety of facilities 
have been used productively for spectroscopic redshift determination, including
the largest telescopes on Earth (e.g., Keck, the Very Large Telescope, and Subaru)
for the faintest sources in deep surveys. Enormous progress has been made with
spectroscopic redshift determination for \xray\ survey sources, particularly 
when multi-object spectrographs can be utilized to target efficiently large 
numbers of \xray\ sources simultaneously; the field sizes of these spectrographs 
often match the sizes of the \xray\ survey fields well. Large samples of \xray\
sources with reasonably complete spectroscopic identification down to 
\hbox{$I=22$--24} are now available for statistical studies
(\citealp[e.g.,][]{barger2003,fiore2003,szokoly2004,eckart2006,dellaceca2008,trouille2008,
trump2009,silverman2010,kochanek2012,trichas2012}). This being said, 
much work is still needed to improve the spectroscopic completeness for some
key fields, in some cases by systematically publishing spectra already
acquired. At fainter fluxes of \hbox{$I=24$--28} where many \xray\ sources are 
found, particularly in the deepest \xray\ surveys, the spectroscopic completeness 
drops rapidly (e.g., even in the intensively studied \hbox{CDF-S}, only 
$\approx 65$\% of the \xray\ sources overall have spectroscopic redshifts). 
This bottleneck remains one persistent driver for the construction of future 
Extremely Large Telescopes in the optical/NIR (see Section~\ref{followup}). 

It is often possible to derive photometric redshifts of reasonable quality 
for \xray\ AGNs when spectroscopic redshifts are unavailable, although 
these are generally of lower quality than those for comparable 
non-AGN galaxies (e.g., see Fig.~\ref{fig-luo2010-photoz}). 
Photometric redshifts also provide a useful cross-check 
even when spectroscopic redshifts are available; e.g., if 
only one emission line is clearly detected then there can be
significant uncertainty in the correct spectroscopic redshift
determination. 
In the best cases, photometric redshifts are derived 
from $\simgt 20$ bands of MIR-to-UV photometric data, 
utilize dedicated templates suitable for \xray\ sources such as AGN/galaxy hybrids, 
utilize dedicated sets of medium-band filters, and/or 
allow for AGN optical/NIR/MIR variability effects between different filters 
observed non-simultaneously
(\citealp[e.g.,][]{salvato2009,cardamone2010,luo2010,xue2012,hsu2014}). 
Photometric redshift derivations for \xray\ sources can reach much
fainter optical magnitudes than can be reached spectroscopically 
(e.g., to $I\approx 28$). When compared with relatively bright 
optical sources having spectroscopic redshifts, they have a 
(magnitude-dependent) typical accuracy in $\Delta z/(1+z)$ 
of \hbox{1--10}\% (using $\sigma_{\rm NMAD}$, see Fig.~\ref{fig-luo2010-photoz}) 
with an outlier fraction of catastrophically incorrect redshifts 
of \hbox{3--20}\%. Further improvements in statistical redshift
estimation for optically faint \xray\ sources should be possible in 
the near future using clustering-based techniques
(\citealp[e.g.,][]{matthews2010,menard2014}). 

In addition to counterpart identification and redshift determination, 
multiwavelength observations play many further key roles in the
effective investigation of sources from cosmic \xray\ surveys. These 
include basic characterization of the nature of the detected sources 
(see Section~\ref{agn-selection}), measurements of broad-band AGN
spectral energy distributions (SEDs) to determine more reliable 
bolometric luminosities and investigate accretion 
processes, and measurements of AGN host-galaxy properties 
[e.g., stellar mass, star-formation rate (SFR), morphology, interaction
status, and large-scale environment; see Section~\ref{ecology}]. 
Furthermore, these multiwavelength data have been used to identify 
AGNs and AGN candidates missed by the \xray\ selection technique 
(e.g., Compton-thick and/or intrinsically \xray\ weak AGNs; see
Section~\ref{agns-missed}). 


\subsection{AGN selection from the general X-ray source population}
\label{agn-selection}

The main classes of extragalactic \xray\ sources detected in cosmic 
surveys include AGNs, starburst galaxies, normal galaxies, galaxy clusters, 
and galaxy groups. We will not review the non-AGN classes here and instead refer
readers to the other reviews cited in Section~\ref{structureofthisreview}
for relevant details (e.g., see Section~2.2 of 
\citealp[][]{brandt2005}). Instead, consistent with our focus in 
this review, we will describe how AGNs are selected from the 
general \xray\ source population. 

Multiple methods can be used to derive a highly reliable 
sample of AGNs, including obscured and low-luminosity
systems, from a sample of \xray\ survey point sources. Those relying 
upon direct use of the \xray\ data include the following: 

\begin{enumerate}

\item
{\it X-ray luminosity.\/} 
Sources with \hbox{0.5--10~keV} luminosities above 
$3\times 10^{42}$~erg~s$^{-1}$ are predominantly AGNs. 
Only rare extreme starburst galaxies in the distant universe, 
such as luminous submillimeter galaxies, can exceed this threshold 
without an AGN being present; caution is needed in applying the 
luminosity threshold when such sources are under study. 

\item 
{\it X-ray luminosity vs.\ SFR.\/}
Many researchers have established relations between \xray\ luminosity
and SFR for starburst/normal galaxies that lack
AGNs (\citealp[e.g.,][]{bauer2002,ranalli2003,persic2007,lehmer2010,mineo2014}). 
\xray\ sources lying well above (typically, $\simgt 5$ times is 
used) such relations are found to be AGNs. When quality data 
capable of constraining host-galaxy SFR are available
(e.g., in the radio, infrared, and/or UV), this method is 
both more reliable and more complete than the straight \xray\ 
luminosity cut of method~1 above. If host-galaxy stellar 
mass is also available, then the expected \xray\ luminosity 
can be estimated as a function of both SFR and stellar mass. 

\item 
{\it X-ray-to-optical/NIR flux ratio.\/} 
Consistent with the low dilution of \xray\ emission by 
host-galaxy starlight noted in Section~\ref{generalutility}, 
AGNs tend to have higher X-ray-to-optical/NIR flux ratios 
than starburst/normal galaxies. Typically thresholds of 
$\log (f_{\rm 0.5-10\,keV}/f_R)>-1$ using the observed-frame $R$ band or 
$\log (f_{\rm 0.5-10\,keV}/f_{\rm 3.6 \mu m})>-1$ using 
the observed-frame 3.6~$\mu$m band from \spitzer\ 
(\citealp[e.g.,][]{werner2004}) serve to select samples 
that are \hbox{90--95}\% AGNs (other bands can also be used in 
a similar fashion, although the requisite threshold will vary).
Ideally these ratios would be derived using rest-frame rather 
than observed-frame bands, since the \xray\ and optical/NIR fluxes 
have significantly different $k$-corrections. However, this is 
often not practical or possible, and when using observed-frame 
bands it is generally best to use the reddest optical/NIR band 
available (e.g., to at least minimize the effects of dust 
extinction in high-redshift sources). 

\item
{\it X-ray spectral shape.\/}
\xray\ sources with flat effective power-law photon indices in 
the \hbox{0.5--10}~keV band of $\Gamma_{\rm eff}<1.0$ are generally
obscured AGNs (obscured AGNs can also have larger $\Gamma_{\rm eff}$ 
values, but use of a larger $\Gamma_{\rm eff}$ threshold can lead
to uncertain AGN identifications). The effective 
photon index from a simple power-law fit
is a useful first-order indicator of spectral shape
even when the observed spectrum does not precisely have a power-law
form. It can be estimated based upon direct \xray\ spectral 
fitting or, in cases of limited counts, based upon a hardness/band 
ratio. In $\Gamma_{\rm eff}<1.0$ cases, the effective photon index is 
generally flat owing to \xray\ absorption and/or Compton reflection. 
The \xray\ binary populations that dominate the emission 
from starburst/normal galaxies are empirically found to produce 
steeper \xray\ spectra with $\Gamma_{\rm eff}\simgt 1.5$. 

\item 
{\it X-ray variability.\/} 
Rapid \xray\ variability by large amplitudes is commonly seen among 
those AGNs where the direct continuum produced close to the SMBH 
is observable. This variability is generally stronger than that
seen from collections of \xray\ binaries in starburst/normal 
galaxies (\citealp[e.g.,][]{young2012}). Furthermore, as noted 
above, some of the \xray\ surveys in Table~\ref{surveystable} 
allow variability studies over much longer timescales. Significantly 
variable sources that also have \xray\ luminosities larger than 
$\approx 10^{41}$~erg~s$^{-1}$ are likely to be AGNs, although
there are rare notable exceptions (\citealp[e.g.,][]{kaaret2001,webb2010}). 

\item 
{\it X-ray position.\/}
The \xray\ positions of AGNs are generally coincident with
the apparent nuclei of their host galaxies (as opposed to, e.g., 
\xray\ binaries, which can be identified across the extent of
the host galaxy often as off-nuclear sources). This 
positional coincidence can be checked 
for relatively low-redshift objects ($z\simlt 0.5$) when 
high-resolution \xray\ (e.g., \chandra) and optical/NIR 
(e.g., \hst) imaging are available (\citealp[e.g.,][]{lehmer2006}). 

\end{enumerate}

\begin{figure}
\includegraphics[scale=0.37,angle=0]{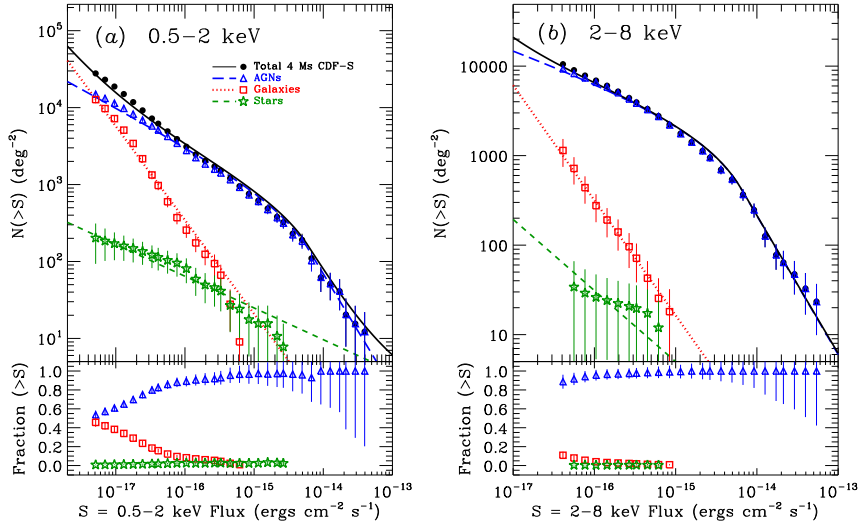}
\caption{Cumulative number counts for the 4~Ms \hbox{CDF-S} in the 
(a) \hbox{0.5--2~keV} and (b) \hbox{2--8~keV} bands. The total 
number counts (black) have been apportioned by source class, as 
labeled, into AGNs (blue), galaxies (red), and stars (green). 
The bottom portions of each panel show the fractional contributions
of each source class to the cumulative number counts. 
Note that AGNs remain the numerically dominant source 
population down to faint fluxes, although at still-fainter 
\hbox{0.5--2~keV} fluxes galaxies will become numerically dominant. 
The AGN number counts reach $\approx 14,900$~deg$^{-2}$ at the 
faintest \hbox{0.5--2~keV} fluxes, and this is the highest sky 
density of reliably identified AGNs found at any wavelength. 
Taken from \citet{lehmer2012}.}
\label{fig-lehmer2012-numcounts}       
\end{figure}

\noindent
Some of these methods have a long history (\citealp[e.g.,][]{maccacaro1988} 
for method~3) while others have been developed/refined more recently. A few 
in-depth applications of these methods include 
\citet[][]{alexander2005},
\citet[][]{brusa2010}, 
\citet[][]{laird2010}, 
\citet[][]{xue2011}, 
\citet[][]{lehmer2012}, 
\citet[][]{civano2012}, and 
\citet[][]{wang2013}. 
Note that some of these methods rely upon having fairly precise redshift
information available while others depend much less upon redshift;
AGN samples can often be selected reasonably well using methods 
\hbox{3--6} together prior to redshift determination. 
AGNs are generally found to make up \hbox{75--95}\% of the sources by
number in current \xray\ surveys, with their percentage contribution 
dropping with survey depth as many starburst/normal galaxies are 
detected at faint fluxes (primarily at low \xray\ energies). 
The precise fractional contribution from AGNs as a function of 
survey depth has been quantified in number counts apportioned by 
source type (see Fig.~\ref{fig-lehmer2012-numcounts}; 
\citealp[e.g.,][]{bauer2004,civano2012,lehmer2012}). 

In addition to the approaches above relying upon the direct use of
\xray\ data, approaches relying upon independent 
multiwavelength data can also be used for AGN selection/confirmation 
from a sample of \xray\ sources. These include the detection of  
broad and/or high-ionization emission lines in optical/NIR spectra, 
high surface brightness radio core emission or extended radio jets/lobes, 
strong infrared emission from hot dust heated by an AGN continuum, and 
distinctive optical variability. 
In well-studied \xray\ and multiwavelength survey fields, multiple 
independent methods have been applied to cross-validate AGN candidates, 
leading to the most reliable and pure samples of distant AGNs available. 



\section{AGN demographics}
\label{demographics}


\subsection{The status of AGN evolution studies before \chandra\ and \xmm}
\label{pre-2000}

The number-density evolution of the AGN population over cosmic time has 
been a topic of intense interest since the 1960's (\citealp[e.g.,][]{schmidt1968}). 
Early work focused on the evolution of luminous quasars and radio
galaxies detected in wide-field optical/radio surveys
(\citealp[see, e.g.,][]{hartwick1990,hewett1994,boyle2001,osmer2004,merloni2013} 
for reviews), and such wide-field studies have continued to advance until the 
present (\citealp[e.g.,][]{wall2005,richards2006,massardi2010,ross2013}). 
Wide-field optical surveys remain largely constrained to the relatively
rare systems where the AGN significantly outshines the host galaxy, owing 
to the selection techniques employed (typically based upon source colors
measured in photometric data). 

Prior to the launches of \chandra\ and \xmm, the population of luminous 
quasars had been well established to evolve strongly over cosmic time, 
peaking in number density at \hbox{$z\approx 2$--3}. Most analyses up
to the year $\approx 2000$ found that the form of the evolution at $z\simlt 2.5$ 
could be fit acceptably with pure luminosity evolution models, although 
there were reports of more complex evolution at bright magnitudes
(\citealp[e.g.,][]{hartwick1990,boyle2000}). Many researchers 
reasonably expected that the basic evolutionary behavior derived for
luminous quasars would also apply at lower luminosities. 

The German-USA-UK \rosat\ mission was the most effective 
surveyor of the distant \xray\ \hbox{(0.1--2.4~keV)} 
universe prior to \chandra/\xmm, contributing to a number of fundamental 
results on AGN demographics. The \rosat\ Deep Survey in the Lockman Hole, 
resolving \hbox{70--80}\% of the \hbox{0.5--2~keV} CXRB,
directly showed that AGNs produce most of the background in this band
(\citealp[e.g.,][]{hasinger1998,schmidt1998}). The responsible 
AGNs showed a range of optical spectral types, including broad-line
and narrow-line spectra, and \xray-to-optical flux 
ratios [$-1\simlt \log (f_{\rm 0.5-2\,keV}/f_R)\simlt 1$]. 
Results on the \xray\ luminosity 
function (XLF) derived from \rosat\ surveys spanning a range of depths 
indicated that luminosity-dependent density evolution (LDDE) described the
data better than pure luminosity evolution or pure density evolution
(\citealp[e.g.,][]{miyaji2000,miyaji2001}; but see also \citealp[][]{page1997}), 
importantly indicating a luminosity dependence of AGN evolution. 
At higher energies, the Japanese-USA \asca\ and Dutch-Italian 
\sax\ missions were able to resolve $\approx 30$\% 
of the \hbox{2--10~keV} CXRB, and follow-up studies indicated 
that the majority of the sources in this band were also AGNs 
(\citealp[e.g.,][]{ueda1998,fiore1999}). Many of these
sources had hard \xray\ spectra with effective power-law photon indices
of \hbox{$\Gamma_{\rm Eff}=1.3$--1.7}, consistent with longstanding
expectations that obscured AGNs make much of the CXRB which 
has \hbox{$\Gamma_{\rm Eff}=1.4$} from \hbox{2--10~keV}
(\citealp[e.g.,][]{setti1989,comastri1995}). 

The demographics of high-redshift AGNs at \hbox{$z=3$--6} were highly 
uncertain when \chandra\ and \xmm\ began operation. Optical and 
radio surveys of luminous quasars both indicated a consistent strong 
decline in number density above $z\approx 3$ 
(\citealp[e.g.,][]{schmidt1995,shaver1999}). However, \xray\ surveys 
of somewhat less luminous quasars suggested a lack of any strong 
decline (\citealp[e.g.,][]{miyaji2000}) and were consistent with a 
constant number density at $z>2$. Theoretical considerations offered
few further constraints, allowing extremely high quasar space 
densities in principle (\citealp[e.g.,][]{haiman1999}). Given the
available constraints, it was feasible that the Universe was
reionized at \hbox{$z\approx 5$--7} by AGNs. 


\subsection{X-ray luminosity functions and the luminosity dependence of AGN evolution}
\label{lum-funcs}

Owing to the advantages of \xray\ surveys detailed in 
Section~\ref{generalutility}, \chandra\ and \xmm\ have allowed the effective 
selection of distant AGNs, including obscured systems, that are up to $\approx 100$ 
times less bolometrically luminous than those found in wide-field quasar surveys
such as the SDSS (\citealp[e.g.,][]{ross2012}; the vast majority of SDSS broad-line 
quasars are straightforwardly detected in moderate-depth \xray\ surveys).
Objects similar to local moderate-luminosity Seyfert 
galaxies can be identified out to $z\approx 5$. 
The sky density of \xray\ selected AGNs in the deepest \chandra\ surveys 
has now reached $\approx 14,900$~deg$^{-2}$
(see Fig.~\ref{fig-lehmer2012-numcounts}; \citealp[][]{lehmer2012}), 
making them $\approx 500$ times more numerous on the sky than 
SDSS quasars. As another comparison, the deepest optical photometric AGN surveys reaching 
$B\approx 24.5$ have delivered AGN sky densities of $\approx 400$~deg$^{-2}$
(\citealp[e.g.,][]{wolf2004,beck2007}; also see \citealp[][]{palan2013}).
Furthermore, the $\approx 14,900$~deg$^{-2}$ sky density is $\approx 15$ times larger 
than that from the \rosat\ Deep Survey (970~deg$^{-2}$; \citealp[][]{hasinger1998}), 
the deepest \xray\ survey conducted prior to \chandra\ and \xmm. 
\nustar\ is now further broadening the parameter space of discovery, allowing 
improved identification of highly obscured AGNs at relatively bright flux 
levels (\citealp[e.g.,][]{delmoro2014}).

\begin{figure}
\includegraphics[scale=0.4,angle=0]{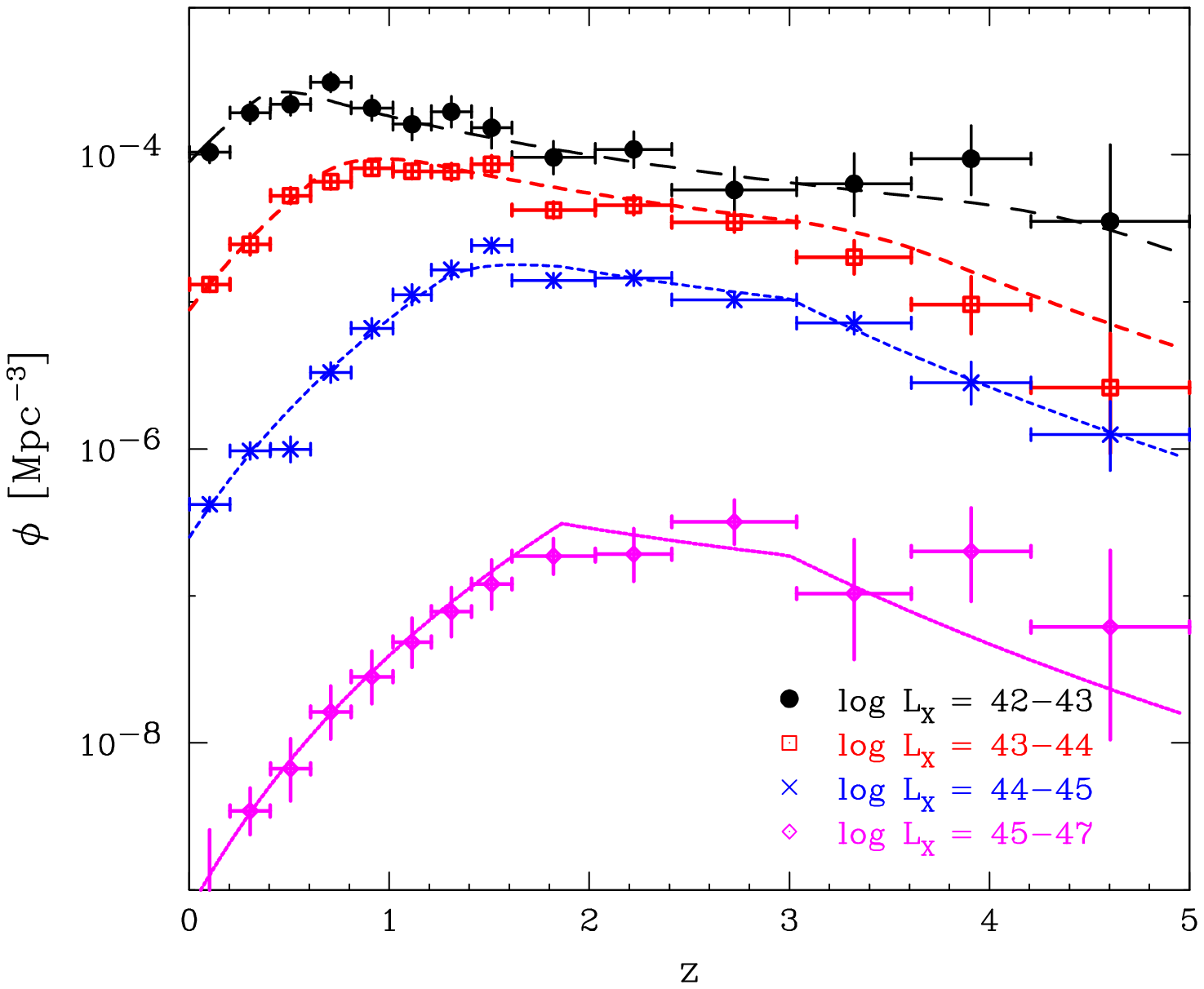}
\includegraphics[scale=0.4,angle=0]{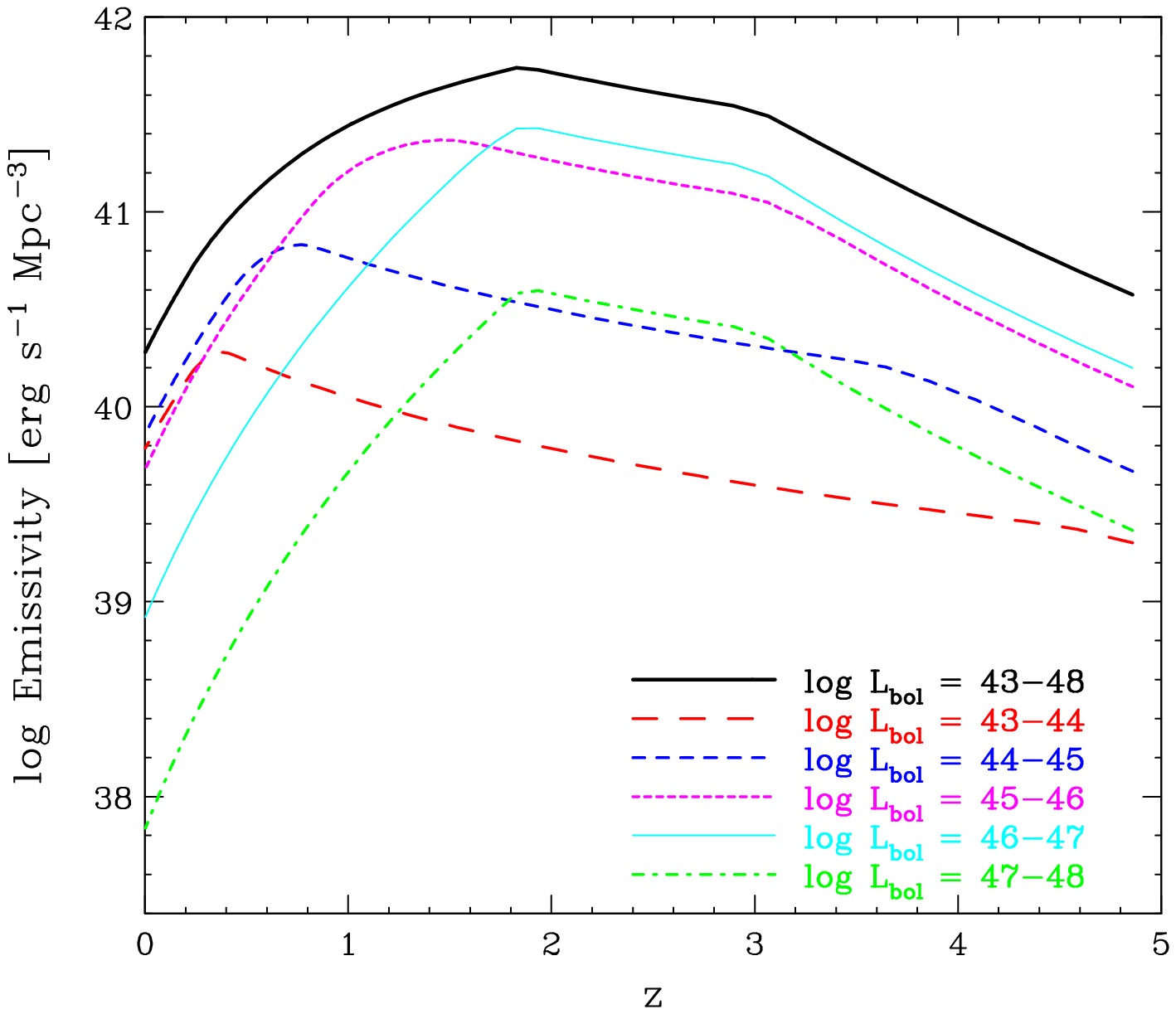}
\caption{(a) Comoving number density vs.\ redshift for AGNs, 
selected from multiple \xray\ surveys, in four rest-frame \hbox{2--10~keV} 
luminosity classes [as labeled; in units of log(erg~s$^{-1}$)]. 
Note that the number density of 
moderate-luminosity AGNs peaks later in cosmic time than that of 
powerful quasars (i.e., AGN cosmic downsizing). 
(b) Comoving bolometric luminosity density vs.\ redshift for the 
same AGN sample in six bolometric luminosity classes 
[as labeled; in units of log(erg~s$^{-1}$)]. 
Note the peak luminosity density at $z\approx 1.8$
for AGNs over the broad range of 
$L_{\rm Bol}=10^{43}$--10$^{48}$~erg~s$^{-1}$.
Taken from \citet{ueda2014}.}
\label{fig-ueda2014}       
\end{figure}

AGN samples from \chandra\ and \xmm\ are now thought to be sufficiently 
complete over a broad part of the luminosity-redshift plane to allow many 
fundamental issues regarding AGN evolution to be addressed (though further 
improvements are still critical; e.g., see Section~\ref{agns-missed}). A key
finding, first reported by \citet{cowie2003} and now a subject of many 
studies, is a notable ``anti-hierarchical'' luminosity dependence of AGN 
evolution, such that the number density of lower-luminosity AGNs peaks 
later in cosmic time than that of powerful quasars 
(see Fig.~\ref{fig-ueda2014}; 
\citealp[e.g.,][]{barger2005,hasinger2005,lafranca2005,silverman2008,ebrero2009,
yencho2009, aird2010,ueda2014}).
This qualitative behavior is also broadly seen in star-forming 
galaxies (\citealp[e.g.,][]{cowie1996}) and is often referred to as 
``cosmic downsizing'', although this term has developed a number of 
usages with respect to galaxies 
(\citealp[e.g.,][]{bundy2006,cimatti2006,faber2007,fontanot2009}). 
AGN downsizing was not widely anticipated prior to its observational
discovery; AGN population synthesis models for the CXRB at the time 
generally adopted the basic evolutionary behavior of luminous quasars 
for AGNs of all luminosities. The most \xray\ 
luminous AGNs with \hbox{$L_X=10^{45}$--$10^{47}$~erg~s$^{-1}$} 
peak in number density at \hbox{$z\approx 2$--3}, consistent with the behavior 
of optically selected quasars, while more common AGNs with 
\hbox{$L_X=10^{43}$--$10^{44}$~erg~s$^{-1}$} peak at 
\hbox{$z\approx 0.8$--1.5}.
Fig.~\ref{fig-ueda2014} shows that the
luminosity dependence of AGN evolution appears sufficiently important
to shift the overall peak of cosmic SMBH power production to
lower redshifts ($z\approx 1.5$--2) than would be expected solely from the 
study of luminous quasars ($z\approx 2$--3). Roughly, at 
$z<1$, \hbox{$z=1$--2}, and $z>2$ the integrated fractions of 
SMBH growth (i.e., the total accreted mass onto SMBHs) 
are broadly comparable at \hbox{25--35\%}, \hbox{37--47\%}, 
and \hbox{23--33\%}, respectively. 
AGNs with \hbox{$L_X=10^{44}$--$10^{45}$~erg~s$^{-1}$} 
dominate SMBH power production at \hbox{$z=1.5$--4}, while those with 
\hbox{$L_X=10^{43}$--$10^{44}$~erg~s$^{-1}$} dominate at lower redshifts
(at very low redshifts of $z\simlt 0.5$, AGNs with 
\hbox{$L_X=10^{41}$--$10^{43}$~erg~s$^{-1}$} also make large fractional
contributions).  
Ongoing \nustar\ surveys at higher \xray\ energies up to $\approx 24$~keV, while 
providing valuable insights, do not suggest any qualitative revisions to this 
basic picture (\citealp[e.g.,][]{alexander2013,delmoro2014}; 
Civano et~al, in prep; Mullaney et~al, in prep); e.g., the vast 
majority of the \nustar\ survey sources were previously detected
by \chandra\ and/or \xmm. 
The downsizing behavior of AGNs has also now been found in 
optically selected (\citealp[e.g.,][]{bongiorno2007,croom2009,shen2012,ross2013})
and radio-selected (\citealp[e.g.,][]{massardi2010,rigby2011}) samples, 
confirming the generality of this phenomenon, although \xray\ selected 
samples remain the most effective at constraining the downsizing behavior. 

Measurement of the quantitative details of the downsizing behavior for \xray\ 
AGNs depends upon many challenging issues, including 
initial detection completeness (and corrections for missed AGNs), 
multiwavelength counterpart identification, 
completeness in redshift determination, 
\xray\ spectral modeling (e.g., \xray\ absorption corrections 
to luminosity estimates), and 
statistical methodology in XLF calculation. 
Thus, while the basic downsizing phenomenon appears securely 
established, there is still some remaining debate over the
precise form of the XLF and its evolution. The most recent
in-depth studies have proposed either LDDE (\citealp[e.g.,][]{ueda2014}) 
or luminosity and density evolution (LADE; \citealp[e.g.,][]{aird2010}); 
in the latter the shape of the XLF is constant with redshift, 
but it undergoes strong luminosity evolution at $z\simlt 1$, and 
overall negative density evolution toward increasing redshift.  
Especially at high redshifts of $z\simgt 3$, the LDDE and LADE models
predict quite different numbers of AGNs. The latest high-redshift 
constraints appear to favor the LDDE model, but further testing
is needed. 

The observed AGN downsizing behavior seen via the measured XLF
could arise due to changes in the mass of the typical active SMBH and/or
changes in the typical accretion rate. A variety of modeling efforts have 
been made to understand the physical nature of AGN downsizing, including 
analytic models, semi-analytic models, and large-scale numerical simulations 
(\citealp[e.g.,][]{hopkins2008,degraf2010,fanidakis2012,
hirschmann2012,hirschmann2014,menci2013}); 
these efforts sometimes also attempt to model simultaneously the growth of 
AGN host galaxies and their downsizing behavior. The numerical simulations 
continue to advance rapidly and include many of the mechanisms relevant 
to SMBH fueling and growth, including galaxy interactions, disk instabilities,
and gas cooling. They have had genuine success in plausibly reproducing the 
apparent basic anti-hierarchical behavior of SMBH growth within the context of the 
hierarchical paradigm for cosmic structure formation. This being said, 
even the most intensive simulations to date lack the spatial/mass resolution 
to model in detail the still uncertain but essential accretion and feedback 
processes operating on small scales, and approximate ``sub-grid'' approaches 
are often adopted for these processes. 

Given the immense modeling challenges, it is understandable that the 
model predictions for XLF evolution and the nature of downsizing end up 
differing in detail. Broadly, and as would initially be expected from the 
XLF results, the models generally indicate that more massive SMBHs grew 
earlier in cosmic time. Some furthermore quantitatively predict a decline 
in average Eddington ratio with redshift
(\citealp[e.g.,][]{fanidakis2012,hirschmann2012,hirschmann2014}). 
The strong early drop in number density 
of the luminous AGN population is predicted to result from 
the exhaustion of cold gas in massive halos 
due to strong early star formation and feedback, as 
well as a decline over time in merging activity. Less-luminous AGNs 
evolve more mildly and can remain numerous later in cosmic time, 
however, as the gas content of their generally lower mass halos 
evolves more mildly. A significant fraction of the less-luminous AGNs 
are also remnants of formerly luminous AGNs that have faded; 
i.e., objects with low Eddington ratios, perhaps intermittently
triggered, whose massive SMBHs are no longer rapidly growing (in
a fractional sense) but can still appear as AGNs. Additional 
observational evidence consistent with this basic picture 
includes estimates of SMBH masses and Eddington 
ratios for distant AGNs in \xray\ surveys (see Section~\ref{agn-fraction}) 
and observations of the mass-dependent growth timescales of local SMBHs 
(\citealp[e.g.,][]{heckman2004}). 

\chandra\ and \xmm\ have also greatly clarified the demographics
of AGNs at \hbox{$z=3$--6}, although there is still scope for
significant improvements. In contrast to the earlier suggestions 
from \rosat\ surveys (see Section~\ref{pre-2000}), the \xray\ data now 
clearly support an exponential decline in the number density of luminous AGNs 
above $z\approx 3$ (\citealp[e.g.,][]{barger-highz2003,cristiani2004,fontanot2007,
silverman2008,brusa2009,civano2011,
fiore2012,hiroi2012,vito2013,vito2014,kalf2014,ueda2014}), ruling 
out some of the more exotic early predictions (\citealp[e.g.,][]{haiman1999}) 
by $\approx 2$ orders of magnitude. Furthermore, quantitative comparisons of space 
densities for optically selected quasars (\citealp[e.g.,][]{mcgreer2013}) 
and \xray\ selected quasars indicate statistical agreement to
within factors of \hbox{2--3}. At lower AGN luminosities, the situation 
is significantly less clear, largely owing to the small solid angles
with sufficiently sensitive \xray\ data as well as substantial challenges with 
spectroscopic/photometric follow-up studies. The current data generally
suggest a decline in space density for moderate-luminosity AGNs
at $z>3$ (\citealp[e.g.,][]{fiore2012,vito2013,vito2014,kalf2014,ueda2014}). 
Some recent work has found this decline may be less pronounced at 
the lowest luminosities, and if this trend holds
then ``cosmic upsizing'' would apply in this regime, perhaps consistent 
with expectations for hierarchical structure formation in the early
universe (i.e., given the young age of the Universe, most SMBHs would not
yet have sufficient masses to generate high luminosities). Improved 
measurements are required for clarification, especially 
for the most highly obscured AGNs at high redshift (\citealp[e.g.,][]{gilli2011}). 
The available space-density estimates indicate that, in the absence of 
dramatic changes in the XLF at very low luminosities, SMBHs probably did not 
produce sufficient power to reionize the Universe at \hbox{$z\approx 5$--7}; 
they likely have secondary effects upon reionization 
(\citealp[e.g.,][]{grissom2014}; but see \citealp[][]{giallongo2012}). 
Stacking analyses of high-redshift galaxies have also set upper limits on the
accreted mass density in black holes out to $z\approx 8$, providing
useful inputs into models of early SMBH formation
(\citealp[e.g.,][]{cowie2012,fiore-proc2012,basuzych2013,treister2013}). 


\subsection{AGNs missed in cosmic X-ray surveys and their importance}
\label{agns-missed}

As noted in Section~\ref{generalutility}, \xray\ surveys do have (small) 
shortcomings, and thus it is essential to utilize well-matched multiwavelength 
surveys to find AGNs missed by the \xray\ technique (as well as to characterize
more reliably the underlying bolometric luminosities of \xray\ detected AGNs; see 
Section~\ref{followup-multiwavelength}). Particularly important are
missed AGNs that still have sufficiently large bolometric luminosities 
to contribute materially to cosmic SMBH growth. The primary way that luminous 
AGNs can be missed in \xray\ surveys involves heavy obscuration, and that will be
the focus of this section below. Additionally, however, there is growing evidence 
that a small fraction of the luminous AGN population is intrinsically \xray\ weak
(\citealp[e.g.,][]{leighly2007,wu2011,luo2014,teng2014}), thereby mildly
challenging the rule of universal luminous \xray\ emission from 
AGNs (see point~1 of Section~\ref{generalutility}). Even in the absence of 
obscuration, such AGNs will often be difficult to detect, but thankfully 
current assessments indicate that intrinsically \xray\ weak AGNs are 
sufficiently rare that they should not substantially impact demographic 
studies (\citealp[e.g.,][]{gibson2008,wu2011,luo2014}). 

Heavy obscuration with \hbox{$N_{\rm H}=(5$--$50)\times 10^{23}$~cm$^{-2}$} or
more, as is commonly seen in the local universe, can diminish the measurable 
\xray\ emission from an AGN to below the limits of detectability, even 
for intrinsically luminous AGNs in deep \xray\ surveys. The 
factor of diminution depends upon the sampled rest-frame energy 
band, the absorption column density, and the absorption geometry (the latter 
setting the level of flux that is Compton reflected/scattered around the 
absorber); the factor can be \hbox{10--100} or more in the \hbox{0.5--10~keV} band 
for large column densities (\citealp[e.g.,][]{comastri2004,burlon2011}). 
Furthermore, owing to the largely energy independent nature of Compton 
scattering below a couple hundred keV, observations at tens of keV offer 
only limited improvements for sources with highly Compton-thick absorption column 
densities. 

\begin{figure}
\includegraphics[scale=0.32,angle=0]{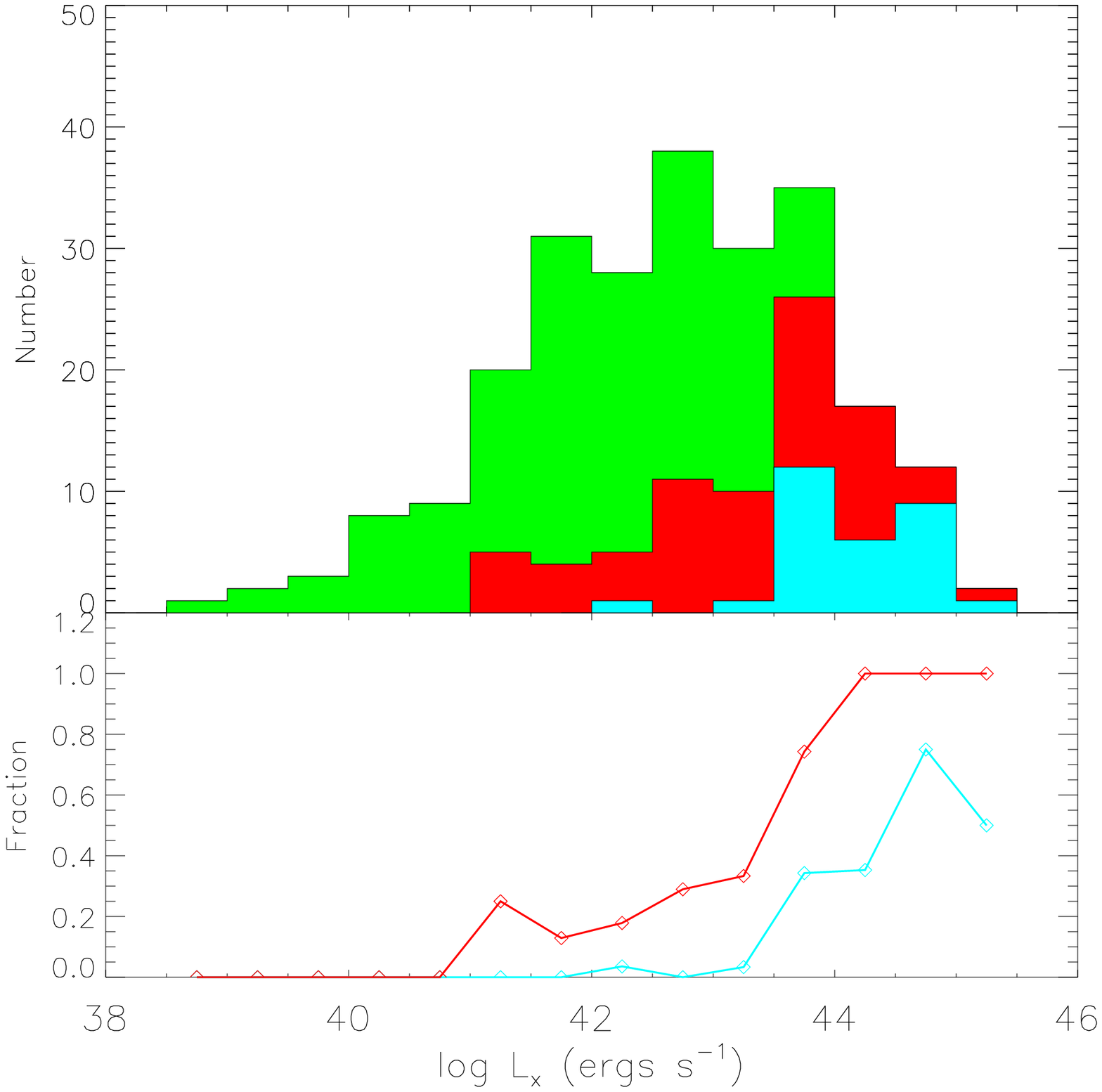}
\hspace{-0.8cm}
\includegraphics[scale=0.32,angle=0]{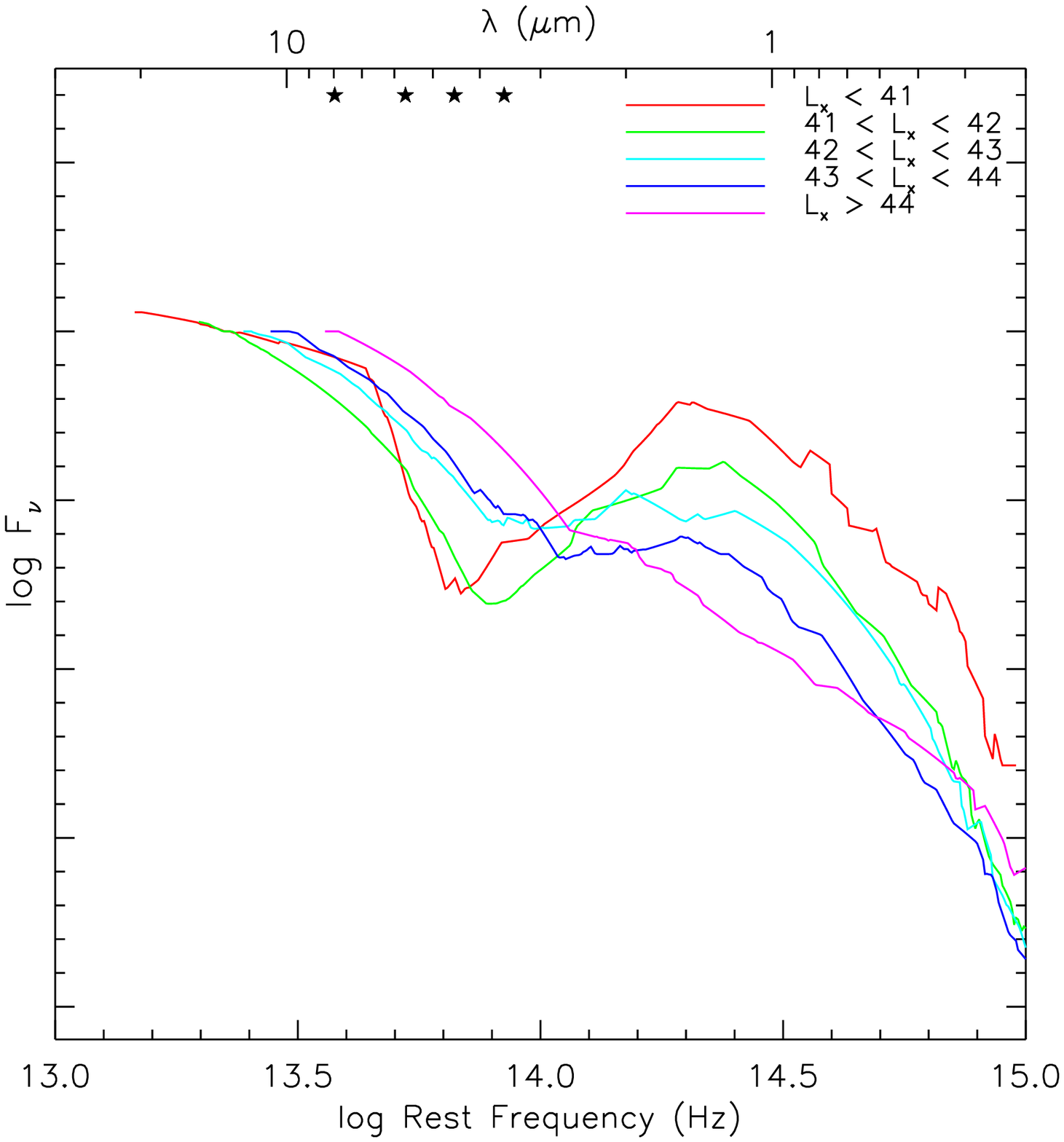}
\caption{(a) Distributions of \xray\ luminosity for \xray\ detected
\spitzer\ IRAC sources in the \hbox{CDF-N} (green histogram), 
AGNs selected by the \citet{lacy2004} MIR color-color criteria 
(red histogram), and MIR power-law galaxies (cyan histogram). 
The bottom panel shows the fraction of \xray\ sources meeting the 
Lacy et~al. (red) or power-law (cyan) criteria. These MIR criteria 
are the most effective at selecting luminous AGNs and do not 
perform as well for moderate-to-low luminosity AGNs. 
(b) Median best-fit optical-to-MIR SEDs of the \xray\ detected
\spitzer\ IRAC sources separated into groups by \xray\ luminosity 
[as labeled; in units of log(erg~s$^{-1}$)]. 
At high \xray\ luminosities the SEDs show a distinctive 
power-law shape, while moving toward lower luminosities the stellar 
bump increasingly dominates. The observed-frame wavelengths of the 
four IRAC bands are shown near the top of this panel as stars. 
Adapted from \citet{donley2007}.}
\label{fig-donley2007-spitzeragn}       
\end{figure}

Multiple methods have been applied to identify AGN candidates that are
undetected in the \xray\ band, and tens of papers have now been written
on this subject. Arguably the most successful methods have 
utilized infrared observations, largely from \spitzer, that
home in on AGN ``waste heat'' (i.e., AGN-heated dust emission) 
and have reduced extinction bias. Such infrared selection techniques 
include the following: 

\begin{enumerate}

\item
{\it Color-color selection from MIR photometric data\/}
(\citealp[e.g.,][]{lacy2004,stern2005,polletta2006,hickox2007,
cardamone2008,donley2008,donley2012,mateos2013}). 
See Fig.~\ref{fig-donley2007-spitzeragn}. 

\item
{\it Power-law selection from MIR photometric and/or MIR spectroscopic data\/}
(\citealp[e.g.,][]{alonsoherrero2006,donley2007,alexander2008,donley2008,park2010}). 
See Fig.~\ref{fig-donley2007-spitzeragn}. 

\item
{\it Excess MIR emission relative to SFR expectations\/}
(\citealp[e.g.,][]{daddi2007,donley2008,treister-integral2009,alexander2011,luo2011}). 

\item
{\it High 24~$\mu$m-to-optical flux ratios and red optical/NIR colors\/}
(\citealp[e.g.,][]{fiore2008,fiore2009,treister-agn-sfgal2009}). 

\item
{\it MIR spectroscopic selection based on deep 9.7~$\mu$m Si features\/}
(\citealp[e.g.,][]{georgantopoulos2011}). 
\
\end{enumerate}

\noindent
Optical, radio, and other techniques have also been utilized to 
identify AGN candidates, including the following: 

\begin{enumerate}

\item
{\it Optical spectroscopic selection based on high-ionization emission lines\/}
(\citealp[e.g.,][]{steidel2002,hunt2004,alexander2008,trump-wfc32011,
juneau2013,juneau2014,vignali2014}). 

\item
{\it Optical variability in multi-epoch photometric data\/}
(\citealp[e.g.,][]{boutsia2009,villforth2010,villforth2012,sarajedini2011}). 
   
\item
{\it Excess radio emission relative to SFR expectations\/}
(\citealp[e.g.,][]{donley2005,delmoro2013}). 

\item
{\it Radio morphology with jet/lobe structure and/or 
radio spectral properties\/}
(\citealp[e.g.,][]{muxlow2005,barger2007}). 

\item
{\it Selected populations in the color-mass diagram\/}
(\citealp[e.g.,][]{xue2012}). 

\end{enumerate}

\noindent
All of these methods have their strengths and weaknesses, in terms
of completeness and reliability 
(e.g., see Fig.~\ref{fig-donley2007-spitzeragn}), and they are best 
utilized together to enable cross-checking of candidates. \xray\ stacking 
analyses using samples of AGN candidates derived with the methods above 
often show hard average \xray\ spectral shapes, indicating that at least
some highly obscured AGNs are indeed present. However, large uncertainties 
remain about the fraction of candidates that are bona-fide AGNs
and the corresponding AGN luminosities, particularly among the 
infrared-selected samples, and much further 
candidate characterization is required; e.g., with deeper or harder 
\xray\ observations and high-quality optical/NIR/MIR spectroscopy. 
Furthermore, many of these methods, on their own, are not effective
at distinguishing between highly obscured and moderately 
obscured AGNs. 

The \xray\ undetected obscured AGNs presumably produce the $\approx 25$\% of the
\hbox{6--8~keV} CXRB that remains unresolved 
(see Section~\ref{main-surveys}), and this 
provides one integral limit upon their overall importance to 
cosmic SMBH growth (\citealp[e.g.,][]{xue2012,ueda2014}); also see
Section~\ref{soltan}. If they have
low intrinsic luminosities, as some analyses suggest, they might
increase the current AGN number counts ($\approx 14,900$~deg$^{-2}$;
\citealp[][]{lehmer2012}) by 50\% or more. 


\subsection{The \soltan\ argument for X-ray selected AGNs}
\label{soltan}

The established relations between the bulge properties of local 
galaxies and the masses of the SMBHs at their centers 
(\citealp[e.g.,][]{kormendy2013,shankar2013}; and references therein)
allow estimation of the total local mass density of SMBHs
($\rho_{\rm \bullet,loc}$, in M$_\odot$~Mpc$^{-3}$). Following 
\citet{soltan1982}, this quantity then serves 
as an integral constraint upon the allowed amount of
total cosmic SMBH growth.
Multiple authors have performed this local SMBH mass-density estimation; e.g., 
\citet{marconi2004} estimate $\rho_{\rm \bullet,loc}=(4.6^{+1.9}_{-1.4})\times 10^5$~M$_\odot$~Mpc$^{-3}$,  
\citet{vika2009} estimate $\rho_{\rm \bullet,loc}=(4.8\pm 0.7)\times 10^5$~M$_\odot$~Mpc$^{-3}$, and 
\citet{li2011} estimate $\rho_{\rm \bullet,loc}=(5.8\pm 1.2)\times 10^5$~M$_\odot$~Mpc$^{-3}$. 
The $\rho_{\rm \bullet,loc}$ values derived over about the past decade have 
generally been in acceptable, though not perfect, agreement. However,  
\citet{kormendy2013} have recently argued that some past SMBH mass estimates 
need to be revised upward by a factor of about \hbox{2--4}, owing to 
improvements in data, improvements in modeling, and the identification
of downward biases in emission-line based SMBH masses (see their Section 6.6). 
This leads to an increase in $\rho_{\rm \bullet,loc}$ by a factor 
of $\approx 2$ (L.C. Ho 2014, private communication). 
Given that this change would be well in excess of the error bars of past 
$\rho_{\rm \bullet,loc}$ estimates, it is clear that one must presently tread 
cautiously when using the \soltan\ argument! The size of the 
required upward revision of $\rho_{\rm \bullet,loc}$ is still a matter of
debate (A. Marconi 2014, private communication). 

X-ray AGN demographers have integrated AGN bolometric luminosity functions, 
derived from XLFs via bolometric corrections, to estimate the total 
amount of cosmic SMBH growth, $\rho_{\rm \bullet,XLF}$,
for comparison to $\rho_{\rm \bullet,loc}$
(\citealp[e.g.,][]{marconi2004,shankar2013,ueda2014}; also see 
\citealp[][]{2014MNRAS.439.2736D} for a recent infrared-based 
perspective). Acceptable 
agreement can be achieved, broadly supporting the idea that radiatively 
efficient gas accretion (i.e., AGN phases) has driven much of cosmic SMBH growth
(significant SMBH merging is allowed, provided the merging progenitors 
grew via radiatively efficient accretion). 
For example, \citet{ueda2014} find 
$\rho_{\rm \bullet,XLF}=(3.9\pm 0.6)\times 10^5 \eta^{-1}_{0.1}$~M$_\odot$~Mpc$^{-3}$, where
$\eta_{0.1}$ is the average mass-to-energy conversion efficiency of 
accretion divided by 0.1. $\eta_{0.1}\approx 0.8$ would give consistency 
with $\rho_{\rm \bullet,loc}$ estimates from \hbox{2000--2012}, while
$\eta_{0.1}\approx 0.4$ would be required if the recent upward revision 
of \citet{kormendy2013} is adopted. For comparison, accretion onto
a Schwarzschild SMBH would give $\eta_{0.1}\approx 0.57$ or higher
(with the exact $\eta_{0.1}$ depending upon the role of magnetic
stress at the innermost stable circular orbit around the SMBH), 
while (prograde) accretion onto a Kerr SMBH would give $\eta_{0.1}$ 
as high as $\approx 3.6$ 
(\citealp[e.g.,][]{agol2000,noble2009,noble2011}; 
J.H. Krolik 2014, private communication). The 
$\rho_{\rm \bullet,XLF}$ vs.\ $\rho_{\rm \bullet,loc}$
agreement is thus fair, though the implied accretion 
efficiency would appear low for upward revisions of 
$\rho_{\rm \bullet,loc}$ as 
suggested by \citet{kormendy2013}. This could
be remedied if XLFs still suffer from incompleteness 
due to, e.g., highly obscured and/or intrinsically \xray\ weak AGNs 
(see Section~\ref{agns-missed}).\footnote{Additional (likely smaller)
relevant factors to consider include 
(1) XLF incompleteness due to uncertainties in the masses and 
growth processes of the high-redshift ``seeds'' of SMBHs (see 
Section~\ref{lum-funcs}), and 
(2) the ejection of SMBHs from galactic nuclei due to gravitational-wave 
production in SMBH merger events. 
\citet{volonteri2013} and \citet{gilfanov2014} provide further
discussion of these factors.}
Such objects are very difficult to detect in \xray\ surveys, 
making this hypothesis challenging to assess. 
Detailed considerations also 
indicate that $\eta$ may depend significantly upon SMBH mass and 
redshift (\citealp[e.g.,][]{li2012,shankar2013,ueda2014}).

The \soltan\ argument can be utilized in a
differential, rather than an integral, manner to investigate
the evolution of the mass function of SMBHs 
(\citealp[e.g.,][]{marconi2004,shankar2013,ueda2014}). The most 
robust conclusion arising from such work is that more massive 
SMBHs generally grew earlier in cosmic time
(also see Section~\ref{lum-funcs}). Unfortunately, present 
uncertainties in bolometric corrections, $\eta$, $\rho_{\rm \bullet,loc}$, 
and XLFs limit the precision of \soltan-argument constraints 
for, e.g., constraining the amount of SMBH growth
missed by \xray\ surveys (see Section~\ref{agns-missed}).  
\soltan-argument constraints are currently a good 
first-order ``sanity check'' but should not be over-interpreted. 


\subsection{The environmental dependence of AGN evolution}
\label{environmental}

Theoretical models of structure formation predict that galaxy growth
is environmentally dependent (i.e., it is accelerated in high-density
regions;
\citealp[e.g.,][]{1996MNRAS.281..487K,2006MNRAS.366..499D}). Observational
support for this hypothesis comes from the finding that the most
evolved and massive spheroids reside in galaxy clusters (high-density
regions) by the present day
\citep[e.g.,][]{baldry2004,2009MNRAS.392.1265S}. How is the growth of
SMBHs influenced by the large-scale environment? A powerful way to
address this question is to compare the fraction of galaxies hosting
AGNs in galaxy clusters (and their high-redshift progenitors,
protoclusters) to that in the field, as measured from 
blank-field \xray\ surveys.

To first order, the fraction of galaxies in clusters hosting \xray\ 
luminous AGNs ($L_{\rm X}\simgt10^{43}$~erg~s$^{-1}$) is found to
evolve strongly with redshift out to $z\approx$~1, qualitatively
similar to what is seen for the field-galaxy population. 
However, the cluster AGN fraction is lower than that in the field
by about an order of magnitude at low redshift, and it appears to
rise more rapidly with redshift
(\citealp[e.g.,][]{2007ApJ...664L...9E,2009ApJ...701...66M,2013ApJ...768....1M}). 
This evolution of the AGN fraction with redshift also broadly tracks that
seen for the star-forming galaxy population in galaxy
clusters. There is evidence that the relative suppression of
AGNs in galaxy clusters is dependent on the richness of the
cluster, the cluster-centric radius explored, and the adopted
luminosity threshold for AGN activity
(\citealp[e.g.,][]{kocevski2009,2012MNRAS.425.1215K,2014MNRAS.442..314K,
2014MNRAS.437.1942E,2014arXiv1407.8181E,2014A&A...567A..83K}),
complicating comparisons between different studies. In less extreme
large-scale environments, such as galaxy groups, there is no clear
evidence for a decrease in the AGN fraction when compared to the field
out to at least $z\approx$~1
(\citealp[e.g.,][]{2008MNRAS.391..183G,2009ApJ...695..171S,2013A&A...552A.111P,2014ApJ...790...43O}).

The source statistics are more limited at $z\simgt 1$, but, on the
basis of current results, the AGN fraction in galaxy clusters at
\hbox{$z=1.0$--1.5} appears broadly similar to that in the field
\citep[e.g.,][]{2013ApJ...768....1M}. At yet higher redshifts there is
a deficit of galaxy cluster systems, although studies of AGNs in 
protoclusters (large-scale overdense regions at $z\simgt 2$) have found an
order of magnitude {\it increase} in the AGN fraction when compared to
the field
\citep[e.g.,][]{2009ApJ...691..687L,2013ApJ...765...87L,2010MNRAS.407..846D},
an apparent reversal of what is seen at $z<1$. There is also tentative
evidence that the AGN fraction in protoclusters increases in regions of
higher galaxy density
\citep[e.g.,][]{2009ApJ...691..687L,2013ApJ...765...87L}, potentially
the opposite to what is seen in massive lower-redshift galaxy clusters
\citep[e.g.,][]{2014MNRAS.437.1942E}. Overall, the current results
suggest that the large-scale environment has an effect on the growth
of SMBHs, which is presumably driven by the availability of a cold-gas
supply in the vicinity of the SMBH and may be sparse in massive,
evolved (i.e., virialized) galaxy clusters but is prevalent in 
less-evolved galaxy clusters and protoclusters
\citep[e.g.,][]{2009MNRAS.395...11V}. However, there are potentially
conflicting results between the suite of published studies, which may
be driven by a number of effects, including the analyses employed to
measure AGN activity, the approach adopted when comparing to non
AGNs (e.g., luminosity and mass thresholds when calculating the AGN
fraction), and the selection of the galaxy clusters and protoclusters.

The large-scale environment can also be quantified with the two-point
correlation function, which measures the clustering strength of
selected populations and provides an estimate of the dark-matter halo
mass. Clustering-strength measurements for \xray\ selected AGNs over
the broad redshift range of $z\approx$~0--3 imply a characteristic
halo mass of $\approx10^{12}$--$10^{13}$~$M_{\odot}$
\citep[e.g.,][]{2009ApJ...701.1484C,hickox2009,2010ApJ...716L.209C,2013MNRAS.428.1382K},
with no highly significant trends with obscuration or luminosity
\citep[e.g.,][]{2006A&A...457..393G,2008ApJ...674L...5P,
2009ApJ...701.1484C,2009A&A...494...33G,allevato2011,2012ApJ...746....1K,2013MNRAS.428.1382K}. This
range in halo mass is in agreement with that found for luminous
star-forming galaxies, optically selected quasars, and massive
galaxies, but it is lower than that found for radio-selected AGNs
(\citealp[e.g.,][]{2009ApJ...701.1484C,hickox2009,2011ApJ...731..117H,2012MNRAS.421..284H,2014MNRAS.443.3327G};
see Fig.~5 of \citealp[][]{alexander2012}). On the basis of
dark-matter halo models, $\approx10^{13}$~$M_{\odot}$ broadly
corresponds to the maximum mass where a halo can support a large
cold-gas supply---the gas in halos significantly more massive than
this is mostly in a hot form, which is less easily accreted onto the
SMBH
\citep[e.g.,][]{2006MNRAS.370.1651C,2009MNRAS.394.1109C,2009MNRAS.395..160K}. The
dark-matter halo may, therefore, have a strong controlling influence on
the fuelling of AGNs
\citep[e.g.,][]{2010MNRAS.405L...1B,2011ApJ...737...50V}.

Further useful constraints can be placed by measuring how \xray\ AGNs
are distributed in dark-matter halos using the halo occupation
distribution (HoD;
e.g.,\ \citealp[][]{2002ApJ...575..587B,2005ApJ...633..791Z})
formalism. The main parameters of the HoD are the fraction of central
and satellite galaxies hosting AGN activity as a function of the halo
mass. Current constraints suggest a preference for
\xray\ AGNs to reside in central galaxies, with $<5$\% identified
with satellites
\citep[e.g.,][]{2011ApJ...741...15S,2013ApJ...774..143R}; however,
models with \xray\ AGNs solely hosted in satellite galaxies can 
also fit the observed clustering in some cases
\citep[e.g.,][]{2011ApJ...726...83M}.



\section{AGN physics}
\label{physics}


The large and relatively complete samples of AGNs detected in 
\xray\ surveys can provide unique insights, usually of a 
statistical character, into the physical processes that shape
their emission (i.e., ``AGN physics''). 
These physical processes span scales from 
the immediate vicinity of the SMBH (light minutes to light
hours) to that of the obscuring material (light days to light
years). Such survey-based physical investigations critically
complement the in-depth targeted \xray\ studies of individual and small 
samples of AGNs that are another major thrust of \xray\ astronomy
\citep[e.g.,][]{mushotzky1993,reynolds2003,fabian2006,turner2009,done2010,gilfanov2014}. 
In this section, we will briefly review results on AGN physics 
coming from three key areas of survey-based investigation: 
the basic properties, luminosity dependence, and 
redshift dependence of AGN \xray\ obscuration
(Section~\ref{obscuration}), 
\xray-to-optical/UV SEDs (Section~\ref{aox-results}), and  
the \xray\ continuum shape and its connection to 
Eddington ratio (Section~\ref{xray-continuum}). 


\subsection{AGN X-ray obscuration: Basic properties, 
luminosity dependence, and redshift dependence} 
\label{obscuration}

Understanding of the nature of the obscuring material in AGNs, often referred 
to generally as the obscuring ``torus'' of orientation-based unification models
\citep[e.g.,][]{antonucci1993}, continues to improve rapidly. In recent years, 
for example, it has become possible to obtain direct estimates of the 
(luminosity dependent) extent of 
the torus \citep[e.g.,][]{burtscher2013,koshida2014} and to ascertain its 
apparently clumpy nature \citep[e.g.,][]{elitzur2006,mor2009}. Studies based on \xray\ 
surveys have advanced understanding of AGN obscuration in several regards. 
First, \xray\ surveys provide arguably the clearest evidence that the majority 
of the AGNs in the Universe are obscured. Basic \xray\ spectral analyses of the 
detected AGNs in deep \xray\ surveys find that the majority show evidence for
obscuration \citep[e.g.,][]{dwelly2006,tozzi2006,merloni2014}, 
even before accounting for biases against detecting heavily
obscured objects (see Section~\ref{agns-missed}). The inferred column-density 
distribution of the underlying obscured population, before separation into 
luminosity and redshift bins, appears to peak roughly around 
$N_{\rm H}\approx 10^{23}$~cm$^{-2}$ with an approximately log-normal shape
having a logarithmic $\sigma \approx 1$. This result implies that a substantial
fraction of AGNs will be highly obscured and even Compton-thick. Additionally,
there is a significant minority of low-obscuration systems with $N_{\rm H}$ values
consistent with zero, in agreement with expectations from unification models.  

Another key advance, where recent \xray\ surveys covering a broad part of 
the luminosity-redshift plane have contributed substantially, has been in 
clarifying the long-suspected \citep[e.g.,][]{lawrence1982,lawrence1991}
luminosity dependence of the fraction of obscured AGNs 
\citep[e.g.,][]{treister2006,hasinger2008,burlon2011,brightman2014,merloni2014,ueda2014}. 
The fraction of AGNs showing \xray\ obscuration 
drops strongly with rising \xray\ luminosity, from 
$\approx 70$\% at $L_{\rm 2-10~keV}=10^{43}$~erg~s$^{-1}$ to 
$\approx 20$\% at $L_{\rm 2-10~keV}=10^{45}$~erg~s$^{-1}$ (for $z<1$). 
Results from recent 
infrared \citep[e.g.,][]{ballantyne2006,treister2008,assef2013,lusso2013,toba2014} 
and optical \citep[e.g.,][]{simpson2005} surveys broadly confirm this basic 
revision of orientation-based unification models
(but see \citealp[][]{lawrence2010}). AGNs of higher luminosities 
may be able to evacuate or destroy, via radiative feedback, circumnuclear 
gas and dust more effectively, leading to a so-called ``receding torus''
with larger opening angle \citep[e.g.,][]{menci2008,nenkova2008}. 
Alternatively, luminosity-dependent changes 
in the underlying \xray-to-optical/UV SED 
(see Section~\ref{aox-results}) may drive changes in the absorption 
properties; e.g., if the obscuring material is often in the form of a 
radiatively driven wind. The precise numerical values for obscured fractions 
differ between different authors and have remaining systematic
uncertainties owing to, e.g., selection biases against highly obscured 
AGNs and limited photon statistics in absorption spectral modeling; at 
faint \xray\ fluxes currently only crude hardness-ratio based absorption 
estimates can be effectively used, and these cannot appropriately characterize
complex \xray\ absorption (\citealp[e.g.,][]{mayo2013,buchner2014}). 
Moreover, wide-field infrared surveys suggest, 
somewhat surprisingly, that the fraction of highly 
obscured AGNs may rise upward substantially again at the highest 
luminosities ($L_{\rm Bol}\sim 10^{47}$~erg~s$^{-1}$)
reaching $\approx 50$\% \citep[e.g.,][]{assef2014,stern2014}. 
At the highest AGN luminosities, the nature of the material typically 
providing the obscuration may change from the standard small-scale torus 
to something else, perhaps more distributed gas and dust that has been 
perturbed by galaxy major mergers leading to high Eddington-ratio AGNs
\citep[e.g.,][]{draper2010}. 

The fraction of AGNs showing \xray\ obscuration, after allowing for
luminosity effects, also appears to rise with redshift
\citep[e.g.,][]{treister2006,hasinger2008,hiroi2012,iwasawa2012,
vito2013,vito2014,brightman2014,merloni2014,ueda2014}. Recent studies find 
this rise can be parameterized as proportional to $(1+z)^{0.4-0.6}$ at least 
up to $z\approx 2$, beyond which the uncertainties become substantial. 
Such evolution of the obscured fraction might arise due to the
generally greater availability of nuclear gas and dust in galaxies at
earlier cosmic times. There remains debate regarding whether this 
redshift evolution applies for all AGNs or primarily for the 
most-luminous ones
(\citealp[e.g.,][]{iwasawa2012,vito2013,merloni2014,ueda2014}); 
a luminosity dependence of the evolution might arise if low-to-moderate 
luminosity AGNs and high-luminosity AGNs have different fueling
mechanisms (see Section~\ref{host-morphs}). 
This evolution of AGNs on the physical scale of
the absorbing medium is notably not accompanied by apparent
small-scale evolution of the accretion disk and its corona
(see Section~\ref{aox-results}). 

Finally, we note briefly that this section has primarily focused 
on \xray\ obscuration. The relations between \xray\ obscuration 
and that at other wavelengths (e.g., optical/UV reddening, blocking
of line emission from the Broad Line Region, UV absorption lines) 
remain an extremely complex issue requiring much further work,
and understanding such relations thoroughly will elucidate the 
structure and kinematics of the obscuring material
(\citealp[e.g.,][]{mainieri2007,trouille2009,
lawrence2010,luo2010,merloni2014} and references therein).  
   

\subsection{Basic X-ray-to-optical/UV spectral energy distribution properties} 
\label{aox-results}

One of the most effective ways to investigate the accretion physics 
of the SMBHs in AGNs is to study their overall broad-band SEDs.
The part of the AGN SED spanning from the \xray\ to the 
optical/UV (hereafter, the ``\xray-to-optical/UV SED'')
is the spectral region where direct accretion emission dominates for 
relatively unobscured systems (after subtraction/removal of the 
light from the AGN host galaxy; see Section~\ref{broad-band-emission}). 
Studies of \xray-to-optical/UV SEDs have a 
long history \citep[e.g.,][]{avni1986}, and they continue to 
deliver fundamental insights as improved data, analysis techniques, 
and theoretical modeling become available. In addition to 
providing constraints upon accretion physics, these studies 
also are critically important, e.g., when

\begin{enumerate}

\item
Assessing the universality of AGN \hbox{X-ray} emission and
selecting remarkable AGNs that deviate from the typical 
\xray-to-optical/UV SED (see Section~\ref{generalutility} 
and \ref{agns-missed}).

\item 
Making bolometric corrections in the \soltan\ argument 
(see Section~\ref{soltan}). 

\item 
Modeling the winds responsible for many of the absorption-line properties 
of AGNs and likely feedback; the UV and extreme-UV (EUV) continuum largely 
drives these winds, which must also be protected from overionization by
the underlying \xray\ emission \citep[e.g.,][]{murray1995,proga2000,luo2014}. 

\end{enumerate}

While the \xray-to-optical/UV portion of the AGN SED is complex
\citep[e.g.,][]{vasudevan2009,trump2011,elvis2012,jin2012,scott2014}, its 
first-order properties can be described in large samples with 
the use of simple parameters. The most common such parameter 
is \aox, defined to be the slope of a nominal power law 
connecting the rest-frame 2500~\AA\ and 2~keV monochromatic 
luminosities; i.e., 
$\alpha_{\rm ox}=0.3838 \log(L_{\rm 2~keV}/L_{2500~\mathring{\rm{A}}})$.
For systems where the direct accretion emission is dominant, 
\aox\ compares the relative amounts of power coming from the 
optically thick accretion disk (at rest-frame 2500~\AA) and 
the accretion-disk corona (at rest-frame 2~keV).

\begin{figure}
\includegraphics[scale=0.31,angle=0]{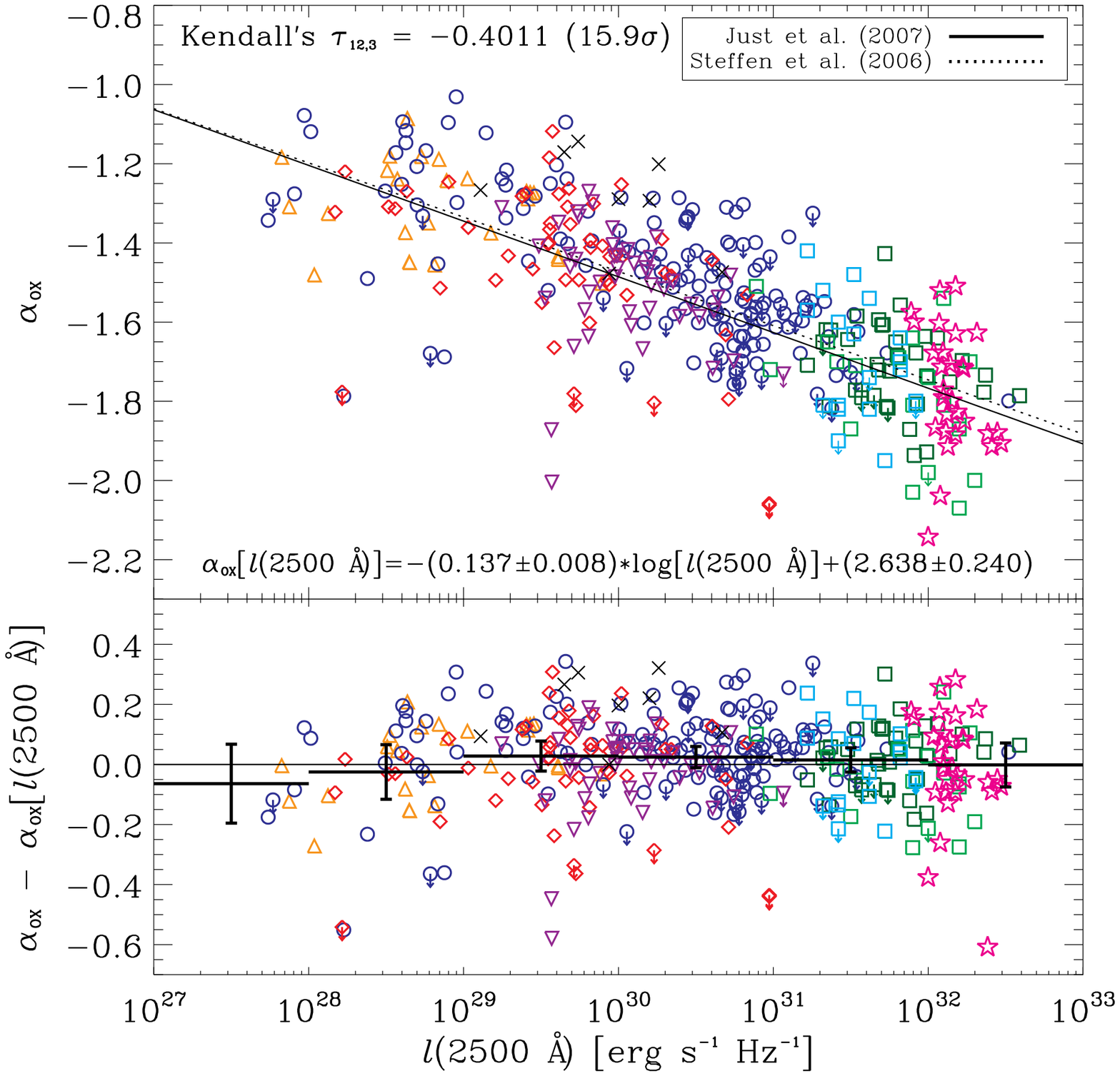}
\includegraphics[scale=0.29,angle=0]{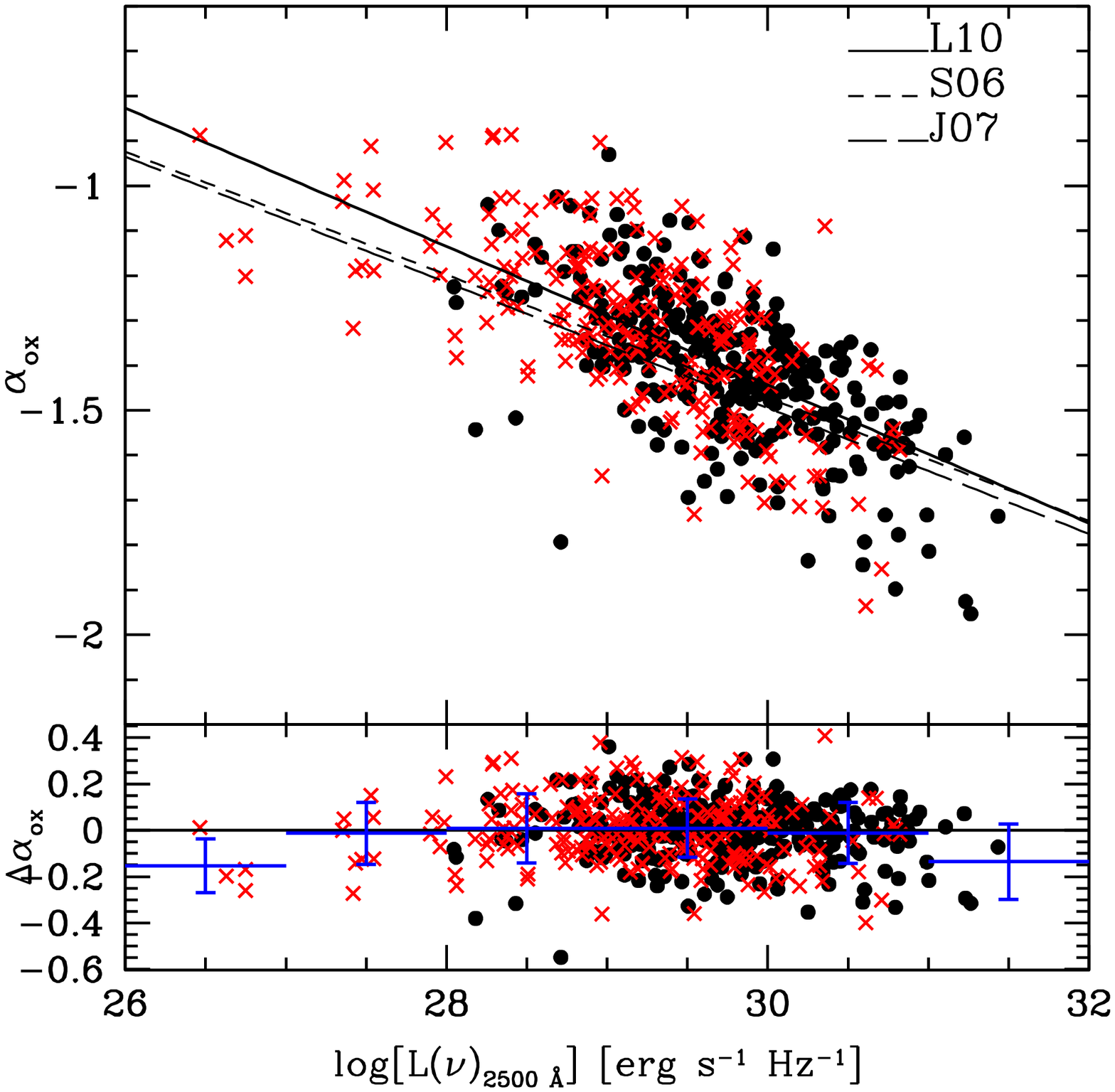}
\caption{The \aox\ parameter vs.\ 2500~\AA\ monochromatic luminosity 
for (a) optically selected AGNs from \citet{steffen2006} and 
\citet{just2007}, and (b) \xray\ selected AGNs from
\citet{lusso2010}. Larger negative values of \aox\ correspond to
weaker \xray\ emission relative to optical/UV emission. The bottom 
portions of each panel show the residuals relative to the best-fit relation. 
In panel (a), the Kendall's $\tau$ coefficient and its significance are
presented, along with the functional form of the best-fit relation. In 
panel (b), the abbreviated references for the plotted fits refer 
to \citet{steffen2006}, \citet{just2007}, and \citet{lusso2010}. 
The symbol colors/types in each panel denote the AGN samples
utilized; see the cited papers for details. 
AGNs become relatively \xray\ weaker with increasing
optical/UV luminosity, over a very wide range of this luminosity.
Taken from \citet{just2007} and \citet{lusso2010}.}
\label{fig-aox-luminosity}       
\end{figure}

X-ray surveys have now provided sensitive and often relatively 
uniform coverage for substantial numbers of unobscured or mildly 
obscured AGNs that span a broad part of the AGN luminosity-redshift 
plane. These have been used, often in conjunction with other AGN 
samples, to extend studies of \xray-to-optical/UV SEDs using 
\aox\ \citep[e.g.,][]{steffen2006,just2007,kelly2007,green2009,lusso2010,young2010}; 
importantly, sensitive \xray\ surveys have allowed the majority population 
of moderate-luminosity AGNs at high redshift to be included in such 
analyses. \aox\ shows a clear correlation with optical/UV luminosity 
($L_{2500~\mathring{\rm{A}}}$) at all investigated redshifts \hbox{($z=0$--6)}, such that 
AGNs with higher $L_{2500~\mathring{\rm{A}}}$ produce less \xray\ emission per unit 
optical/UV emission (see Fig.~\ref{fig-aox-luminosity}). 
This finding is qualitatively in agreement with  
earlier studies \citep[e.g.,][]{avni1986,anderson1987,wilkes1994}, 
and the correlation with $L_{2500~\mathring{\rm{A}}}$ appears stronger
and tighter than that with estimates of $L_{\rm Bol}/L_{\rm Edd}$. The 
correlation is now well established over at least five orders of magnitude 
in $L_{2500~\mathring{\rm{A}}}$, and the observed range of \aox\ corresponds 
to a substantial range of $\approx 20$ in \xray\ vs.\ optical/UV luminosity. 
There is considerable object-to-object scatter, corresponding 
to a factor of $\approx 3$ in \xray\ vs.\ optical/UV luminosity in 
either direction around the \aox-$\log (L_{2500~\mathring{\rm{A}}})$ 
relation. Some of this scatter simply arises 
from (generally) non-simultaneous observations of AGNs that vary both
in the \xray\ and optical/UV, but the majority appears to be genuine 
intrinsic scatter \citep[e.g.,][]{gibson2012,vagnetti2013}. 
Additional physical parameters beyond $L_{2500~\mathring{\rm{A}}}$, such  
as $L_{\rm Bol}/L_{\rm Edd}$, likely can explain much of this scatter
\citep[e.g.,][]{kelly2008,shemmer2008,lusso2010,young2010,jin2012}. 
The \aox-$\log (L_{2500~\mathring{\rm{A}}})$ relation is also likely nonlinear
\citep[e.g.,][]{steffen2006,maoz2007,green2009,vagnetti2013}, 
appearing flatter at low luminosities and steeper at high luminosities, 
but detailed constraints on the form of the nonlinearity remain limited. 

The functional form and parameters of the \aox-$\log (L_{2500~\mathring{\rm{A}}})$ 
relation provide fundamental constraints that any successful model of the 
SMBH disk-corona system must be able to reproduce. For example, as large-scale 
numerical magnetohydrodynamic simulations of black-hole accretion flows 
continue their rapid advances \citep[e.g.,][]{schnittman2013}, it is expected 
that researchers will soon be able to determine which physical parameters 
($L_{\rm Bol}/L_{\rm Edd}$, SMBH mass, SMBH spin, magnetic-field structure) set 
the disk-to-corona power balance and the observed \xray-to-optical/UV SED.
Given this importance for accretion models, it is of concern that the 
recent improved studies of the \aox-$\log (L_{2500~\mathring{\rm{A}}})$ relation, 
while agreeing on its existence, sometimes disagree about its slope/intercept 
and functional form; e.g., fitted parameters disagree by more than is 
allowed by their statistical uncertainties. Until such systematic uncertainties 
in the observational results are resolved, it will be difficult to use them 
to inform physical disk-corona models with high fidelity. 
Furthermore, \aox\ was defined at rest-frame 2500~\AA\ and 2~keV largely 
for observational convenience, rather than for fundamental reasons, and 
it is not obvious that these are the optimal wavelength/energy choices 
to consider for characterization of \xray-to-optical/UV SEDs. In this
vein, \citet{young2010} have considered the effects of varying these 
choices. The slope of the \aox-$\log (L_{\rm Opt})$ relation does depend 
significantly upon chosen \xray\ energy, generally becoming steeper/flatter 
as the definition is moved to lower/higher energies. On the other hand, the 
slope does not appear to depend strongly upon the chosen optical/UV wavelength. 

Most recent studies of \hbox{X-ray}-to-optical/UV SEDs using \aox\ find
no significant evolution with redshift
\citep[e.g.,][]{steffen2006,just2007,green2009,lusso2010}; 
the tightest limits require the \xray-to-optical/UV luminosity ratio 
to change by $\simlt 30$\% out to \hbox{$z=5$--6}. However, there are
counterclaims finding evidence for redshift evolution 
\citep[e.g.,][]{kelly2007}, and the issue requires further scrutiny
with even further improved samples. The current consensus is that, in 
spite of the dramatic evolution of AGN number densities over cosmic time
(see Section~\ref{lum-funcs}), the inner accretion properties of the 
individual AGN unit are notably stable. 


\subsection{X-ray continuum shape as an estimator of Eddington ratio} 
\label{xray-continuum}

As the primary indicator of SMBH growth rate, the Eddington ratio is of
critical importance in studies of SMBH 
demographics (see Section~{\ref{lum-funcs}), 
physics (see Section~{\ref{aox-results}), and 
ecology (see Section~\ref{agn-fraction}). 
$L_{\rm Bol}/L_{\rm Edd}$ is typically derived by estimating the mass of 
a SMBH (and thus its Eddington limit) and also its bolometric 
luminosity. Unfortunately, mass estimates and bolometric corrections
for SMBHs in the distant universe generally have substantial uncertainties
(e.g., see \citealp[][]{shen2013} and \citealp[][]{peterson2014} for
discussion of virial SMBH mass estimators and \citealp[][]{hao2014}
for discussion of bolometric corrections; also see
Section~\ref{agn-fraction}). Thus, it is 
important to have as many methods as possible for $L_{\rm Bol}/L_{\rm Edd}$ 
estimation, so that different approaches can be cross checked. 

\begin{figure}
\hspace{1.5cm}
\includegraphics[scale=0.8,angle=0]{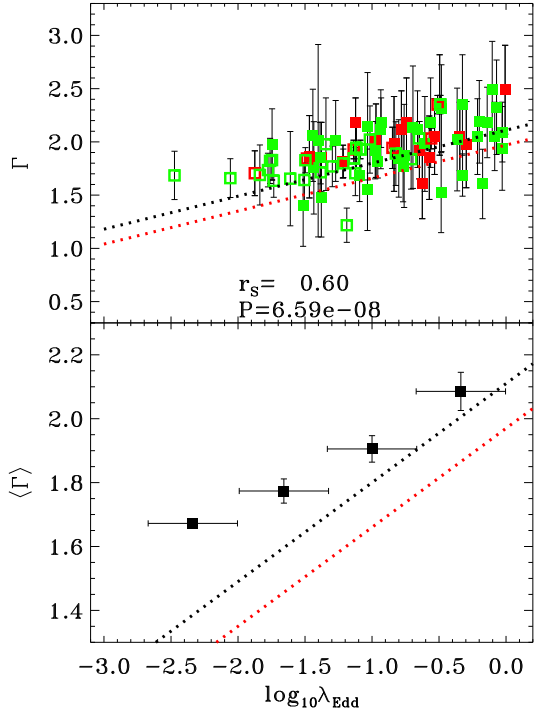}
\caption{Hard \xray\ power-law photon index vs.\ Eddington ratio 
($\lambda_{\rm Edd}=L_{\rm Bol}/L_{\rm Edd}$) for a sample of AGNs from 
COSMOS and \hbox{E-CDF-S}. The top panel shows results for the
individual AGNs, where red and green points indicate $L_{\rm Bol}/L_{\rm Edd}$
measurements based upon H$\alpha$ and Mg~{\sc ii} virial estimators,
respectively. A correlation is clearly present (the Spearman-test correlation 
coefficient and probability are listed), albeit with significant 
object-to-object scatter. The bottom panel shows binned average 
values for groups of individual AGNs from the top panel. The 
black and red dotted lines show earlier $\Gamma-L_{\rm Bol}/L_{\rm Edd}$ 
correlations from \citet{shemmer2008} and \citet{risaliti2009}, 
respectively. The apparent systematic offsets of the three correlations
may be due to differences in SMBH mass estimation and spectral fitting
methodology. Taken from \citet{brightman2013}.} 
\label{fig-brightman2013-gamma-eddington}       
\end{figure}

It has long been suspected that the intrinsic hard \xray\ 
\hbox{(rest-frame 2--10~keV)} power-law photon index ($\Gamma$)
of a radio-quiet AGN can be used as an estimator of $L_{\rm Bol}/L_{\rm Edd}$ 
\citep[e.g.,][]{pounds1995,brandt1997}. Higher $L_{\rm Bol}/L_{\rm Edd}$ 
is expected to lead to increased Compton cooling of the accretion-disk
corona, and thus steeper \hbox{2--10~keV} power-law spectra (i.e., larger values 
of $\Gamma$; higher $L_{\rm Bol}/L_{\rm Edd}$ also often leads to the production
of a strong ``soft \xray\ excess'' affecting the spectrum below rest-frame 
2~keV). The early suspicions have now been confirmed via both 
targeted \citep[e.g.,][]{shemmer2006,shemmer2008} and 
\xray\ survey-based \citep[e.g.,][]{risaliti2009,jin2012,brightman2013,fanali2013}
studies, which find clear $\Gamma-L_{\rm Bol}/L_{\rm Edd}$ correlations 
across a broad range of luminosity and redshift (and no apparent redshift
dependence); see Fig.~\ref{fig-brightman2013-gamma-eddington}. 
The $\Gamma-L_{\rm Bol}/L_{\rm Edd}$ correlation does have significant 
object-to-object scatter (a factor of $\approx 3$, when high-quality 
$\Gamma$ measurements are available; \citealp[e.g.,][]{shemmer2008}) and thus, 
as with other methods of $L_{\rm Bol}/L_{\rm Edd}$ estimation, it is best used in 
a statistical sense to characterize the average Eddington ratio of a 
sample of objects. The $\Gamma-L_{\rm Bol}/L_{\rm Edd}$ technique, of course, 
requires a reliable measurement of the {\it intrinsic\/} power-law photon 
index; i.e., corrected for \xray\ absorption and Compton-reflection effects. 
The penetrating nature of \hbox{2--10~keV} \xrays\ naturally  
mitigates absorption effects, and this technique should be effective for 
moderate column densities up to $N_{\rm H}\approx 10^{22.5}$~cm$^{-2}$ where some other 
techniques fail (e.g., due to reddening of optical line emission from the
Broad Line Region). It may be possible, with broad-band \xray\ coverage and
in-depth modeling, to recover the intrinsic photon index for even larger
values of $N_{\rm H}$ \citep[e.g.,][]{arevalo2014,puccetti2014}.
Finally, the $\Gamma-L_{\rm Bol}/L_{\rm Edd}$ technique, 
once calibrated, is a {\it direct\/} $L_{\rm Bol}/L_{\rm Edd}$ estimator that does
not require intermediate estimation of SMBH mass. In fact, it can be 
utilized to serve as another SMBH mass-estimation technique via the
dependence of $L_{\rm Edd}$ on SMBH mass
\citep[e.g.,][]{shemmer2008}.  



\section{AGN ecology}
\label{ecology}

Ecology refers to the relationship between ``organisms'' and their
physical surroundings. From the point of view of this review, the
ecology of AGNs refers to the relationship between the AGN and the
host-galaxy environment.\footnote{In this review we use the term
host-galaxy environment to describe the properties of the host
galaxy (e.g.,\ mass, color, morphology, SFR) rather
than the large-scale environment in which the galaxy resides.}
Until comparatively recently AGNs were considered an exotic phenomenon
with no close connection to their host galaxies. However, over the
past 20~years, the identification of tight relationships between the
mass of the SMBH and various host-galaxy properties for nearby
galaxies (e.g., the bulge luminosity, mass, and velocity dispersion;
\citealp[][]{korm95,mago98,gebh00,kormendy2013}) has comprehensively
dismissed this view by implying that (1) most (if not all) massive
galaxies have hosted AGN activity at some time over the past
$\approx$~13~Gyr of cosmic evolution and (2) the growth of SMBHs and
galaxies is connected, potentially via a link between AGN activity and
star formation (the two primary processes whereby SMBHs and galaxies
are grown). Since the majority of the growth of SMBHs has occurred at
$z\simgt0.2$ (see Section~\ref{lum-funcs} and
Fig.~\ref{fig-ueda2014}), \xray\ surveys of distant AGNs provide key
insight into when, where, and how these SMBHs grew, and can shed light
on any connections between AGN activity and star formation (when
combined with other multiwavelength observations).

In this section we review research on the host galaxies of distant
\xray\ AGNs and explore if anything ``special'' is happening in the
AGN-hosting galaxies by making comparisons to galaxies that lack AGN
activity. We start with a brief overview of the broad-band emission
from galaxies and AGNs to illustrate how the properties of an AGN host
galaxy can be measured without significant contamination from the AGN
itself. To provide the broadest redshift baseline to search for trends
in the host-galaxy properties of distant AGNs we often include
comparisons to the host-galaxy properties of AGNs in the nearby
universe, principally those detected by the {\it Swift}-BAT survey
\citep[e.g.,][]{tuel10bat,2013ApJS..207...19B}.

\subsection{The broad-band emission from galaxies and AGNs}
\label{broad-band-emission}

\begin{figure}
\centerline{
  \includegraphics[width=0.9\textwidth]{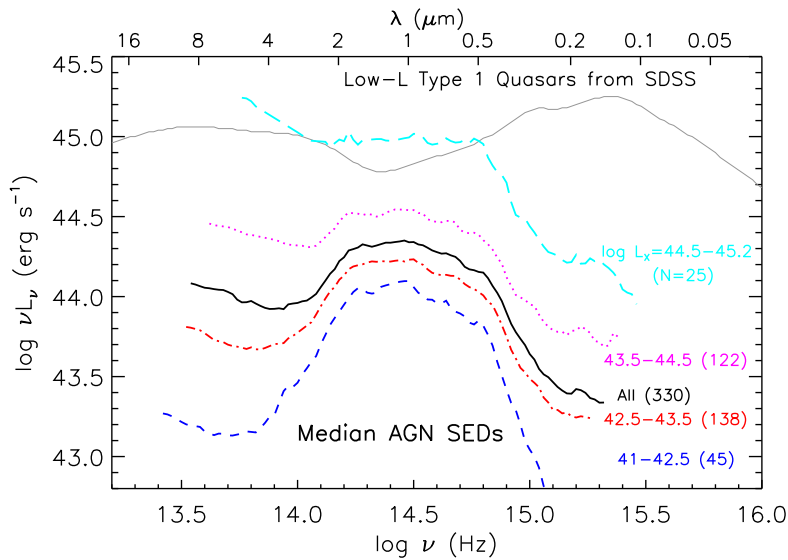}}
\caption{The median rest-frame UV--MIR SEDs of \xray\ AGNs at
  \hbox{$z\approx 0$--4}. The range in \xray\ luminosity [in units of
    log(erg~s$^{-1}$)] for subsets of the \xray\ AGNs are
  shown in addition to the number of \xray\ AGNs used to
  produce the median SEDs. These SEDs are compared to the mean SED of
  the low-luminosity quasars from the SDSS, with the host-galaxy
  contribution removed to indicate the expected SED from a ``pure
  AGN'' (gray curve; \citealp[][]{2006ApJS..166..470R}). The emission
  at rest-frame $\approx$~0.2--4~$\mu$m from the \xray\ AGNs
  is predominantly due to starlight from the host galaxy since the
  majority of the AGNs are obscured or intrinsically weak at optical
  wavelengths. Taken from \citet[][]{luo2010}.}
\label{luo}       
\end{figure}

The bulk of the emission from galaxies is produced at
UV--submillimeter wavelengths (\hbox{$\approx$~0.1--1000~$\mu$m};
\citealp[e.g.,][]{2012ARA&A..50..531K}) and is primarily due to the
radiation produced by populations of stars as well as AGN activity,
when present. The intrinsic emission from these stellar populations
peaks at UV--NIR wavelengths and corresponds to the (approximate)
black-body radiation from stars over a range of masses and ages
(\citealp[][]{1979ApJS...40....1K}). This intrinsic emission is
typically modified by the presence of dust, particularly in regions of
young and forming stars, which are generally optically thick to
short-wavelength UV--NIR radiation. Consequently, the emission from
young and forming stars is often most efficiently measured using
infrared observations since the starlight will heat the dust and
thermally reradiate the emission at FIR wavelengths
(\hbox{$\lambda\approx$~30--300~$\mu$m}; typical temperatures of
\hbox{$T\approx$~10--100~K}). The majority of the emission from
galaxies undergoing intense star formation is, therefore, produced at
FIR wavelengths, while the majority of the emission from quiescent
galaxies (i.e., those with little on-going star formation) is produced
at UV--NIR wavelengths.

A significant fraction of the AGNs detected in \xray\ surveys are
obscured or intrinsically weak at UV--NIR wavelengths; see
Section~\ref{generalutility} and \ref{obscuration}. While this makes
it challenging to determine the properties of the AGN over that band
pass, it allows for convenient measurements of the host-galaxy
properties without significant contamination from the AGN
\citep[e.g.,][]{simmons2008,silv09xcosmos_env,2010MNRAS.405..718P,xue10xhost,lusso2011}.
See Fig.~\ref{luo} for the SEDs of distant \xray\ AGNs, showing the
host-galaxy emission peaking at $\approx$~0.2--4~$\mu$m. Reliable
host-galaxy measurements for the population of luminous and unobscured
AGNs are also possible provided the AGN emission can be accurately
constrained (and therefore removed) at UV--NIR wavelengths
\citep[e.g.,][]{2010ApJ...713..970A,2010ApJ...708..137M,bongiorno2012,2012A&A...540A.109S}.
The most basic host-galaxy properties that we can measure at UV--NIR
wavelengths are luminosity, color, and morphology. The host-galaxy
color provides a basic characterization of the galaxy integrated
stellar populations and, when combined with the luminosity, a reliable
estimate of the mass of the host galaxy
\citep[e.g.,][]{2001ApJ...550..212B,2009MNRAS.400.1181Z,2013ARA&A..51..393C}. The
morphology can provide clues to the formation and evolution of
galaxies and, when spatially resolved color information is available,
basic constraints on the stellar populations across the galaxy. We
explore the host-galaxy masses, colors, and morphologies of \xray\ AGNs
in Section~\ref{host-masses} and \ref{host-morphs}.

AGNs are often luminous at infrared wavelengths due to the thermal
emission from dust in the vicinity of the accretion disk (e.g., the
putative obscuring torus; \citealp[][]{antonucci1993}), potentially
contaminating infrared-based SFR estimates. However, since the
accretion disk is hotter than young stars, the dust is typically
heated to higher temperatures ($\approx$~100--1000~K) and, therefore,
the majority of the infrared emission from the AGN is shifted to
shorter NIR--MIR wavelengths than that from star formation
($\approx$~3--30~$\mu$m;
\citealp[e.g.,][]{netz07qsosf,mull11agnsed}). Consequently, for many
AGN studies, the FIR emission is taken as a measurement of the SFR;
however, to provide the most accurate SFR constraints it is necessary
to decompose the infrared emission into the AGN and star-formation
components, which is essential for reliable SFR measurements from
intrinsically luminous AGNs
\citep[e.g.,][]{netz07qsosf,syme10spitzer,mull11agnsed,2012ApJ...759..139K,delmoro2013,2014MNRAS.439.2736D}. We
explore the SFRs of \xray\ AGNs in Section~\ref{star-formation}.

\subsection{Host-galaxy masses and colors}
\label{host-masses}

The large numbers of optically obscured and intrinsically weak AGNs
detected in \chandra\ and \xmm\ surveys provided the first detailed
measurements of the host-galaxy properties of distant AGNs without
significant contamination from AGN activity at UV--NIR
wavelengths. Initial studies noted that the AGN host galaxies were
typically luminous ($\simgt{L_{\rm *}}$; i.e., the knee of the galaxy
luminosity function; \citealp[][]{1976ApJ...203..297S}) and have red
optical colors, suggesting massive early type systems (Hubble types
Sa--E;
\citealp[e.g.,][]{2001AJ....122.2156A,2002AJ....123.1149A,barger2001ssa13,2002AJ....124.1839B,2001ApJ...551L...9C,2003A&A...406..483S,2004MNRAS.348..529G}). More
recent studies estimated the host-galaxy stellar masses by fitting the
UV--NIR emission with stellar-population models and applied
appropriate mass-to-light conversions to the host-galaxy
luminosities. These analyses were significantly helped by the launch
of \spitzer\ in 2004, which provided the first extensive rest-frame
NIR data for distant sources using the IRAC instrument (3.6--8~$\mu$m;
\citealp[][]{2004ApJS..154...10F}). As expected, given the results
from the initial studies, the majority of the distant \xray\ AGNs were
found to be hosted by massive galaxies ($>3\times10^{10}$~$M_{\odot}$;
\citealp[e.g.,][]{akiy05xhost,2006ApJ...640...92P,alonsoherrero2008,bund08quench,brus09xhost,xue10xhost,2011A&A...535A..80M,lusso2011}),
with the average stellar mass comparable to that of $M_{\rm *}$
($\approx10^{11}$~$M_{\odot}$; i.e., the knee of the stellar-mass
function; \citealp[][]{cole2001,marchesini2009,ilbert2010}). Distant
AGNs are also identified in lower-mass host galaxies (down to
$\approx10^{7}$--$10^{9}$~$M_{\odot}$) and are most prevalent in the
deepest \chandra\ surveys
\citep[e.g.,][]{shi08xhost,brus09xhost,xue10xhost,xue2012,schramm2013}
but they always comprise a minority of the AGN population in
\xray\ surveys. As discussed in Section~\ref{agn-fraction} this is
more due to challenges in detecting these sources rather than their
intrinsic rarity. No strong trend for an increase or decrease in the
average stellar mass of \xray\ AGNs with redshift is found down to
$z\approx$~0.2. However, by contrast, the average stellar mass of {\it
  Swift}-BAT selected AGNs at $z<0.05$ is found to be substantially
lower than the distant \xray\ AGNs
($\approx2\times10^{10}$~$M_{\odot}$;
\citealp[e.g.,][]{koss11bathost}). This lower average stellar mass is
not obviously due to selection effects since the majority of the
sources have \xray\ luminosities comparable to the distant AGNs
(14--195~keV luminosities of $\approx10^{43}$--$10^{44}$~erg~s$^{-1}$;
\citealp[][]{tuel10bat,koss11bathost}). Therefore, the differences in
the average stellar masses of distant and nearby \xray\ AGNs appear to
be due to either a genuine decrease in the host-galaxy masses over the
narrow redshift range of $z\approx$~0.05--0.2 or different approaches
in the estimation of the host-galaxy masses.


A common diagnostic to characterize the properties of galaxies is the
color-magnitude diagram (CMD), which plots rest-frame optical colors
vs. absolute magnitude and provides insight into the integrated
stellar populations \citep[e.g.,][]{stra01galcol,baldry2004}. The CMD
for galaxies is found to be bimodal out to at least $z\approx$~1--2
\citep[e.g.,][]{bell2004,brammer2009,xue10xhost}, with the majority of
galaxies either falling on the ``red sequence'' or the ``blue cloud'',
which are believed to correspond broadly to quiescent and star-forming
galaxies, respectively. Intense star-forming galaxies can also lie on
the red sequence due to the presence of dust-obscured star formation
rather than quiescent stellar populations
\citep[e.g.,][]{cardamone2010,bongiorno2012,rosario2013colors,wang2013}. Consequently,
the CMD is not, in isolation, a reliable indicator of the degree of
on-going star formation and other analyses are often required to
determine which red-sequence galaxies are quiescent and which are
intensely forming stars \citep[e.g.,][]{rosario2013colors}; see
Section~\ref{broad-band-emission}. A modest fraction of the galaxy
population lies in a narrow ``green valley'' between these two
dominant regions, which is likely to comprise galaxies with a mix of
red and blue stellar populations (e.g.,\ a galaxy with both
significant ongoing star-formation and a significant old stellar
population) in addition to galaxies transiting from the blue cloud to
the red sequence due to the (potentially rapid) shut down of star
formation in the host galaxy
\citep[e.g.,][]{faber2007,martin2007col,hasinger2008,scha09agn,schawinski2014,
cimatti2013,2014ApJ...783...40G,pan2014}.

The first CMD analyses of distant \xray\ AGNs showed that many lie in
the green valley \citep[e.g.,][]{nand07host,silv08host,hickox2009},
where only a minority of the optically selected galaxy population is
found. The distinct difference between the locations of the AGNs and
the optically selected galaxy populations in the CMD implies that
distant AGNs are found in a subset of the galaxy population and,
tentatively, suggests that they are the catalysts for the transition
of galaxies from the blue cloud to the red sequence (e.g.,\ through
the suppression of star formation via energetic winds, outflows, and
jets; see
\citealp[][]{2005ARA&A..43..769V,alexander2012,2012ARA&A..50..455F}
for reviews). However, later studies showed that clear distinctions
between the host-galaxy colors of the AGN and the galaxy populations
mostly disappear when the galaxy sample is matched in mass to the AGN
sample
\citep[e.g.,][]{silv09xcosmos_env,xue10xhost,2010MNRAS.408..139P,rosario2013colors},
with broadly similar fractions of coeval galaxies and AGNs found in
the red sequence, green valley, and blue cloud out to at least
$z\approx$~3.\footnote{Given the evident mass dependence on the colors
  of galaxies, color-{\it mass} diagrams are now often used in
  preference to color-magnitude diagrams for host-galaxy analyses.}
The lack of significant differences in the host-galaxy colors appears
to suggest that, in general, distant \xray\ AGNs are not found in
``special'' host-galaxy environments (at least in terms of the
color-mass plane). This general conclusion is in contrast to that
found for \xray\ AGNs in the nearby universe, where the hosts of {\it
  Swift}-BAT AGNs at $z\simlt$~0.05 are found to have bluer colors
than the coeval galaxy population (when matched in mass to the AGN
sample; \citealp[e.g.,][]{koss11bathost}), suggesting a connection
between the presence of young stars and AGN activity. These apparent
disagreements are not necessarily in contradiction since there is
evidence that the host galaxies of \xray\ AGNs out to $z\approx$~1
have experienced more recent star formation than the coeval galaxy
population (e.g., from the resolved host-galaxy colors and the depth
of the 4000~\AA\ break;
\citealp[e.g.,][]{ammons2009,silv09xcosmos_env,2010MNRAS.408..139P,ammons2011,rosario2013resolvedcols,2014MNRAS.443.3538H});
see Section~\ref{star-formation} for a more detailed discussion of the
star-formation properties of distant AGNs.

\subsection{Host-galaxy morphologies}
\label{host-morphs}

Many of the cosmic \xray\ survey fields have extensive coverage at
UV--NIR wavelengths from {\it Hubble Space Telescope (HST)}
observations. The excellent spatial resolution of {\it HST}
($\approx0.1^{\prime\prime}$) allows moderately detailed
characterization of the rest-frame optical host-galaxy properties of
distant AGNs through kpc-scale measurements of the host-galaxy
morphology and the identification of potential triggers of AGN
activity (e.g., galaxy merger and galaxy interaction signatures). A
number of different approaches for measuring host-galaxy morphologies
have been developed (e.g., automated classifications based on the
distribution of light in the galaxy, two-dimensional fits to the
surface brightness profiles to provide disk/bulge decompositions, and
human ``eyeball'' classification), which appear to give broadly
similar results \citep[e.g.,][]{Huertas-Company2014}. The contribution
of the AGN to the optical emission can significantly bias
morphological measurements
\citep[e.g.,][]{simmons2008,2010MNRAS.408..139P} and, therefore, the
majority of the studies focus on optically obscured or intrinsically
weak AGNs.

Distant \xray\ AGNs are found to reside in a broad range of
host-galaxy types out to at least $z\approx$~3, from disk-dominated to
bulge-dominated systems
\citep[e.g.,][]{grog05xhost,pier07morph,2010MNRAS.408..139P,gabo09xmorph,geor09xmorph,cist11agn,cisternas2014,kocevski2012,bohm2013,fan2014}.
See Fig.~\ref{kocevski} for a comparison of the morphologies of AGNs
and mass-matched galaxies at $1.5<z<2.5$. The first studies found that
AGNs typically reside in more bulge-dominated systems than the coeval
galaxy population \citep[e.g.,][]{grog05xhost,pier07morph}. However,
as with the CMD analyses, clear differences mostly disappeared when
the galaxy sample was matched in mass to the AGN sample
\citep[e.g.,][]{kocevski2012,bohm2013,fan2014,villforth2014}; we note
that recent evidence from a systematic morphological analysis of AGNs
and mass-matched galaxies over the broad redshift range of
$z=$~0.5--2.5 has found evidence that AGNs at $z\approx$~1 are
preferentially hosted in disk-dominated galaxies when compared to the
galaxy population, although the differences are comparatively subtle
\citep[][]{2014arXiv1409.5122R}. More significant morphological
differences are found by the present day, with {\it Swift}-BAT AGNs
$\approx$~2 times more likely to reside in spiral (disk dominated)
galaxies than comparably massive inactive galaxies
\citep[e.g.,][]{koss11bathost}, suggesting that not all host-galaxy
environments are capable of hosting significant AGN activity by
$z\approx$~0.

\begin{figure}
\centerline{
  \includegraphics[width=0.9\textwidth]{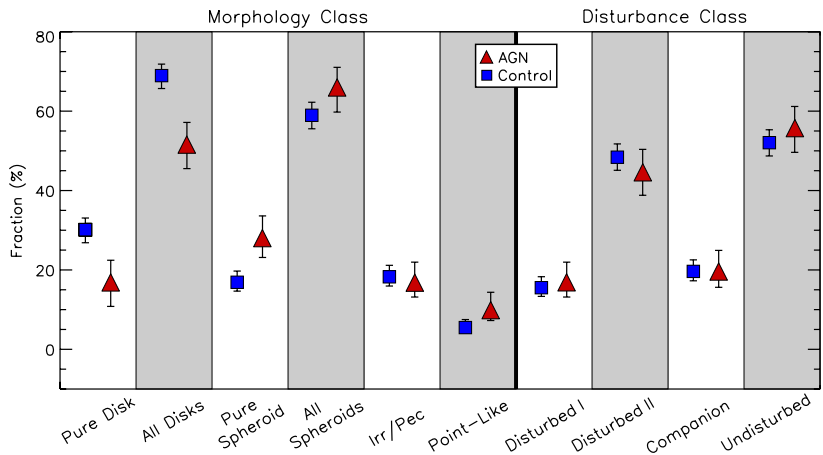}}
\caption{Fraction of AGN host galaxies (AGN: red triangles) and
  non-AGN host galaxies (control: blue squares) at $1.5<z<2.5$ in a
  given morphological and disturbance class. The non-AGN sample is
  matched in mass to the AGN host-galaxy sample. In terms of both
  morphology and disturbance classifications the AGN and the non-AGN
  sample appear very similar. Taken from \citet[][]{kocevski2012}.}
\label{kocevski}       
\end{figure}

How is the AGN activity triggered? Prior to the launches of
\chandra\ and \xmm, it was widely predicted that distant AGNs are
triggered by gas-rich major mergers, violent events where two
similar-mass galaxies interact and merge, torquing the gas and driving
it toward the central SMBH, where it can be accreted
\citep[e.g.,][]{sanders1988,barn92merge,1996ApJ...464..641M,hopkins2008}. Contrary
to these expectations, only a minority ($\simlt20$\%) of the AGN
population over the broad redshift range of $z\approx$~0.2--2.5 were
found to have the clear signatures expected for major mergers
(disturbed host-galaxy morphologies and tidal tails;
\citealp[][]{silv11agninter_aph,cist11agn,kocevski2012,bohm2013,villforth2014}).
The fraction of {\it Swift}-BAT AGNs at $z<0.05$ with major-merger
signatures is also $\approx$~20\%, consistent with that found for the
distant AGNs \citep[][]{koss10batagn,cotini2013}. These results,
therefore, suggest that other processes such as galaxy interactions,
minor mergers, and secular processes (e.g., galaxy bars, disk
instabilities, and clumpy cloud accretion) may be responsible for
triggering the majority of AGN activity
\citep[e.g.,][]{silv11agninter_aph,cist11agn,cisternas2014,bournaud2012,kocevski2012,bohm2013,cheung2014,trump2014,villforth2014}.
However, we note that there is evidence that the most luminous AGNs
are preferentially triggered by major mergers, as also found for the
most powerful star-forming galaxies
\citep[e.g.,][]{treister2012,kartaltepe2012}, which could indicate
that major mergers are required to drive sufficient quantities of gas
into the central regions of galaxies to power the most luminous
systems.

In general, the fraction of the coeval galaxy population with major
merger signatures is comparable to that found for the (majority of)
distant AGNs, when the galaxy sample is matched in mass to the AGN
sample; however, see \citet[][]{2014arXiv1409.5122R} for tentative
evidence of a factor $\approx$~2 decrease in the fraction of
$z=$~0.5--1.0 galaxies with major-merger signatures when compared to
the coeval AGN population. More significant differences are found by
the present day, with a $\approx$~5 times lower fraction of the galaxy
population hosted in major mergers when compared to {\it Swift}-BAT
selected AGNs ($\approx$~4\% vs. $\approx$~20\%;
\citealp[][]{cotini2013}). The emerging picture, therefore, suggests
that, while the absolute fraction of moderate-luminosity AGNs with
major merger signatures is comparatively constant at $\approx$~20\%
over the broad redshift range of $z\approx$~0--3, there are
significant differences between the major-merger fraction of the AGN
and coeval galaxy populations by $z\approx$~0. As we discuss in
Section~\ref{star-formation}, these differences may be related to the
greater availability of a cold-gas supply in the distant universe when
compared to the local universe (i.e.,\ the ubiquity of gas-rich
galaxies may be a key factor).

\subsection{AGN fraction and Eddington-ratio distribution}
\label{agn-fraction}

On the basis of the suite of studies explored in the previous
sections, there are few significant differences between the
host-galaxy properties of distant \xray\ AGNs and the coeval galaxy
population, when the samples are matched in mass. More significant
differences are evident by $z\approx$~0 and some differences may
already be in place by $z\approx$~1. However, to first order, these
results suggest that any galaxy with similar properties to the AGN
host galaxies is also capable of hosting an AGN and, therefore, the
fraction of galaxies with AGN activity provides a basic measurement of
the AGN duty cycle (i.e.,\ how often mass accretion onto the SMBH
switches on and off).


As revealed from a large number of studies to date, the fraction of
galaxies hosting AGN activity above a fixed \xray\ luminosity
threshold rises steeply with host-galaxy mass out to at least
$z\approx$~2--3
\citep[e.g.,][]{bund08quench,brus09xhost,xue10xhost,georgakakis2011,bluck2011,aird2012,bongiorno2012,mullaney2012ms}. See
Fig.~\ref{xue} for the AGN fraction as a function of stellar mass
across different redshift and \xray\ luminosity ranges. For example,
the fraction of $z\approx$~1 galaxies hosting AGN activity with
$L_{\rm X}>10^{43}$~erg~s$^{-1}$ increases from $\approx$~0.3\% at a
stellar mass of $\approx10^{10}$~$M_{\odot}$ to $\approx$~3\% at a
stellar mass of $\approx10^{11}$~$M_{\odot}$; the AGN fraction
globally increases by a further factor of $\approx 5$ for a lower
luminosity threshold of \hbox{$L_{\rm X}>10^{42}$~erg~s$^{-1}$}. The
AGN fraction does not appear to have a strong dependence on
host-galaxy color \citep[e.g.,][]{xue10xhost,georgakakis2011},
although it is a strong function of SFR; see
Section~\ref{star-formation}. There are no comparably detailed
analyses for \xray\ AGNs in the nearby universe; however, on the basis
of optically selected AGNs from the SDSS
\citep[e.g.,][]{2003MNRAS.346.1055K,best05}, the AGN fraction is found
to rise with increasing stellar mass but then flattens out at masses
of $\simgt10^{11}$~$M_{\odot}$. No statistically significant evidence
for a flattening in the \xray\ AGN fraction at high stellar masses is
found for distant AGNs, leaving some uncertainty over whether this
result is specific to optically selected AGNs or whether a significant
decrease in the duty cycle of \xray\ AGN activity has occurred in the
most massive galaxies at $z<0.2$ that is not seen at $z>0.2$. It is
also possible that the observational signature for a decrease in the
duty cycle of \xray\ AGN activity at high stellar masses is too subtle
to identify in the current studies.

\begin{figure}
\centerline{
  \includegraphics[width=1.0\textwidth]{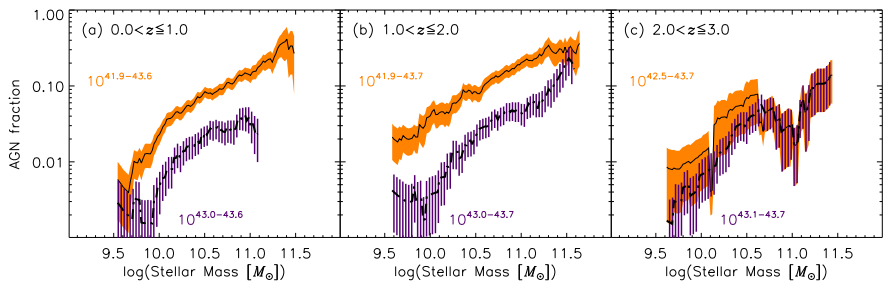}}
\caption{The fraction of galaxies hosting an AGN as a function of
  stellar mass for different redshift and \xray\ luminosity ranges [as
    labeled; luminosity ranges in log(erg~s$^{-1}$)]. A strong
  stellar-mass dependence on the AGN fraction is found for all
  redshifts. Adapted from \citet[][]{xue10xhost}.}
\label{xue}       
\end{figure}

Overall these results suggest that AGNs are more common in massive
galaxies than less massive galaxies, suggesting that the SMBHs in
massive galaxies are growing more rapidly than the SMBHs in less
massive galaxies. However, we must be careful in our physical
interpretation of this result as there are strong biases against
detecting AGNs {\it at a fixed luminosity threshold} in lower-mass
galaxies than higher-mass galaxies since a lower-mass SMBH would need
to be accreting at a higher Eddington ratio than a higher-mass SMBH to
be detected.\footnote{Here it is assumed that higher-mass galaxies
  have more massive SMBHs than lower-mass galaxies, which is
  reasonable since (1) there is a broad relationship between
  host-galaxy mass and SMBH mass and (2) any evolution in the
  stellar--SMBH mass relationship with redshift appears to be modest
  \citep[e.g.,][]{jahn09bhgal,2011ApJ...742..107B,2013ApJ...767...13S}.}
To directly explore whether there is a mass dependence to the growth
rates of SMBHs requires measuring the Eddington ratios of SMBHs across
a broad range in mass.

The calculation of an Eddington ratio (referred to here as
$\lambda_{\rm Edd}$) relies on knowing a number of uncertain
quantities, including the SMBH mass and the bolometric AGN luminosity,
which typically has to be indirectly estimated from a
single-wavelength measurement of the AGN luminosity (e.g., from the
\xray\ band, which represents only a few percent of the bolometric
luminosity;
\citealp[][]{elvi94,marconi2004,hopk07qlf,hao2014}).\footnote{Ideally
  the bolometric luminosity would be directly measured from the
  primary AGN continuum over the optical--X-ray waveband. However, it
  is expected to peak at far-UV wavelengths, which is unobservable due
  to absorption from the Galaxy. See, for example,
  \cite{vasudevan2009} and \cite{2012MNRAS.420.1825J} for some
  observational approaches to estimating the primary AGN continuum of
  nearby AGNs.}  Consequently, it is challenging to derive accurate
Eddington ratios, particularly for large numbers of distant AGNs where
the majority of the sources lack direct SMBH mass measurements and
indirect methods are required to provide mass constraints (e.g., a
proxy for the SMBH mass such as the host-galaxy mass, luminosity,
velocity dispersion, or bulge luminosity). To remove the uncertainty
on the SMBH mass, an approach adopted by some researchers is to
calculate the ``specific accretion rate'', where the SMBH mass is
replaced with the stellar mass, which is a more directly measured
quantity \citep[e.g.,][]{brus09xhost,aird2012,bongiorno2012}. In this
review we will refer to Eddington ratios but we caution that this is
not a directly measured quantity and any Eddington-ratio measurements
are subject to considerable (factor of a few) systematic
uncertainties.

The Eddington ratios estimated for distant \xray\ AGNs cover a broad
range: $\lambda_{\rm Edd}\approx10^{-5}$--1 for implied SMBH masses of
$\approx10^6$--$10^9$~$M_{\odot}$, with the majority at $\lambda_{\rm
  Edd}\approx10^{-4}$--$10^{-1}$
\citep[e.g.,][]{babic2007,ballo2007,rovi07xhost,alonsoherrero2008,hickox2009,raim10nh,simmons2011,trump2011,lusso2012,nobuta2012,matsuoka2013}. The
distribution of Eddington ratios implied from these studies is, at
least partially, dictated by limitations in the data (i.e.,
incompleteness effects such as the \xray\ sensitivity limits and the
\xray\ luminosity threshold adopted to identify AGN activity in these
studies) and, consequently, does not provide a reliable measurement of
the intrinsic Eddington-ratio distribution. The first studies to
correct for these limitations and construct an intrinsic
Eddington-ratio distribution revealed striking results: the
Eddington-ratio distribution can be characterized by a power law with
a slope that is independent of {\it both} host-galaxy mass and
redshift out to $z\approx$~2--3
(\citealp[][]{aird2012,2013ApJ...775...41A,bongiorno2012}); however,
we note that the current data would also be consistent with a broad
log-normal distribution, if the peak of the distribution lies below
current sensitivity limits. The slope of the power law is steep and
corresponds to a $\approx$~5--10 increase in the probability to detect
an SMBH that is accreting at a $\approx$~10 times lower Eddington
ratio. See Fig.~\ref{aird} for the derived Eddington-ratio
distribution for a range of stellar masses at $0.2<z<1.0$. A power-law
Eddington-ratio distribution is qualitatively consistent with the
latest constraints for distant optically selected quasars
\citep[e.g.,][]{2010A&A...516A..87S,shen2012,2013ApJ...764...45K};
however, more quantitative comparisons are limited by the different
approaches adopted in estimating SMBH masses between the
\xray\ studies (scaled from the stellar mass) and the optical quasar
studies (virial SMBH mass) and the different AGN selection
approaches. No comparably detailed measurements of the Eddington-ratio
distribution have been produced for \xray\ AGNs in the nearby
universe. Nevertheless, on the basis of optically selected AGNs from
the SDSS, \citet[][]{kauf09modes} have argued that there are two
regimes of growth in nearby AGNs: for systems with significant star
formation the Eddington-ratio distribution is characterized by a broad
log-normal distribution while for quiescent systems with little or no
star formation the Eddington-ratio distribution is characterized by a
power-law distribution. Further work is required to understand the
differences between these results and those found for the distant
\xray\ AGNs.

\begin{figure}
\centerline{
  \includegraphics[width=0.95\textwidth]{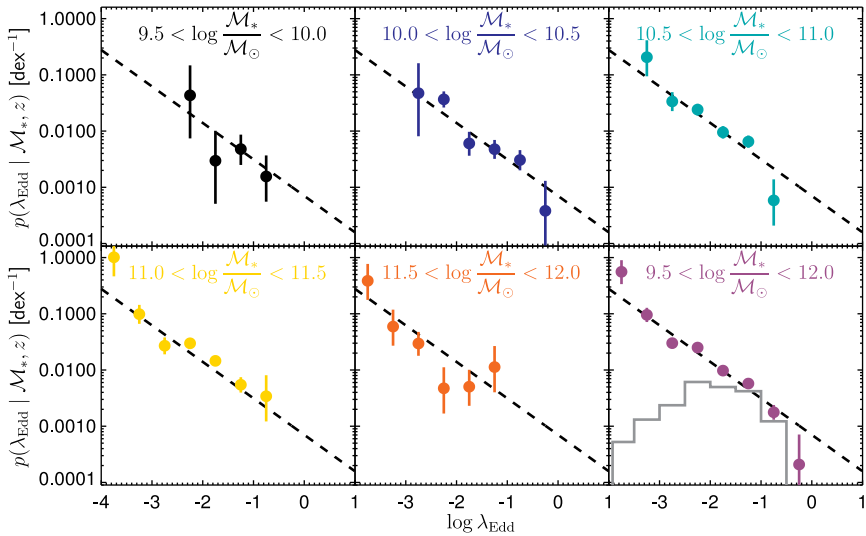}}
\caption{Eddington-ratio distribution (i.e., the probability that a
  galaxy will host an AGN at a given Eddington ratio) for different
  ranges in stellar mass for galaxies at $0.2<z<1.0$. The data have
  been corrected for \xray\ sensitivity incompleteness effects to
  provide an accurate measurement of the intrinsic Eddington-ratio
  distribution. The final panel compares the observed Eddington-ratio
  distribution (gray histogram) to the intrinsic Eddington-ratio
  distribution measured across the full range in mass and demonstrates
  the need to account for the effects of incompleteness in the
  \xray\ data. The dashed line is a best-fit power-law model to the
  Eddington-ratio distribution, evaluated at $z=$~0.6. The
  Eddington-ratio distribution is consistent with being independent of
  stellar mass over $9.5<log(M_{\rm *}/M_{\odot})<12.0$ at
  $0.2<z<1.0$. Taken from \citet[][]{aird2012}.}
\label{aird}       
\end{figure}

The fraction of galaxies hosting AGN activity at a given Eddington
ratio is found to increase with redshift out to at least
$z\approx$~2--3 \citep[e.g.,][]{aird2012,bongiorno2012}; i.e.,\ the
normalization of the Eddington-ratio distribution increases with
redshift. The current constraints suggest that this redshift
dependence is strong [$\approx(1+z)^{3.5-4.0}$;
  \citealp[][]{aird2012,bongiorno2012}], indicating that AGN activity
was about an order of magnitude more common in galaxies at
$z\approx$~1 for a fixed Eddington ratio than at $z\approx$~0;
however, it is not clear whether this evolution is independent of mass
across all redshifts. This strong redshift evolution is consistent
with the evolution in SFR found for star-forming galaxies (see
Section~\ref{star-formation}).

What insight do these results provide on ``AGN downsizing'' (see
Section~\ref{lum-funcs})? The strong increase in the AGN fraction with
redshift can be explained by either (1) an increase in the duty cycle
of AGN activity at a given Eddington ratio or (2) an increase in the
characteristic Eddington ratio with redshift. The majority of the
theoretical models predict the second scenario as being a major driver
of AGN downsizing (see Section~\ref{lum-funcs}); however, since the
observed Eddington-ratio distribution is a power law it is not
possible to distinguish between these different scenarios with the
current data (i.e., there are no distinctive features to measure a
shift in the characteristic Eddington ratio). Many of the theoretical
models also predict a redshift dependence in the characteristic mass
of accreting SMBHs, which is not clearly observed but may be due to
limitations in the current data (see
Section~\ref{question-downsizing}).

\subsection{Star formation and specific star formation rates}
\label{star-formation}

The suite of studies explored in this review have revealed when,
where, and how SMBHs have grown in the distant universe. However, we
have not yet investigated how the growth of the SMBH relates to the
growth of the host galaxy; i.e., the connection between AGN activity
and star formation. We would expect at least a broad connection
between AGN activity and star formation since (1) the volume averaged
star formation and SMBH mass accretion rates track each other (with a
3--4 orders of magnitude offset) out to at least $z\approx$~2
\citep[e.g.,][]{heckman2004,merl04bhsf,silverman2008,aird2010,mullaney2012volavg}
and (2) there is a tight relationship between the SMBH mass and spheroid mass
for galaxies in the nearby universe
\citep[e.g.,][]{korm95,mago98,gebh00,kormendy2013}, which provides
``archaeological'' evidence for past joint SMBH--galaxy
growth. However, these results only afford broad integrated
constraints on the overall SMBH--galaxy growth and do not provide
clear clues on how the SMBH and galaxy have grown in individual
systems, which requires more direct SFR measurements.

The first studies to constrain the SFRs of distant \xray\ AGNs used a
broad variety of star-formation indicators, including optical
spectroscopy, MIR data, and submillimeter (submm) observations
\citep[e.g.,][]{2005Natur.434..738A,poll07agnsed,silv09xcosmos_env,tric09agnsf,mull10agnsf,lutz10agnsf,xue10xhost,2011ApJ...742....3R}. These
initial studies showed that AGNs of a fixed \xray\ luminosity can have
a broad range of SFRs (up to $\approx$~5 orders of magnitude variation
between individual sources; see Fig.~14 in
\citealp[][]{2011ApJ...742....3R}) and provided evidence that the
average SFRs of AGNs increase with redshift. However, significant
uncertainties remained in the SFR measurements, such as (1) potential
contamination from AGN activity to the SFR estimates, (2) potential
underestimation of the SFR due to obscuration by dust, and (3)
uncertain extrapolation from the observed wavelength to the total
SFR. These issues are best addressed by FIR observations, which trace
the peak of the star-formation emission with less contamination from
AGN activity; see Section~\ref{broad-band-emission}. Consequently, the
launch of the \herschel\ observatory in 2009, the first observatory
with high sensitivity at FIR wavelengths (six photometric bands over
$\approx$~70--500~$\mu$m; \citealp[][]{2010A&A...518L...1P}), offered
the potential to make the first accurate SFR measurements of distant
AGNs (see \citealp[][]{2014ARA&A..52..373L} for a recent review of
\herschel\ survey results).

On the basis of a large number of studies using \herschel\ data, it is
now abundantly clear that the average SFRs of \xray\ AGNs increase
strongly with redshift out to at least $z\approx$~3
\citep[e.g.,][]{shao10agnsf,harrison2012,mullaney2012ms,rosario2012,rosario2013colors,rovilos2012,2012A&A...540A.109S},
confirming the trend found from the first SFR studies. See
Fig.~\ref{sfr}a for an example of the average SFR as a function of
redshift and AGN luminosity. The majority of the studies used stacking
analyses to provide average FIR constraints since only a modest
fraction of the \xray\ AGNs are detected by \herschel; however, these
results are also broadly reproduced using more detailed SED fitting
analyses taking account of photometric upper limits and calculating
average constraints using survival analysis techniques (F.~Stanley
et~al. in preparation).\footnote{In stacking analyses, the images of a
  selected source population are combined (i.e., stacked) and the flux
  is measured from the combined image, providing a constraint on the
  average flux of the source population. An advantage of stacking is
  that individually undetected sources can be included in the analysis
  and average constraints can even be placed on the fluxes of source
  populations when none of the sources are individually detected.} In
general, most studies showed no clear evidence for any luminosity
dependence on the average SFR for moderate-luminosity AGNs ($L_{\rm
  X}=10^{42}$--$10^{44}$~erg~s$^{-1}$) and the average SFR was found
to be broadly constant over this luminosity range, at any given
redshift; however, we note that some \xray\ luminosity dependence on
the average SFR is often seen at $z\simlt$~1 and is most prominent at
$z\approx$~0 (see Fig.~\ref{sfr}).

By contrast, a broad range of results have been found for
high-luminosity AGNs ($L_{\rm X}\simgt10^{44}$~erg~s$^{-1}$) at
$z\simgt$~1, with researchers arguing that either the average SFR
increases with both redshift and \xray\ luminosity, increases only
with redshift (following the trend seen for the moderate-luminosity
AGNs), or decreases with \xray\ luminosity
\citep[e.g.,][]{harrison2012,2012Natur.485..213P,rosario2012,rosario2013colors,rovilos2012,2012A&A...540A.109S}. Some
of the variation between the different results for high-luminosity
AGNs is due to two practical factors: (1) the SFRs of luminous AGNs
are more difficult to measure reliably since the AGN can contribute
significantly to the FIR emission and (2) luminous AGNs are less
common than moderate-luminosity AGNs, limiting the statistical power
of studies restricted to small-area fields. Indeed, the studies
performed in large-area fields with good source statistics all find
that the average SFR of luminous AGNs is either constant with
\xray\ luminosity (extending the trend seen for the
moderate-luminosity AGNs) or rises with \xray\ luminosity, with the
change from a rising trend to a flat trend found to be a function of
redshift
(\citealp[e.g.,][]{harrison2012,rosario2012,rosario2013colors},
F.~Stanley et~al. in preparation); see Fig.~\ref{sfr}.

\begin{figure}
\centerline{
  \includegraphics[width=0.48\textwidth]{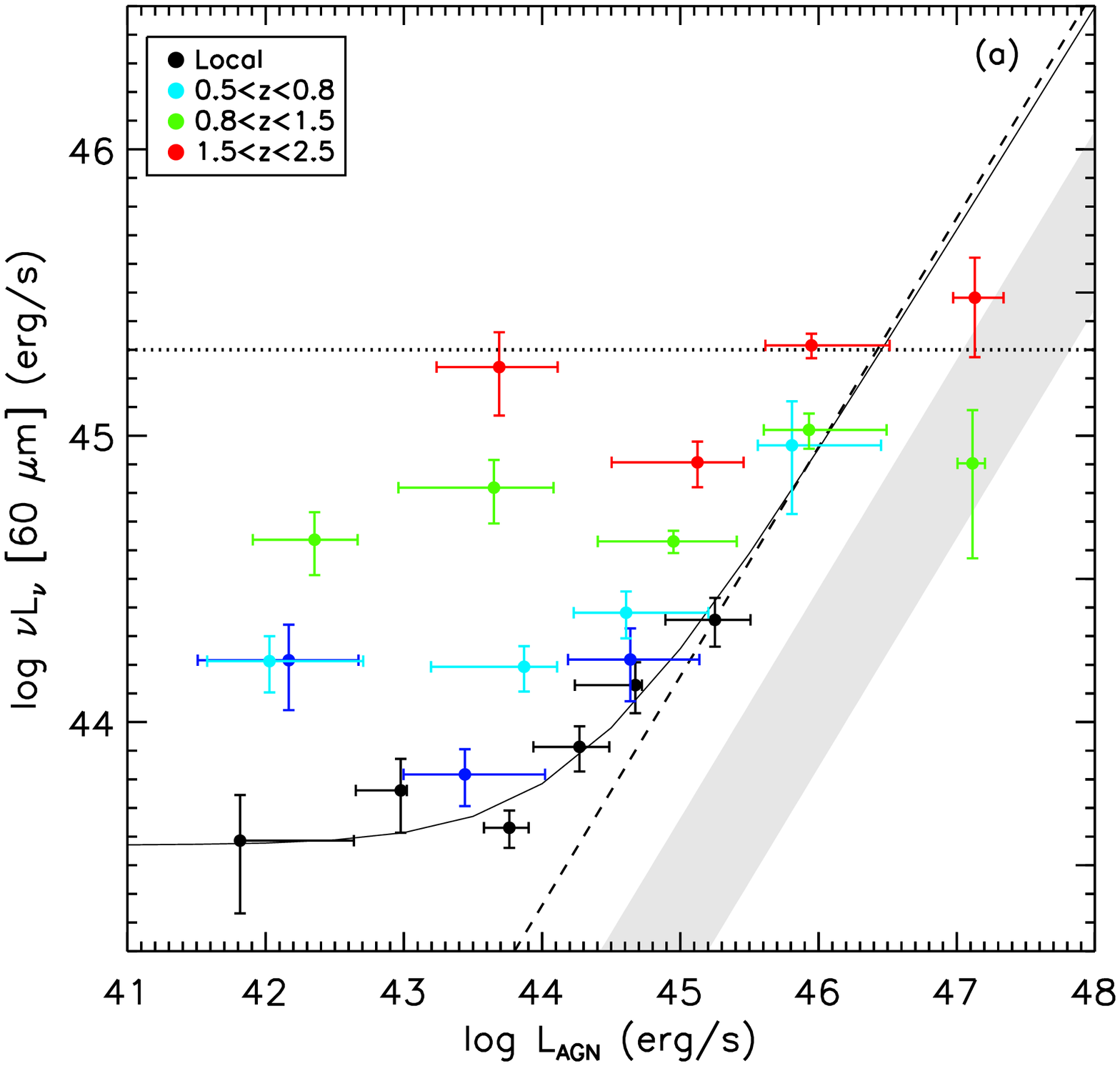}
  \includegraphics[width=0.5\textwidth]{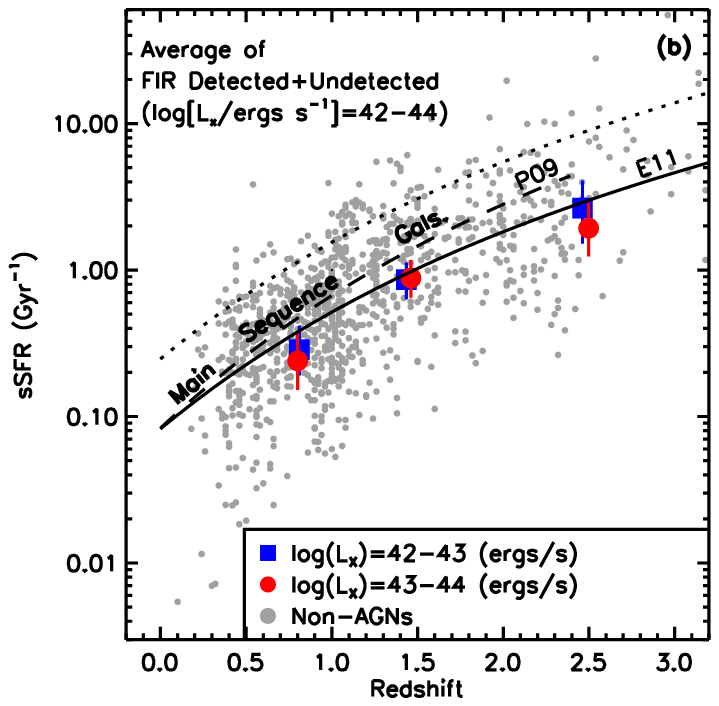}}
\caption{(a) Median 60~$\mu$m (FIR) luminosity vs. AGN luminosity for
  \xray\ AGNs over $z\approx$~0--2.5 (as labeled), showing the
  observed relationship between AGN activity and star formation. The
  solid curve shows the two-component functional fit to the local AGN
  data from {\it Swift}-BAT, the dotted line is the expected FIR
  luminosity for typical star-forming galaxies at $z=2$ (see also
  panel b), the dashed line is the AGN--star formation luminosity
  relationship for AGN-dominated systems from
  \citet[][]{2009MNRAS.399.1907N}, and the shaded region corresponds
  to the estimated 1~$\sigma$ range for an AGN SED (see Section 3.1 of
  \citealp[][]{rosario2012}). $L_{\rm AGN}$ corresponds to the
  bolometric AGN luminosity: $L_{\rm 2-10 keV}=10^{42}$~erg~s$^{-1}$
  and $L_{\rm 2-10 keV}=10^{44}$~erg~s$^{-1}$ correspond to $L_{\rm
    AGN}=5.7\times10^{42}$~erg~s$^{-1}$ and $L_{\rm
    AGN}=3.4\times10^{45}$~erg~s$^{-1}$, respectively; (b) Average
  sSFR vs. redshift for \xray\ AGNs with $L_{\rm
    X}=10^{42}$--$10^{44}$~erg~s$^{-1}$ (as labeled) over
  $z=$~0.5--3. The AGNs are compared to FIR-detected star-forming
  galaxies not hosting AGN activity (non AGNs) and the tracks trace
  the evolution in sSFR found for star-forming galaxies with redshift,
  as defined by \citet[][]{2009ApJ...698L.116P} and
  \citet[][]{elbaz2011}. Overall, the \xray\ AGNs broadly trace the
  evolution in SFR and sSFR found for the star-forming galaxy
  population. However, the observed relationship between the AGN and
  star-formation luminosity is complex and is probably, at least
  partially, driven by the different timescales of stability between
  star formation and AGN activity; see
  Section~\ref{star-formation}. Adapted from
  \citet[][]{mullaney2012ms} and \citet[][]{rosario2012}.}
\label{sfr}       
\end{figure}

The strong increase in the average SFR for the \xray\ AGNs with
redshift tracks the increase seen in the overall star-forming galaxy
population [\hbox{$\approx(1+z)^4$};
  \citealp[e.g.,][]{daddi2007ms,noes07,rodighiero2010,elbaz2011}]. The
increase in SFR with redshift for star-forming galaxies is found to be
independent of galaxy mass, such that the specific star formation rate
(sSFR; the ratio of stellar mass to SFR) evolves strongly with
redshift across all stellar masses, and is thought to be driven by the
availability of a cold-gas supply (i.e., the distant galaxies are more
gas rich than the nearby galaxies;
\citealp[e.g.,][]{dadd10fgas,genzel2010,tacconi2013}). The tightness
of the sSFR at any given redshift and the lack of a strong mass
dependence is often referred to as the ``main sequence'' of star
formation. The average SFRs and sSFRs of the \xray\ AGNs are in good
quantitative agreement with those found for the star-forming galaxy
population
\citep[e.g.,][]{xue10xhost,2011A&A...535A..80M,mullaney2012ms,rosario2012,rosario2013colors},
suggesting that the same factors that drive star formation also drive
AGN activity. See Fig.~\ref{sfr}b for a comparison of sSFR between
\xray\ AGNs and star-forming galaxies over $z=$~0.5--3. This
connection between AGN activity and star formation is further
strengthened by the following additional results:

\begin{enumerate}

\item The average SFRs of distant \xray\ AGNs are systematically
  higher than the {\it overall} galaxy population, which includes
  quiescent galaxies in addition to star-forming galaxies
  \citep[e.g.,][]{2012A&A...540A.109S,2013MNRAS.433.1015S,2014MNRAS.441.1059V},
  demonstrating that AGN activity is more closely connected to
  star-forming galaxies than the general galaxy population. We note
  that AGNs are also detected in quiescent galaxies but they appear to
  comprise a minority of the population
  \citep[e.g.,][]{azadi2014,2014MNRAS.440..339G}.

\item The fraction of galaxies hosting \xray\ AGN activity at a fixed
  luminosity threshold increases strongly with SFR
  \citep[e.g.,][]{2010ApJ...709..572K,syme10spitzer,xue10xhost,2011ApJ...742....3R,juneau2013}.

\item The SFR is found to correlate tightly with AGN luminosity when
  averaged over all galaxies in a cosmological volume, irrespective of
  whether an AGN is detected or not (e.g., when calculated as a
  function of galaxy mass or SFR;
  \citealp[e.g.,][]{2011ApJ...742....3R,mullaney2012volavg,chen2013}).

\item The host galaxies of \xray\ AGNs at $z\simlt$~1 appear to,
  preferentially, have younger stellar populations than the coeval
  galaxy population
  \citep[e.g.,][]{silv09xcosmos_env,2010MNRAS.408..139P,rosario2013resolvedcols,2014MNRAS.443.3538H};
  see Section~\ref{host-masses}.

\item The strong increase in the fraction of galaxies hosting AGN
  activity at a fixed Eddington ratio is consistent with the increase
  in the average SFR with redshift [$\approx(1+z)^{4}$;
    \citealp[e.g.,][]{aird2012,bongiorno2012}; see
    Section~\ref{agn-fraction}], suggesting that the duty cycle of AGN
  activity increases in accordance with the cosmological increase in
  SFR.

\end{enumerate}

These results imply a general connection between AGN activity and star
formation. However, the observed relationship between AGN and
star-formation luminosity does not clearly indicate a {\it
correlation}, particularly at moderate AGN luminosities where the
average SFR is flat across at least 2 orders of magnitude in AGN
luminosity; see Fig.~\ref{sfr}a. As shown by several models
\citep[e.g.,][]{2013MNRAS.434..606G,hickox2014,2014MNRAS.437.3373N,2014MNRAS.443.1125T},
these apparently uncorrelated data can be explained by differences in
the timescales of stability between star formation and AGN activity:
star formation is assumed to be relatively stable over long periods
(of order $\approx$~100~Myr) while AGNs are assumed to vary
significantly on short timescales ($\simlt$~Myrs). As an example, the
\citet[][]{hickox2014} model assumes that the long-term average of AGN
activity and star formation is constant (i.e., when averaged over many
duty cycles of mass accretion) but allows the {\it observed}
\xray\ luminosity to vary by orders of magnitude on short timescales
on the basis of an assumed Eddington-ratio distribution. This model,
and the other models referenced above, reproduce the broad trends seen
between \xray\ luminosity and average SFR and demonstrate how
short-term variations can significantly disguise an underlying
correlation over longer timescales. It, therefore, seems likely that the
general hypothesis that AGN activity varies significantly on short
timescales while the star formation is comparatively stable is broadly
correct. However, it is not yet clear which model provides the best
physical description of the observed trends between \xray\ luminosity
and average SFR, and further observational diagnostics beyond a simple
average SFR (e.g.,\ the distribution of SFRs as a function of
\xray\ luminosity, the fraction of quiescent galaxies hosting AGNs as
a function of \xray\ luminosity, the stellar-mass dependence) are
required to provide greater diagnostic power.

We finally conclude our discussion of the connection between AGN
activity and star formation by considering whether AGNs have
significant impact on the star formation in the host galaxy (e.g.,\ by
driving energetic winds, outflows, and jets, commonly referred to as
``AGN feedback''; see
\citealp[][]{2005ARA&A..43..769V,alexander2012,2012ARA&A..50..455F}
for reviews). On the basis of many AGN-feedback models we would expect
a decrease (i.e.,\ a suppression) in the SFR of \xray\ AGNs when
compared to the overall galaxy population, particularly at the highest
\xray\ luminosities. However, depending on how quickly the star
formation is suppressed, the signatures of suppressed star formation
could be comparatively subtle, particularly when averaged over AGN
populations (e.g.,\ as a function of \xray\ luminosity; see Section~4
of \citealp[][]{harrison2012} for a further discussion of some
potential limitations). More sensitive SFR constraints for individual
systems, in addition to the measurement of sensitive SFR distributions
as a function of, e.g., \xray\ luminosity and redshift, are required
to provide more sensitive tests of the impact of AGN activity on star
formation.

\subsection{The cosmic balance of power: SMBH mass accretion vs. stellar radiation}
\label{cosmic-power}

How much does SMBH mass accretion contribute to the balance of power
in the cosmos? As mentioned in Section~\ref{main-surveys} and
\ref{agns-missed}, AGNs dominate the CXRB. However, the CXRB only
comprises a minority of the overall cosmic background radiation, which
has broadly equal contributions from the cosmic UV--optical background
and the cosmic infrared background (CIRB;
\citealp[][]{2001ARA&A..39..249H}), when the strongly dominant cosmic
microwave background is excluded (i.e., the relic emission from the
Big Bang; \citealp[e.g.,][]{1965ApJ...142..419P}).

The cosmic UV--optical background is dominated by stellar emission
\citep[e.g.,][]{1992ApJ...389L...1M,2001ApJ...550L.137T,2010ApJ...712..238F}
but, prior to the launch of \chandra\ and \xmm, it was predicted that
AGNs may contribute several tens of percent of the CIRB
\citep[e.g.,][]{1999MNRAS.305L..59A,1999MNRAS.303L..34F,1999astro.ph..9089G}. These
models typically assumed that powerful obscured quasars would
contribute significantly to the CIRB through the re-radiation of the
dust-obscured AGN emission at infrared wavelengths. However, direct
observational measurements from the AGNs detected in the \chandra\ and
\xmm\ surveys showed that the contribution to the CIRB from AGNs is
much more modest at $\approx$~5--10\% (even before subtracting
contributions from star formation to the infrared emission from the
\xray\ AGN host galaxies), with the majority of the CIRB produced by
star formation
(\citealp[e.g.,][]{2002A&A...384..848E,2002A&A...383..838F,2004MNRAS.355..973S,treister2006,2007ApJ...660..988B}). Therefore,
the cosmic background emission since the formation of galaxies is
dominated by stellar radiation processes rather than mass accretion
onto SMBHs. Part of the origin for the discrepancy between the
original predictions and the observations is the assumed AGN evolution
(i.e., the original models did not anticipate the ``AGN downsizing''
results; see Section~\ref{lum-funcs}) and the assumed contribution
from Compton-thick AGNs to the CIRB, a component that is still
relatively poorly constrained but is unlikely to fundamentally change
these broad conclusions.



\section{Future prospects}
\label{future}

In this review we have highlighted how cosmic \xray\ surveys of
distant AGNs have provided key insight into the demographics, physics,
and ecology of growing SMBHs. Thanks to the revolutionary
\xray\ facilities of \chandra, \xmm, and \nustar, we now have a
dramatically improved picture of how SMBHs grew through cosmic time,
their accretion and obscuration physics, and the connection between
SMBHs, their host galaxies and the larger scale environment. However,
despite these great advances, many fundamental questions remain
unanswered. In this final section we focus on some of the key
remaining big questions and discuss how current and future facilities
can be used to address them over the coming decades.

\subsection{Remaining big questions}
\label{questions}

\subsubsection{What drives AGN downsizing?}
\label{question-downsizing}

Excellent progress has been made over the past decade in measuring the
space density and luminosity density evolution of \xray\ AGNs, and it
is now clear that lower-luminosity AGNs peaked at lower redshifts than
higher-luminosity AGNs, often referred to as ``AGN downsizing''; see
Section~\ref{lum-funcs} and Fig.~\ref{fig-ueda2014}. However, what is
not yet clear from the observational data are the factors that drive
this luminosity dependence. Theoretical models broadly predict that
this behavior is driven by a decrease in the characteristic Eddington
ratio and/or a decrease in the characteristic active SMBH mass with
decreasing redshift. The current observational constraints
\citep[e.g.,][]{aird2012,bongiorno2012} suggest a decrease in the
characteristic Eddington ratio with decreasing redshift but they do
not find any clear SMBH mass dependences (modulo that the stellar mass
is often used as a proxy for the SMBH mass in these studies); see
Section~\ref{agn-fraction}. However, on the basis of optical studies,
at $z\simlt$~0.2 there is clear evidence for a strong mass dependence
on the volume-average growth rates of SMBHs, where the growth times of
the most massive SMBHs are orders of magnitude longer than those of
lower-mass SMBHs, indicating that the most massive SMBHs must have
been growing more rapidly in the past
\citep[e.g.,][]{heckman2004}. Testing this result with \hbox{X-rays}
is important to verify that the mass dependence of SMBH growth is not
specific to optical studies and will require sensitive \xray\ data and
good source statistics at both $z\simlt$~0.2 and
$z\simgt$~0.2. Connecting our picture of SMBH growth at $z\simlt$~0.2
to the emerging picture of SMBH growth at $z\simgt$~0.2 will be key to
a greater understanding of AGN downsizing and the cosmological growth
of SMBHs.

\subsubsection{How did the first SMBHs form and grow?}
\label{question-firstsmbh}

There are several theoretical models for the formation of the first
SMBHs (remnants of population III stars; direct collapse of primordial
gas clouds; BH merging in dense stellar clusters; \citealp[e.g.,
  see][]{volo10bhform} for a recent review), which predict different
initial SMBH masses (typically referred to as ``seed'' SMBHs) of
$\approx$~100--100,000~$M_{\odot}$. Since the mass of a SMBH dictates
the maximum luminosity that can be produced through accretion (i.e.,
the Eddington luminosity), high-redshift XLFs can help distinguish
between these different formation scenarios and, from the evolution of
the XLFs, constrain the growth of the first SMBHs. The small bias
against absorption makes \hbox{X-rays} a powerful probe of accretion
in the early growth of SMBHs (e.g.,\ rest-frame 14--70~keV at $z>6$
for observed-frame 2--10~keV), particularly since the early SMBH
growth phases were likely to be heavily absorbed [as may be expected
  given the gas-rich environment and small physical sizes of the first
  SMBHs (M.~Volonteri et~al. in preparation) and implied by the
  redshift-dependence of \xray\ absorption; see
  Section~\ref{obscuration}]. One clear observational signature of
early SMBH growth is ``cosmic upsizing''
\citep[e.g.,][]{ueda2014}, which would be revealed by a change in the
relative ratio between the number density of high-luminosity and
lower-luminosity AGNs (i.e.,\ a change in the shape of the XLF) at
$z\simgt$~3--4; see Section~\ref{lum-funcs}. To model accurately the
early growth of SMBHs it will be necessary to construct XLFs in
several redshift bins down to moderate luminosities over
$z\approx$~4--8, requiring excellent high-redshift source
statistics. The current XLF constraints are weak at $z\simgt$~4, which
is due to the intrinsic rarity of such AGNs as well as source
identification challenges (see
Section~\ref{new-surveys}). Constraining the evolution of the
highest-redshift AGNs is a primary goal of several future
\xray\ observatories (see Section~\ref{new-xrays}), although deep
\chandra\ surveys over large areas, in combination with ambitious
targeted follow-up programs (see Section~\ref{followup}), can better
constrain the faint end of the $z\simgt$~4 XLF.

\subsubsection{How many obscured AGNs are missed in the current \xray\ surveys?}
\label{question-missedagns}

X-ray surveys arguably provide the most efficient selection of
obscured AGNs, particularly at high energies and at high redshifts
where the rest-frame energies allow for penetration of large absorbing
column densities (up to $N_{\rm H}\approx10^{24}$~cm$^{-2}$). However,
many of the most highly obscured AGNs will be missed in the current
\xray\ surveys, restricting our census of the overall AGN population;
see Section~\ref{agns-missed}. The identification of the most highly
obscured AGNs could be more than just a ``book keeping'' exercise
since they may reside in qualitatively different environments than
less obscured AGNs (e.g., in galaxy major mergers and the most intense
starbursts, where there is potentially more gas to obscure the AGN)
and, therefore, may evolve differently, possibly modifying results on
the fraction of obscured AGNs with redshift and luminosity; see
Section~\ref{obscuration}. The majority of the current studies are
based around {\it Chandra} and \xmm\ surveys, but significant progress
will be made using \nustar, where the higher-energy sensitivity
provides a ``cleaner'' selection of AGNs with less absorption bias,
particularly at $z\simlt$~1 where the rest-frame energies probed by
\chandra\ and \xmm\ are modest. On longer timescales \athena\ will
also provide greatly improved identification and characterization of
heavily obscured AGNs from \xray\ spectral fitting (i.e., accurate
$N_{\rm H}$ and reflection measurements; see Section~\ref{new-xrays}).
Multiwavelength observations will also allow for the identification of
\xray\ undetected AGNs that produce luminous emission at other
wavelengths (e.g., at infrared, radio, and optical wavelengths, when
the contaminating emission from the host galaxy is reliably accounted
for using SED decomposition and/or spectroscopy) and will become very
powerful when the {\it James Webb Space Telescope (JWST)\/}, the 
successor to {\it HST\/}, is launched (see Section~\ref{followup}).

\subsubsection{What causes the dependence of \aox\ on optical/UV luminosity, and
why are there intrinsically \xray\ weak outliers?}

Although the basic dependence of \aox\ upon optical/UV luminosity has
been known for about three decades, recent studies have substantially
improved measurements of the form of this relation (see
Section~\ref{aox-results}).  This being said, further improvement is
still needed since the quantitative results of some recent studies
disagree by considerably more than their statistical
uncertainties. Future work must aim to reduce and realistically assess
the inevitable systematic errors that enter such analyses (e.g.,\ AGN
variability effects, detection-fraction effects, absorption effects,
host-galaxy light contamination, and AGN misclassification). The
effects of additional physical parameters, particularly Eddington
ratio, also need better investigation. Furthermore, the few
outstanding claims for a significant dependence of \aox\ upon redshift
need checking given the general consensus against a measurable
redshift dependence (e.g.,\ via re-analysis of the original data used
to claim redshift dependence). At the same time as the observational
situation is advanced, improved numerical simulations of the SMBH
disk-corona system are required so that expectations for the behavior
of the \xray-to-optical/UV SED, including the basic cause of the
\aox-$\log (L_{2500~\mathring{\rm{A}}})$ relation, can be derived from
first-principles physics.

Additionally, outliers from the \aox-$\log
(L_{2500~\mathring{\rm{A}}})$ relation that appear to be intrinsically
\xray\ weak need further investigation (see
Section~\ref{agns-missed}). While these objects seem sufficiently rare
that they should not affect AGN demographic studies materially, they
may nevertheless provide novel insights
(cf. \citealp[][]{eddington1922}) into the SMBH disk-corona system and
emission-line regions. For example, some \xray\ weak outliers may be
systems with extremely high Eddington ratios where radiation-trapping
effects largely prevent \xray\ emission from escaping the accretion
flow. Alternatively, perhaps these outliers are not truly
intrinsically \xray\ weak, and we have simply been tricked by a
complex absorption scenario that is not yet properly understood or
appreciated.

\subsubsection{What host-galaxy environments are conducive to AGN activity?} 
\label{question-hosts}

A somewhat surprising result is how inconspicuous the host galaxies of
distant \xray\ AGNs are when compared to galaxies not hosting AGN
activity. There are no clear differences in the host-galaxy colors and
morphologies (broad-scale galactic structure and galaxy
merger/interaction signatures) of \xray\ AGNs and galaxies when
matched in stellar mass (see Sections~\ref{host-masses} and
\ref{host-morphs}), at least at $z\simgt$~1 (there is evidence for
some differences at $z\simlt$~1). The star-formation properties of
distant \xray\ AGNs also appear to be broadly similar to those of
distant star-forming galaxies and elevated when compared to the
overall galaxy population (see Section~\ref{star-formation}),
suggesting a general connection between AGN activity and star
formation. Clearer differences between the host galaxies of
\xray\ AGNs and galaxies are more evident by the present day, when it
appears that \xray\ AGNs reside in a subset of the overall galaxy
population. However, it is not yet clear when and how these
differences arose, primarily because no study has yet uniformly
measured the host-galaxy properties of both nearby and distant
systems, as required to remove any potential differences between the
methods used in the broad suite of studies published to date. From the
point of view of the star-formation properties, deeper infrared--mm
data are required to provide more powerful tests than a comparison of
average SFRs (see Section~\ref{followup-infrared}); for example, a
comparison of the SFR distributions between AGNs and galaxies and
improved measurements on the fraction of AGNs in quiescent galaxies.


\subsubsection{How does large-scale environment affect AGN activity?}
\label{question-environment}

Large-scale environment appears to have a significant affect on AGN
activity, as demonstrated, for example, from the increased fraction
(decreased fraction) of galaxies hosting AGN activity in protoclusters
(rich galaxy clusters) versus the field; see
Section~\ref{environmental}. The availability of a cold-gas supply in
the $\simlt$~pc vicinity of the SMBH is presumably the essential
requirement for mass accretion but these results suggest that the gas
availability may be controlled by the large-scale environment. For
example, there is no lack of gas in the most massive galaxy clusters,
but it is mostly in a hot form rather than in the cold form that can be
easily accreted by SMBHs. Protoclusters and rich galaxy clusters
represent the most extreme high-density environments but to provide a
fully coherent picture of the role of large-scale environment on the
growth of SMBHs requires measurements of AGN activity across all
environments as a function of redshift (e.g., voids; field; groups;
poor clusters; rich clusters; protoclusters). To achieve this aim
requires \xray\ observations covering a large enough cosmic volume to
detect significant numbers of \xray\ sources across the full range of
large-scale environments down to \xray\ sensitivity limits sufficient
to detect the majority of the SMBH growth. With sufficient data it
would then be possible to construct XLFs and explore the host-galaxy
properties as a function of environment and redshift. Improved
measurements of the clustering of AGNs as a function of physical
parameters (e.g.,\ AGN type, AGN luminosity, redshift) will further
reveal how the dark-matter halo affects the growth of SMBHs,
particularly for HoD analyses, which require excellent source
statistics to more accurately constrain the fractions of \xray\ AGNs
in the satellite and central galaxies within halos (see
Section~\ref{environmental}). Good progress can be made with current
facilities (see Section~\ref{new-surveys}) while the large FOVs of
future planned and proposed \xray\ observatories will make them
ideally suited to addressing this key question (see
Section~\ref{new-xrays}).

\subsection{Additional targeted \xray\ surveys with operating missions}
\label{new-surveys}

The \chandra, \xmm, and \nustar\ observatories are performing well
and, subject to funding considerations, have the capability to
undertake productive observations of the cosmos for at least the next
$\approx$~5~yr. Having operated successfully over the last
$\approx$~15~yr, \chandra\ and \xmm\ have already covered much of the
accessible flux--solid angle plane, from deep pencil-beam surveys to
shallow wide-area surveys; see Fig.~\ref{fig-omega-depth}. Considering
all of the \chandra\ and \xmm\ surveys listed in Table~1, there are an
estimated $\approx$~500,000 unique \xray\ sources detected over
$\approx$~1,000~deg$^2$ of the sky. However, the source statistics and
areal coverage are strongly dominated by the serendipitous surveys,
which are non contiguous and typically have limited spectroscopic
coverage; see Section~\ref{followup} for future large-area/all sky
multiwavelength survey plans. The blank-field surveys comprise of
order $\approx$~45,000 unique \xray\ sources detected over
$\approx$~80~deg$^2$, of which about half are from the on-going, and
currently unpublished, \xmm\ XXL survey.

The current suite of \chandra\ and \xmm\ surveys with good
spectroscopic completeness have covered a broad swathe of the $L_{\rm
  X}$--$z$ plane (e.g., see Fig.~3 of \citealp[][]{ueda2014}). The
majority of the detected sources lie at $z\approx$~0.3--4 and have
$L_{\rm X}\approx10^{43}$--$10^{45}$~erg~s$^{-1}$. The excellent
source statistics over this redshift range allow for the accurate
construction of XLFs in discrete redshift ranges and reliable
inferences about the overall source properties in discrete bins across
the $L_{\rm X}$--$z$ plane. However, only a modest number of AGNs are
detected at $z<0.3$ and $z>4$ in the current \chandra\ and
\xmm\ surveys ($\simlt$~50; e.g., \citealp[][]{kalf2014,vito2014}),
which is at least partially due to the small cosmological volumes
probed at these redshifts; for example, the predicted yield of $z<0.3$
($z>4$) AGNs with 2--10~keV luminosities of
$\simgt10^{43}$~erg~s$^{-1}$ ($\simgt10^{44}$~erg~s$^{-1}$) is
$\approx$~2--3~deg$^{-2}$ ($\approx$~10--65~deg$^{-2}$, depending on
whether there is a high-redshift space-density decline) down to
2--10~keV fluxes of $\approx10^{-14}$~erg~cm$^{-2}$~s$^{-1}$
($\approx10^{-15}$~erg~cm$^{-2}$~s$^{-1}$).\footnote{The \xray\ source
  density predictions are based on the \citet[][]{gilli2007} model for
  AGNs with $N_{\rm H}=10^{20}$--$10^{24}$~cm$^{-2}$; see
  http://www.bo.astro.it/$\sim$gilli/counts.html} The modest number of
AGNs at $z<0.3$ from the \chandra\ and \xmm\ surveys is partially
mitigated by the {\it ASCA} medium sensitivity survey
\citep[e.g.,][]{2003ApJS..148..275A} and the high-energy all-sky
\xray\ surveys, such as {\it Swift}-BAT
\citep[e.g.,][]{tuel10bat,2013ApJS..207...19B} and will be greatly
bolstered by the \xmm\ XXL survey, in addition to the {\it Chandra}
and \xmm\ serendipitous surveys and \erosita\ (see
Section~\ref{new-xrays}). At $z>4$, a major challenge in addition to
the relative rarity of high-redshift AGNs, is obtaining a reliable
redshift constraint, particularly for moderate-luminosity AGNs where
the majority are likely to be optically faint (i.e., $R\simgt$~25
mag). However, improving the source statistics of $z>4$ and $z<0.3$
AGNs is a worthwhile endeavor since $z>4$ corresponds to $<$~1.5~Gyr
after the Big Bang and the era of the first growth and formation of
SMBHs (see Section~\ref{question-firstsmbh}) while the $z=$~0.0--0.3
redshift range corresponds to 1/4 of cosmic history and an epoch that
connects the evolution of AGNs from higher redshifts to the present
day.

Despite the effective exploration of parameter space from the current
\chandra\ and \xmm\ surveys, an area of inquiry that is still
relatively unexplored is the role of environment in the growth of
SMBHs; see Sections~\ref{environmental} and
\ref{question-environment}.  Current results suggest that the
large-scale environment has a significant effect on the growth of
SMBHs. However, the current \chandra\ and \xmm\ surveys have not yet
had the combination of both area and sensitivity sufficient to detect
the majority of the growth of SMBHs (i.e., a factor of $\approx$~10
below the knee of the XLF, $L_{\rm *}$) across the full range of
large-scale structure environments; based on cosmological simulations,
regions of $\simgt$~4~deg$^2$ are required to map out the largest
structures \citep[e.g.,][]{2005Natur.435..629S}. The \xmm\ XXL survey
covers sufficient survey volume ($\approx$~50~deg$^2$ in two fields)
but is only sensitive to the most-luminous AGNs, close to the knee of
the XLF, while the \chandra\ and \xmm\ observations of COSMOS have
sufficient sensitivity but only cover $\approx$~2~deg$^2$ and,
therefore, have a comparatively poor sampling of all large-scale
structure environments. The richest galaxy clusters will typically be
too rare to be included in a blank-field survey of $\approx$~4~deg$^2$
but analyses of the pointed observations of galaxy clusters
\citep[e.g.,][]{2009ApJ...701...66M,2013ApJ...768....1M} could be
included to trace the most extreme large-scale structure
environments. To map out the large-scale structures requires large
contiguous fields and, therefore, this is not a scientific project that
could be undertaken by the serendipitous surveys.

\nustar\ has only been operating for $\approx$~2 yr and, with a
comparatively small FOV, has covered a more limited swathe of the
flux--solid angle plane than \chandra\ and \xmm. The current source
statistics from the combination of all of the {\it NuSTAR} surveys
(see Table~1) are good at intermediate \xray\ fluxes [$\simgt$~100
  sources with 8--24 keV fluxes of
  $(5-50)\times10^{-14}$~erg~cm$^{-2}$~s$^{-1}$], allowing for
statistically significant inferences about the source populations in
this flux range. However, $<10$ sources are detected at both faint and
bright \xray\ fluxes (8--24~keV fluxes of
\hbox{$<5\times10^{-14}$~erg~cm$^{-2}$~s$^{-1}$} and
$>5\times10^{-13}$~erg~cm$^{-2}$~s$^{-1}$), restricting the
reliability of any inferences about the properties of these source
populations. Further \nustar\ surveys with comparable exposure times
to the deepest current surveys in addition to a shallow
\nustar\ survey will improve the source statistics at both the faint
and bright \xray\ flux ends, and also bridge the (approximate) order
of magnitude difference in flux between the brightest {\it NuSTAR}
survey sources and the faintest {\it Swift}-BAT survey sources.

\subsection{Multiwavelength follow-up observations of \xray\ surveys}
\label{followup}


\begin{figure}
\centerline{
  \includegraphics[width=0.7\textwidth]{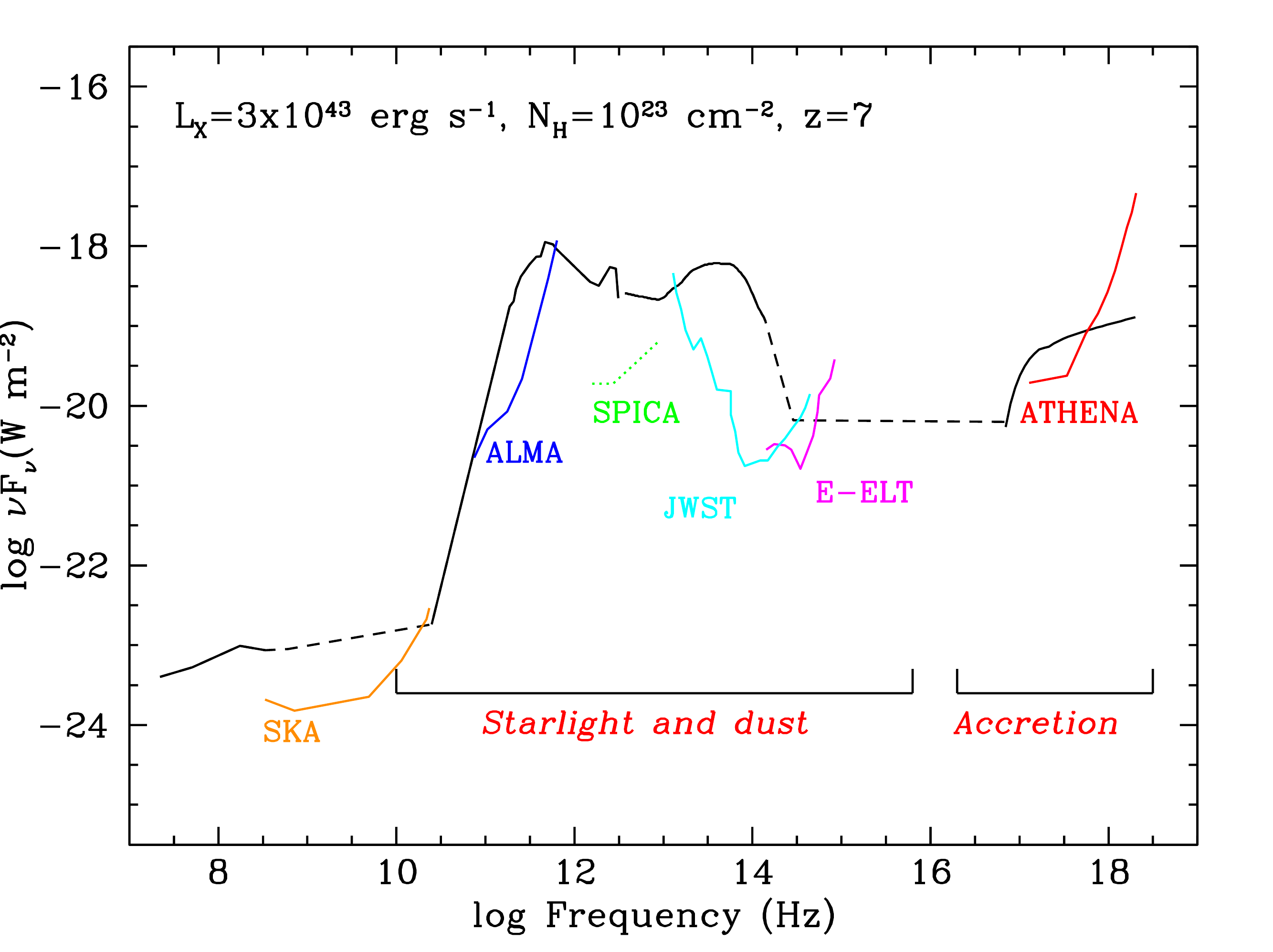}}
\caption{Multiwavelength SED of a $z=7$ obscured AGN with $L_{\rm
    X}=3\times10^{43}$~erg~s$^{-1}$ and $N_{\rm H}=10^{23}$~cm$^{-2}$,
  based on the template SED constructed by \citet[][]{lusso2011}. The
  relative sensitivities of selected future observatories are shown to
  illustrate their potential for characterizing a high-redshift AGN;
  ALMA has started operations but has not yet reached its full
  potential. The 3~$\sigma$ sensitivity limits for a $\approx$~40~ks
  exposure are plotted for all of the observatories except for
  \athena, where a 300~ks exposure was assumed. Adapted from
  \citet{2013arXiv1306.2325A}.}
\label{brusa-sed}       
\end{figure}

The heart of an \xray\ survey is of course the \xray\ data. But the
backbones are the supporting multiwavelength observations, which
provide the necessary data to measure the key properties and
environments of the detected sources and to construct XLFs. The
majority of the extragalactic \xray\ surveys listed in Table~1 are in
well-established survey fields with extensive multiwavelength
data. The available multiwavelength data are more limited for the
serendipitous surveys, which cover non-contiguous areas across the sky
and often have to rely on shallow all-sky survey data. Here we review
current and future plans for multiwavelength follow-up observations of
\xray\ surveys, focusing on both all-sky and large-area survey plans
in addition to deeper targeted multiwavelength follow-up
facilities. As a demonstration of the sensitivity of several selected
future observatories, see Fig.~\ref{brusa-sed}.

\subsubsection{Optical--MIR wavelengths}
\label{followup-optical}

At optical wavelengths, often the band pass of choice for initial
counterpart identification and spectroscopic observations of
\xray\ sources, the entire sky has been observed in three bands down
to an $R$-band magnitude of $R\approx$~21--22 (the SuperCOSMOS Sky
Survey; \citealp[e.g.,][]{2001MNRAS.326.1279H}). Recently completed
and on-going optical large-area surveys will extend the coverage down
to optical magnitudes of 23--24 in five optical bands over practically
the entire sky: for example, the SDSS
\citep[e.g.,][]{york00sdss,2011ApJS..193...29A}, VST ATLAS
\citep[e.g.,][]{2013Msngr.154...38S}, PanSTARRS
\citep[e.g.,][]{2010SPIE.7733E..0EK,2013ApJS..205...20M}, and DES
\citep[e.g.,][]{2005IJMPA..20.3121F}. From 2022--2032, LSST will
increase the optical coverage over $\approx$~18,000--30,000~deg$^2$ in
six bands down to $r\approx$~27.5 with a 10~yr survey on a dedicated
8.4~m telescope \citep[e.g.,][]{2008arXiv0805.2366I}, sufficient to
identify optical counterparts for almost all \xray\ detected sources,
including the $z\simgt$~4 AGNs required to constrain the early phases
of SMBH growth in the universe.

Broader wavelength coverage over optical--MIR wavelengths is required
to extend simple counterpart identification to the measurement of
accurate photometric redshifts and host-galaxy masses of the
\xray\ sources. In the NIR--MIR band pass, 2MASS
\citep[][]{2006AJ....131.1163S} and {\it WISE}
\citep[][]{2010AJ....140.1868W} have provided shallow to
moderate-depth data across the entire sky. While providing a good
general resource, the 1.1--2.3~$\mu$m 2MASS survey is too shallow
($K\approx$~14~mag) to detect many of the sources found in
\xray\ surveys; however, the {\it WISE} \hbox{3.4--22~$\mu$m} survey
is an excellent complement to wide-field or shallow \xray\ surveys,
particularly for the \nustar\ surveys where the majority of the
sources are bright at MIR wavelengths
\citep[e.g.,][]{alexander2013,lansbury2014}. The on-going VISTA
surveys are observing $\approx$~20,000~deg$^2$ of the Southern
hemisphere at NIR wavelengths to $\simgt$~4~mags deeper than 2MASS
(principally the VHS survey;
\citealp[e.g.,][]{2013Msngr.154...35M,sutherland2014}) and the UKIDSS
surveys have observed $\approx$~4,000~deg$^2$ of, predominantly, the
Northern hemisphere to $\simgt$~4~mags deeper than 2MASS (principally
the LAS survey; \citealp[][]{2007MNRAS.379.1599L}), with substantially
deeper data in smaller regions (e.g., in the VST ATLAS survey;
\citealp[e.g.,][]{2007Msngr.127...28A}). These surveys are
particularly useful for characterizing the properties of the
\xray\ sources detected in large-area and serendipitious
\xray\ surveys (e.g.,\ providing photometric redshifts and host-galaxy
masses), where the existing NIR coverage is comparatively shallow. As
described below, deeper optical--MIR coverage over smaller areas of
the sky will be possible over the coming decade with, for example, the
{\it Wide-Field Infrared Survey Telescope (WFIRST)}, 
{\it JWST\/}, and the operation of the first Extremely Large 
Telescopes (ELTs; telescopes with $\simgt$~30~m diameter mirrors).

The deployment of highly multiplexing optical--NIR spectrographs on
large telescopes over the next $\approx$~5 yr (e.g., Blanco-DESpec;
{\it Euclid}; Mayall-DESI; SDSS~IV-eBOSS; Subaru-PFS; VISTA-4MOST;
VLT-MOONS;
\citealp[e.g.,][]{2011arXiv1110.3193L,2012arXiv1209.2451A,2012SPIE.8446E..0TD,2013arXiv1308.0847L,2014arXiv1408.2825S,2014PASJ...66R...1T,2014SPIE.9147E..0NC})
will provide the spectroscopic complement to large-area and
serendipitious \xray\ surveys by yielding redshifts for hundreds to
thousands of \xray\ sources in an individual observation over large
FOVs (typically $\approx$~1--5~deg$^2$) down to optical magnitudes of
$\approx$~24. In addition to providing accurate source redshifts, the
spectroscopic observations allow for characterization of the
emission-line properties of the \xray\ sources (e.g.,\ identification
of broad emission lines; measuring the emission-line gas conditions
from emission-line ratios;
\citealp[e.g.,][]{1987ApJS...63..295V,2013ApJ...774L..10K}) and the
identification of the large-scale structure environment in which they
reside (e.g.,\ identifying ``filaments'' of galaxies and AGNs in
narrow redshift ranges). These analyses are particularly powerful when
using spectrographs with sensitivity out to NIR wavelengths since they
allow for the identification of rest-frame optical emission lines of
distant AGNs at $z\simgt$~1 (e.g.,\ {\it Euclid}; Subaru-PSF;
VLT-MOONS). Many ambitious observing programs are already planned
using these instruments to obtain spectroscopic redshifts for millions
of galaxies and AGNs (e.g.,\ SDSS~IV-eBOSS and 4MOST follow up of {\it
  eROSITA} sources). On slightly longer timescales (mid 2020's), {\it
  WFIRST} is expected to launch and observe $>$~1,000~deg$^2$ of the
sky down to faint NIR depths at {\it HST} resolution and obtain NIR
spectroscopic-grism redshifts for millions of distant galaxies and
AGNs out to $z\simgt$~8
\citep[e.g.,][]{2013arXiv1305.5425S,2014arXiv1411.0313G}.

A non-negligible fraction ($\simlt$~3\%) of the sources detected in
the deepest \xray\ surveys remain undetected at optical--MIR
wavelengths even with the most sensitive current observatories (e.g.,
{\it HST}; 8--10~m class telescopes; \spitzer), and a larger fraction
($\simlt$~40\%) lack spectroscopic redshifts
\citep[e.g.,][]{luo2010,xue2011,hsu2014}. Revolutionary advances in
ultra-faint imaging and spectroscopy will be made over the next
5--10~yr with the launch of {\it JWST} and the first ELTs. {\it
  JWST} \citep[][]{2006SSRv..123..485G} is a NASA--ESA--Canadian Space
Agency satellite hosting a 6.5~m telescope with high sensitivity at
0.6--28~$\mu$m and diffraction-limited spatial resolution
($\approx$~0.1--1~arcsec). {\it JWST} is planned to be launched in the
next $\approx$~5 yr and will provide NIR--MIR imaging and spectroscopy
a factor $\simgt$~10 times below current facilities, providing the
potential to achieve (near) complete spectroscopic redshifts for the
\xray\ sources.\footnote{For further details of {\it JWST}, see
  http://www.jwst.nasa.gov.} Several ELTs are planned over the next
decade, including the European ELT (E-ELT), the Giant Magellan
Telescope, and the Thirty Meter Telescope. The ELTs will complement
{\it JWST} in an analogous manner to how the current 8--10~m
telescopes complement {\it HST}, with innovative instruments and
ultra-deep imaging and spectroscopy over large FOVs.\footnote{For
  further details of the ELTs, see
  http://www.eso.org/public/teles-instr/e-elt/, http://www.gmto.org/,
  and http://www.tmt.org/.}

\subsubsection{Infrared--radio wavelengths}
\label{followup-infrared}

Long-wavelength data at MIR--submm wavelengths are required to measure
the energetics for the dust-obscured star formation and AGN components
of the \xray\ sources. \spitzer\ and \herschel\ have provided an
excellent resource at MIR--FIR wavelengths over the past decade for
the majority of the \xray\ survey fields. However, they have only
covered a comparatively small fraction of the overall sky
($\approx$~1,000~deg$^2$; see \citealp[][]{2014ARA&A..52..373L}),
which often restricts MIR--FIR studies of the \xray\ sources detected
in the serendipitious surveys to the all-sky MIR--FIR coverage. {\it
  WISE} has provided moderate-depth MIR imaging across the whole sky
(see above), although the available all-sky survey data at FIR
wavelengths ($\approx$~30--200~$\mu$m) is limited to the shallow {\it
  IRAS} and {\it AKARI} surveys, which typically only provide useful
FIR measurements for $z\simlt$~0.1 sources, severely restricting SFR
constraints for serendipitous \xray\ sources that lack
\herschel\ coverage. The only currently operating FIR telescope is
SOFIA (\citealp[e.g.,][]{2012ApJ...749L..17Y}; the effective
operational lifetimes of \spitzer\ and \herschel\ at MIR--FIR
wavelengths were limited by their helium cryogen supply), an airborne
observatory with moderate sensitivity over the broad
$\approx$~0.3--1600~$\mu$m band pass. While providing good FIR
constraints for nearby AGNs, only the brightest distant AGNs will be
detected by SOFIA and it, therefore, does not provide significant
improvements over existing SFR constraints for the \xray\ sources
detected in the serendipitious surveys.

On longer timescales ($\approx$~2025--2030), the {\it Space Infrared
  Telescope for Cosmology and Astrophysics (SPICA)} is currently the
leading concept for a next-generation MIR--FIR observatory, although
it is not yet fully funded. The primary aim of {\it SPICA}
\citep[e.g.,][]{2014SPIE.9143E..1IN,2014SPIE.9143E..1KR} is faint
medium-resolution spectroscopy over the broad 20--210~$\mu$m band
pass, which will allow for the first extensive studies of the MIR--FIR
emission lines of distant systems to provide accurate measurements of
the star-formation properties and the interstellar medium of distant
\xray\ AGNs; however, {\it SPICA} will also provide deep broadband
MIR--FIR photometry to sensitivity levels slightly below those of
\spitzer\ and \herschel, allowing for deep MIR--FIR observations
beyond the \xray\ survey regions with existing {\it Spitzer--Herschel}
coverage. Future MIR--FIR space-borne observatories such as {\it
  SPICA} are key to drive forward our understanding of the role that
distant AGNs play in the formation and evolution of galaxies.

The terrestrial atmosphere is mostly opaque at MIR--FIR wavelengths
and the majority of the MIR--FIR telescopes described above are
mounted inside space-borne observatories (SOFIA is the exception but
it is still operated from an aircraft). However, there are discrete
band passes at $\approx$~0.3--3~mm, in the submm--mm wave band, where
the atmosphere is sufficiently transparent to allow for sensitive
ground-based observations (see Fig.~4 of
\citealp[][]{2014PhR...541...45C}). The negative $K$-correction for
typical AGN and galaxy SEDs at these wavelengths means that a distant
AGN or galaxy with a given SFR has a comparable submm--mm flux across
the broad redshift range of $z\approx$~0.5--6
\citep[e.g.,][]{2002PhR...369..111B,2014PhR...541...45C}. Over the
past two decades there have been several ground-based observatories
sensitive at submm--mm wavelengths (e.g.,\ APEX; ASTE; CARMA; IRAM;
JCMT; SMA; see \citealp[][]{2014PhR...541...45C} for a recent review),
which have provided the first clear views of the submm emission
from distant AGNs and galaxies. The most recent submm--mm
observatories, which have either started operations or are planned to
start in the next $\approx$~5 yr are the Atacama Large
Millimeter/Submillimeter Array (ALMA;
\citealp[][]{2009IEEEP..97.1463W}), the Large Millimeter Telescope
(LMT; \citealp[e.g.,][]{2010SPIE.7733E..12H}), and CCAT
\citep[e.g.,][]{2012SPIE.8444E..2MW}. ALMA is an interferometer and
provides high-resolution imaging and spectroscopy down to
sensitivities over an order of magnitude deeper than previous
submm--mm facilities with sub-arcsec resolution. ALMA has the
potential to provide unprecedented insight into the star-formation
properties of distant \xray\ AGNs (e.g.,\ SFR constraints down to
those of quiescent galaxies and potentially resolving the extent of
star formation in some sources); see Fig.~\ref{brusa-sed}. Both the
LMT and CCAT aim to reach a broadly similar submm--mm sensitivity
limit as ALMA but at a lower spatial resolution over larger FOVs
(e.g.,\ up to 1~deg$^2$ for CCAT) and will provide deep submm--mm
measurements of the star-formation properties for large samples of
distant \xray\ AGN, particularly when combined with sensitive MIR--FIR
data.

Radio observations provide independent constraints on the amount of
star formation and (radio bright) AGN activity in distant
\xray\ sources, allowing for a more complete census of AGN activity
and the exploration of the connection between AGN activity and star
formation. Over the past two decades, large area \xray\ surveys and
serendipitous \xray\ surveys have largely relied on the NRAO VLA Sky
Survey (NVSS; \citealp[][]{condon1998}) and the VLA FIRST survey
(\citealp[][]{becker1995}), which cover
\hbox{$\approx$~10,000--33,000~deg$^2$} down to mJy levels at
1.4~GHz. These data are sufficient to detect radio-bright
\xray\ sources, although the majority of the \xray\ source population
is $\approx$~1--2 orders of magnitude fainter than the sensitivity of
NVSS and FIRST. However, substantially deeper radio surveys are now
being undertaken over $\simgt$~1,000~deg$^2$ and many aim to cover the
majority of the visible sky (e.g., at low $\simlt$~200~MHz
frequencies: GMRT-TGSS; LOFAR; MWA; at mid $\approx$~1~GHz
frequencies: Apertif-WODAN; ASKAP-EMU; MeerKAT-MIGHTEE; VLA-VLASS; see
\citealp[][]{2011PASA...28..215N,2013PASA...30...20N,2014PASP..126..196L}
for a recent summary). Many of these surveys are precursors and
pathfinders to the Square Kilometer Array (SKA;
\citealp[e.g.,][]{2009IEEEP..97.1482D,2014arXiv1408.5317C}), an
international project to build the world's largest radio telescope,
planned to start operations over the next $\approx$~10--20~yr. The SKA
is designed to operate over a very broad range of frequencies
($\approx$~50~MHz to 20~GHz) down to sub-$\mu$Jy levels at sub-arcsec
resolution, sufficient to detect essentially all of the \xray\ sources
at radio frequencies and provide an independent method of AGN
selection (i.e.,\ radio bright AGNs undetected at \xray\ energies) in
addition to sensitive SFR constraints; see Fig.~\ref{brusa-sed}. Due
to the high sensitivity and large FOV, SKA will be able to undertake a
near all-sky survey (3~$\pi$ steradians) to sensitivities equivalent
to the deepest current radio surveys ($\approx$~1--2~$\mu$Jy rms at
1.4~GHz; R.~P.~Norris, 2014, private communication), sufficient to
detect the majority of the \xray\ sources in the current
serendipitious and all-sky \xray\ surveys.

\subsection{New \xray\ survey missions}
\label{new-xrays}

Several new \xray\ observatories are planned in the near \hbox{($<5$~yr)} and
long \hbox{($>10$~yr)} term that will extend the great progress that {\it
  Chandra}, \xmm, and \nustar\ have made toward our
understanding of the demographics, physics, and ecology of distant
SMBHs.

In the near term, both \erosita\ and \astroh\ are expected to become
operational, with launch dates of
2015--2016. \erosita\ \citep[][]{merloni2012} is a joint
Russian--German mission and will provide imaging and spectroscopy over
$\approx$~0.5--10~keV. The primary objective of \erosita\ is to
perform a moderate-depth survey of the entire sky within the first 4
yr of launch at relatively low spatial resolution (effective
half-energy width of $\approx$~28--40~arcsec, depending on energy),
detecting $\approx$~3 million AGNs out to $z\approx$~6. The expected
sensitivities of the all-sky survey are
$\approx1\times10^{-14}$~erg~cm$^{-2}$~s$^{-1}$ (0.5--2~keV) and
$\approx2\times10^{-13}$~erg~cm$^{-2}$~s$^{-1}$ (2--10~keV) and will
be $\approx$~4 times more sensitive in the deepest regions at the
ecliptic poles; the all-sky sensitivity limits are $\approx$~20 times
deeper than {\it ROSAT} at 0.5--2~keV \citep[][]{1999A&A...349..389V}
and $\approx$~200 times deeper than HEAO~1~A-2 at 2--10~keV
\citep[][]{1982ApJ...253..485P}. \astroh\ \citep[][]{2012SPIE.8443E..1ZT}
is a joint JAXA-NASA mission and will provide imaging and spectroscopy
of cosmic \xray\ sources over the broad energy range of
$\approx$~0.3--80~keV. \astroh\ should have a comparable angular
resolution and sensitivity limit as \nustar\ at $\simgt$~3~keV but
over a broader energy range. \astroh\ also has a wide-field imaging
spectrometer and a high-resolution micro-calorimeter that will provide
spectroscopy of cosmic \xray\ sources over 0.3--12~keV at a higher
spectral resolution than \chandra, \xmm, and
\nustar\ ($\Delta{E}<7$~eV). The relatively small FOV of \astroh\ for
imaging at hard \xray\ energies ($\approx$~9~$\times$~9~arcmin; see
Table~2 of \citealp[][]{2012SPIE.8443E..1ZT}) means that it is
potentially better suited to observing the well-established survey
fields than undertaking new wide-area surveys at hard \xray\ energies.

In the longer term, many exciting \xray\ observatory concepts have
been proposed that will provide revolutionary advances in our
understanding of the properties and evolution of the sources detected
in cosmic \xray\ surveys.\footnote{For example, see the list of
  proposed \xray\ observatory concepts solicited by NASA in 2012:
  http://pcos.gsfc.nasa.gov/studies/xray/x-ray-mission-rfis.php.} We
do not have the space in this review to discuss all of these concepts
and, to date, only one has been selected for long-term financial
support (\athena; \citealp[][]{2013arXiv1306.2307N}). However, to
provide some flavor of the \xray\ facilities that may be available in
the next $\approx$~10--30~yr, we briefly describe four different
proposed concepts: \athena, \hexp, \smartx, and \wfxt.

\athena\ is an ESA-led mission that is scheduled for launch in
2028. With a large collecting area ($\approx$~2.0--2.5~m$^2$ at
1~keV), large FOV ($\approx$~40~$\times$~40~arcmin), and good spatial
resolution ($\approx$~3--5~arcsec half-energy width), \athena\ will be
an excellent general-purpose \xray\ observatory and a devastatingly
effective survey machine, achieving a given flux--solid angle limit
$\approx$~2 orders of magnitude more quickly than \chandra\ and
\xmm. The large effective area and good angular resolution combination
is achieved from innovative Silicon-pore optic technology. The
ultimate sensitivity limit of \athena\ is dictated by the confusion
limit and will be comparable to that of a $\approx$~2~Ms
\chandra\ observation [0.5--2~keV fluxes of
  $\approx$~(2--3)~$\times10^{-17}$~erg~cm$^{-2}$~s$^{-1}$;
  \citealp[e.g.,][]{alexander2003,luo2008}]. However, \athena\ will be
able to achieve that sensitivity limit in \hbox{$\approx$~200--400~ks}
and, therefore, a 2~Ms \athena\ survey would cover $\approx$~1~deg$^2$
to this depth, as compared to \chandra\ which only reaches this
sensitivity limit over the central region. The large collecting area
of \athena\ will also provide high signal-to-noise \xray\ spectroscopy
of distant \xray\ AGNs at $\approx$~100~eV resolution, allowing for
direct redshift measurements from the identification of iron K$\alpha$
emission lines, and accurate measurements of their spectral properties
(i.e., $N_{\rm H}$, $\Gamma$, reflection, intrinsic
\xray\ luminosity); a high spectral resolution microcalorimeter
(called the X-IFU) will provide $\approx$~2.5~eV resolution but over a
smaller $\approx$~5~$\times$~5~arcmin FOV. With these capabilities,
\athena\ will (among other things) efficiently identify
moderate-luminosity AGNs at $z\simgt$~6 (see Fig.~\ref{brusa-sed}),
potentially constraining the seeds of SMBHs (see
Section~\ref{question-firstsmbh}), and perform a near-complete census
of AGNs out to at least $z\approx$~3, even identifying many
Compton-thick systems from the detection of strong iron K$\alpha$
emission \citep[e.g.,][]{2013arXiv1306.2325A}.

\hexp\ (PI: F.~Harrison) is a natural successor to \nustar\ and
combines an optimized optics design at high energies (half-power
diameter resolution of $\approx$~10--15~arcsec) with a broader energy
bandpass of $\approx$~0.1--200~keV and a larger effective area than
\nustar\ and \xmm. \hexp\ aims to be $\approx$~40 times more sensitive
than \nustar, sufficient to resolve $\approx$~90\% of the CXRB at its
$\approx$~20--40~keV peak and to detect AGNs almost independent of the
presence of absorption out to $z\approx$~6. \smartx\ (PI:
A.~Vikhlinin) is a natural successor to \chandra\ and aims to use
adaptive optics to achieve excellent $\approx$~0.5~arcsec resolution
(half-power diameter) at 0.2--10~keV with $\approx$~30 times the
effective area of \chandra. \smartx\ is predicted to reach the depth
of the 4~Ms \hbox{CDF-S} survey over 5~deg$^2$ in 4~Ms of exposure or
$\approx3\times10^{-19}$~erg~cm$^{-2}$~s$^{-1}$ at 0.5--2~keV in a
single pointing of 4~Ms, sufficient to detect the first growing SMBHs
at $z\approx$~10--20 [$L_{\rm
    X}=$~(0.4--2)~$\times10^{42}$~erg~s$^{-1}$;
  \citealp[e.g.,][]{2013MmSAI..84..805V}]. \wfxt\ (PI: S.~Murray) is a
natural successor to \erosita\ and combines a large
$\approx$~1~deg$^2$ FOV with good angular resolution of
$\approx$~5~arcsec (half-energy width) to provide mapping of large
areas of the sky to faint flux limits at $\approx$~0.5--7~keV.
\wfxt\ is predicted to achieve a sensitivity limit comparable to that
of a 2~Ms \chandra\ observation and with a dedicated 3~yr program
would be able to survey $\approx$~100~deg$^2$ and
$\approx$~3,000~deg$^2$ to \hbox{0.5--2~keV} flux limits of
$\approx4\times10^{-17}$~erg~cm$^{-2}$~s$^{-1}$ and
$\approx5\times10^{-16}$~erg~cm$^{-2}$~s$^{-1}$, respectively,
sufficient to detect $\approx$~5~million AGNs overall and tens of AGNs
even at $z=$~8--10 \citep[e.g.,][]{murray2013}. With high
\xray\ sensitivity, large FOV (with good spatial resolution across the
full FOV), and good positional accuracy, \wfxt\ provides a great
complement to the next-generation optical imaging surveys undertaken
by, for example, LSST (see Section~\ref{followup-optical}).

\vspace{1.0cm}

Ever since the first rocket flights of the 1960's, cosmic
\xray\ surveys have been an essential tool for elucidating the
processes of mass accretion onto SMBHs. With orders of magnitude
improvements in sensitivity over previous generation \xray\ missions,
the \chandra, \xmm, and, most recently, \nustar\ observatories have
(arguably) provided greater leaps forward in our understanding of the
demographics, physics, and ecology of distant growing SMBHs than any
other facility over the past two decades. This is a subject area
strongly driven by technological advances in telescope and instrument
design, and the revolutionary developments in \xray\ and
multiwavelength facilities over the coming decades promise yet greater
advances in our understanding of when, where, and how SMBHs have grown
in the distant universe.



\begin{acknowledgements}
We thank 
J.A. Aird, 
F.E. Bauer, 
P.N. Best,
J.N. Bregman, 
M. Brightman,  
M. Brusa, 
A. Comastri, 
J.L. Donley, 
P. Gandhi,
R. Gilli,  
C.M. Harrison,
R.C. Hickox,  
L.C. Ho, 
D.D. Kocevski, 
J.H. Krolik, 
B.D. Lehmer, 
B. Luo, 
E. Lusso, 
A. Merloni, 
J.R. Mullaney,
R.P. Norris, 
D.J. Rosario,  
A.E. Scott,
O. Shemmer,
Y. Shen, 
F. Stanley,  
M. Sun, 
E. Treister, 
J.R. Trump,
Y. Ueda,
C. Vignali, 
M. Volonteri, and  
Y.Q. Xue
for feedback, helpful discussions, and sharing information. 
We gratefully acknowledge financial support from 
NASA ADP grant NNX10AC99G (WNB), 
Chandra X-ray Center grants AR3-14015X and G04-15130A (WNB), 
the Leverhulme Trust (DMA), and 
the Science and Technology Facilities Council (ST/I001573/1; DMA). 

\end{acknowledgements}


\bibliographystyle{spbasic}      
\bibliography{ba-refs01}   

%
%

\end{document}